\documentclass[twoside,11pt]{article}
\usepackage[abbrvbib, preprint]{jmlr2e}
\usepackage{amsmath} 
\usepackage{amsthm}
\usepackage{amssymb} 
\usepackage{bm}
\usepackage{blindtext}
\usepackage{xcolor}
\usepackage[capitalise]{cleveref}
\usepackage{tikz} 
\usetikzlibrary{shapes.geometric,arrows.meta,positioning, shapes.misc}  
\usetikzlibrary{overlay-beamer-styles,positioning,decorations.pathreplacing,calligraphy} 
\usetikzlibrary{hobby,shapes.misc} 
\usetikzlibrary{positioning} 
\usetikzlibrary{bayesnet} 
\usepackage{algorithm2e}
\usepackage{booktabs} 
\usepackage{arydshln} 
\usepackage{cancel}
\usepackage{multirow}
\usepackage{enumerate}
\usepackage{enumitem}
\usepackage{mathtools}

\newcommand{\BlackBox}{\rule{1.5ex}{1.5ex}}  
\ifdefined\proof
    \renewenvironment{proof}{\par\noindent{\bf Proof\ }}{\hfill\BlackBox\\[2mm]}
\else
    \newenvironment{proof}{\par\noindent{\bf Proof\ }}{\hfill\BlackBox\\[2mm]}
\fi
\newtheorem{theorem}{Theorem}
\newtheorem{assumption}{Assumption}
\newtheorem{lemma}[theorem]{Lemma} 
\newtheorem{proposition}[theorem]{Proposition} 
\newtheorem{remark}[theorem]{Remark}

\newtheorem{definition}[theorem]{Definition}

\theoremstyle{definition}
\newtheorem{example}{Example} 

\newcommand{\mc}{\mathcal}
\newcommand{\bb}{\mathbb}

\newcommand{\pa}{\text{pa}}

\newcommand{\id}{\mathrm{id}}
\providecommand{\argdot}{{\,\vcenter{\hbox{\tiny$\bullet$}}\,}}
\providecommand{\ind}{{\perp\!\!\!\perp}}

\providecommand{\mc}{\mathcal}
\providecommand{\bb}{\mathbb}
\providecommand{\pa}{\textrm{pa}}

\providecommand{\argdot}{{\,\vcenter{\hbox{\tiny$\bullet$}}\,}}

\providecommand{\doInt}{\textsf{do}}
\newcommand{\Aut}{\mathrm{Aut}}

\newcommand{\supp}{\text{supp}}

\crefname{proposition}{Proposition}{Propositions}
\crefname{assumption}{Assumption}{Assumptions}

\usepackage{lastpage}
\jmlrheading{23}{2025}{1-\pageref{LastPage}}{}{}{21-0000}{Hugh Dance and Benjamin Bloem-Reddy}

\ShortHeadings{Counterfactual Cocycles}{Counterfactual Cocycles}
\firstpageno{1}

\begin{document}

\title{Counterfactual Cocycles: A Framework for Robust and Coherent Counterfactual Transports}

\author{\name Hugh Dance \email hugh.dance.15@ucl.ac.uk \\
       \addr Gatsby Computational Neuroscience Unit\\
       University College London\\
       London, United Kingdom
       \AND
       \name Benjamin Bloem-Reddy \email benbr@stat.ubc.ca \\
       \addr Department of Statistics\\
       University of British Columbia\\
       Vancouver, Canada}

\editor{My editor}

\maketitle

\begin{abstract}%
Estimating joint distributions, or couplings, over counterfactual outcomes is central to personalized decision-making and treatment risk assessment. Two emergent frameworks with identifiability guarantees are: (i) bijective structural causal models (SCMs), which are flexible but brittle to mis-specified latent noise; and (ii) optimal-transport (OT) methods, which avoid latent noise assumptions but can produce incoherent counterfactual transports which fail to identify higher-order couplings. In this work, we bridge the gap with \emph{counterfactual cocycles}: a framework for counterfactual transports that uses algebraic structure to provide coherence and identifiability guarantees. Every counterfactual cocycle corresponds to an equivalence class of SCMs, however the cocycle is invariant to the latent noise distribution, enabling us to sidestep various mis-specification problems. We characterize the structure of all identifiable counterfactual cocycles; propose flexible model parameterizations; introduce a novel cocycle estimator that avoids any distributional assumptions; and derive mis-specification robustness properties of the resulting counterfactual inference method.  We demonstrate state-of-the-art performance and noise-robustness of counterfactual cocycles across synthetic benchmarks and a 401(k) eligibility study.
\end{abstract}


\section{Introduction}

In many fields such as medicine, economics and public policy, decision-makers need to predict outcomes under different actions. Common examples include estimating how a higher drug dose would affect a patient’s recovery; forecasting today’s inflation if last year’s interest rates had been higher; or inferring the effect of tax relief on poverty levels. This gap between observed data and counterfactual scenarios lies at the heart of identification in causal inference: we see outcomes under one (observational) regime, but want to know what would have happened under another (counterfactual) regime.

Over recent decades, two complementary frameworks have provided principled ways to identify causal quantities from observables.  
The \emph{potential outcomes} framework \citep{rubin1974estimating} posits latent counterfactual variables whose statistical links to observed outcomes encode causal assumptions. The \emph{causal graphical model} framework \citep{pearl2000models,spirtes2000causation} represents those assumptions in a directed acyclic graph and derives identification by \(d\)-separation and the \emph{do}-calculus. Both formalisms yield identification results for average treatment effects, marginal counterfactual distributions, and many other causal targets, by expressing them as functionals of the observed data distribution.

Yet a fundamentally harder class of causal targets remains: those that demand an explicit joint distribution across counterfactual outcomes, known as a \emph{counterfactual coupling}. To understand the distinction, consider estimating the effect of a binary treatment $X\in\{0,1\}$ on outcomes $Y:=(Y_1,\dots,Y_p)\in\mathbb{R}^p$. The \emph{average treatment effect} (ATE) is
\[
\mathrm{ATE}:=\mathbb{E}[Y(1)-Y(0)]=\mathbb{E}[Y(1)]-\mathbb{E}[Y(0)]\,.
\]
where $Y(x)$ is the potential outcome under treatment state $x \in \{0,1\}$. Since this factorizes into treated and control means, in a randomized controlled trial (RCT) it can be estimated by sample averages. Now, consider instead the \emph{treatment harm rate} (THR) \citep{shen2013treatment}, which measures the probability that treatment worsens outcomes:
\begin{align}
\text{THR} := \mathbb{P}\bigl(Y(1) \preceq Y(0)\bigr)
\;=\;\bb E\bigl [ \mathbf{1}\{Y(1) \preceq Y(0) \}\bigr] \,.\label{eq:THR}
\end{align}
Here $\mathbf{1}\{Y(1)\preceq Y(0)\}=1$ if $Y_j(1)\leq Y_j(0)$ for some index $j\in\{{1,\dots,p}\}$, and $0$ otherwise.
The THR is an important criterion for assessing whether a treatment passes the ``\emph{first, do no harm}'' principle in medical science \citep{young2015first}. In contrast to the ATE, the THR cannot be recovered from the marginal distributions $\bb P_{Y(1)}$ and $\bb P_{Y(0)}$, instead requiring an expectation under the \emph{joint distribution} $\bb P_{Y(1),Y(0)}$, which we call the \emph{counterfactual coupling}. Unfortunately, one cannot identify the counterfactual coupling from the marginals alone, as infinitely many joint distributions can admit the same marginal distributions \citep{villani2021topics}. It also cannot be directly estimated from data, since at most one of $Y(1)$, $Y(0)$ are ever observed per sample or unit. Hence, additional modeling assumptions are needed.  
The THR is just one motivation for the need for counterfactual couplings. Other situations include individualized treatment and decision‐making  \citep{imai2010general}, algorithmic fairness \citep{kusner2017counterfactual}, and other treatment risk assessments \citep{kallus2023treatment}. 

A predominant approach to identifying and estimating counterfactual couplings is to use the framework of \emph{Structural Causal Models (SCMs)} \citep{pearl2000models,peters2017elements}. In the present setting with no confounding, the idea is to posit a structural model  
\[
Y \;=_{\mathrm{a.s.}}\; f\bigl(X,\,\xi\bigr),
\quad \xi \sim \bb P_{\xi}, \quad \xi \,\ind \, X, 
\]
where \(\xi\) captures all unobserved factors affecting \(Y\) (in practice, one may augment $X$ to include measured covariates $Z$). The coupling is then characterized by $f$ and $\bb P_\xi$ as
\[
(Y(1)\, ,\, Y(0)) \; =_{\mathrm{a.s.}} \; (f(1,\xi)\,,\, f(0,\xi))\;, \quad \xi \sim \bb P_\xi \;.
\]
To ensure the pair $(f,\bb P_\xi)$ can be identified up to model-specific automorphisms, present theory requires $f(x,\argdot)$ to be \emph{bijective} \citep{xi2023indeterminacy, javaloy2023causal}. An established, state-of-the-art approach to modeling bijective SCMs is to use \emph{causal normalizing flows} (or flow-based SCMs) \citep{pawlowski2020deep, khemakhem2021causal, nasr2023counterfactual, javaloy2023causal}. This involves specifying a simple base distribution (e.g., \(\widehat{\bb P}_\xi = \mathsf{N}(0,I)\)) and learning  $f$ using flexible classes of conditional diffeomorphisms parameterized by deep neural networks \citep{papamakarios2021flows}. However, as we establish in \cref{sec:background},  this approach is brittle to the choice of base distribution. For instance, if the tails or support of this distribution are mis-specified, the true flow $f$ can be extremely complex  \citep{jaini2020tails}, may not exist \citep{cornish2020relaxing}, and the resulting estimator $\hat f$ can fail to converge (see e.g., \cref{ex:mis-spec} in \cref{sec:background:scms}). Such limitations are known in the normalizing flows literature, but existing solutions either fail to fully address the problems, or sacrifice bijectivity of $f$, losing any identifiability guarantees. 

A recent line of work has instead turned to \emph{optimal transport} (OT) methods
\citep{charpentier2023optimal,de2024transport,balakrishnan2025conservative}, appealing to the notion that counterfactual worlds should be as similar as possible while satisfying the desired change \citep{lewis1973causation}. The basic idea is to specify transports between counterfactuals---e.g., \vspace{-7pt}

\[Y(1)=T_{1,0}(Y(0))\;,\]\vspace{-15pt}

and estimate them by minimizing a transport cost while preserving the marginal distributions. This approach avoids specifying latent noise distributions and guarantees identifiability for continuous variables
\citep{villani2021topics}.
However, when there are more than two treatment values (e.g., $x \in \{0,1,2\}$), there are multiple ways to couple the counterfactuals (e.g., $Y(0) \mapsto (Y(1), Y(2))$ or $Y(0) \mapsto Y(1) \mapsto Y(2)$).
Since OT maps are not generally closed under composition (i.e., $T_{2,0}\neq T_{2,1}\!\circ T_{1,0}$), each path may induce a different coupling, leading to model incoherence and an identifiability problem (see \cref{ex:ot} in \cref{sec:bg:ot}).

\subsection{Our Contributions}

In this work, we propose a transport-based framework for modeling counterfactual couplings that is free of the noise mis-specification biases of SCMs and incoherence of OT methods. Below we summarize the main contributions and plan of the paper. 

In \cref{sec:background}, we cover the necessary background and demonstrate the limitations of SCMs and OT in detail. These limitations serve as motivation for our approach.

In \cref{sec:method}, we develop our framework in a simplified confounding-free setting. We start by asking what properties a set of transports $\{T_{x,x'}\}_{x,x' \in \bb X}$ must satisfy to be able to couple a set of counterfactuals (a.k.a. potential outcomes) under different treatment levels:\vspace{-5pt}
\begin{align}
    Y(x) =_{\mathrm{a.s.}} T_{x,x'}(Y(x')) \quad \forall x,x' \in \bb X  \;. \label{eq:cocycle_intro}
\end{align} \vspace{-18pt}

It turns out that the {necessary} and {sufficient} properties are precisely\vspace{-5pt}
\begin{align}
   T_{x,x} = \text{id} \quad  \text{(Identity)}, \quad\text{and} \quad T_{x'',x'}\circ T_{x',x} = T_{x'',x} \quad  \text{(Path Independence)}\;, \label{eq:cocyclepropsintro}
\end{align} \vspace{-18pt}

on the support of the counterfactuals, along with a marginal-matching property (\cref{thm:cocycle_sufficiency}). These properties make the function 
$T:(x,x',y)\mapsto T_{x,x'}(y)$ a \emph{cocycle} 
in dynamical systems theory \citep{arnold1998random}. We hence call \eqref{eq:cocycle_intro} a \emph{counterfactual cocycle} model.  The path independence property is the key ingredient missing from OT methods. 

In \cref{thm:cocycle_factorization}, we show that {any} counterfactual cocycle can be expressed using a family of injective functions $(f_x)_{x \in \bb X}$ with left inverses $(f_x^+)_{x \in \bb X}$, in the sense that for each $x$ and $x'$,\vspace{-7pt}

\[    T_{x,x'} = f_{x} \circ f_{x'}^+\quad  \text{with probability one under $\bb P_{Y(x')}$}\;.\]\vspace{-13pt}

This gives us a general recipe to construct flexible counterfactual cocycles: (i) specify each $f_x: \bb Y \to \bb Y$ using any parameterized bijection (e.g., a normalizing flow or invertible neural network), and (ii) set $T(x,x',y') := f_{x} \circ f_{x'}^{-1}(y')$. We provide general conditions under which cocycles are identifiable (\cref{thm:iden:group:coboundary}) and show how to specify cocycles that enforce these conditions using knowledge of the causal ordering of outcomes (\cref{thm:tmi_uniqueness}). 
We conclude by showing in \cref{thm:scm_equivalence} that, under standard potential outcomes assumptions, each counterfactual cocycle is equivalent to an SCM with independent noise:
\[Y(x) =_{\mathrm{a.s.}} f_{x} \circ f_{x'}^+(Y(x')) \quad \Longleftrightarrow \quad Y =_{\mathrm{a.s.}} f(X,\xi), \quad \xi \ind X\;,\quad \text{where $f_x = f(x,\argdot)$.}\]
Importantly, the cocycle $T(x,x',y')=f_x\circ f_{x'}^{+}(y')$ does not depend on the distribution of the noise $\bb P_\xi$. This suggests a robustness to directly modeling counterfactual cocycles. In particular, a set of bijections $\mc F$ is well-specified for the cocycle as long as there exists \emph{some} distribution $\bb P_\xi^\star$ and functions $\{\hat f_x\}_{x\in \bb X} \subseteq \mc F$ such that
$\hat f_x(\xi^\star) \sim \bb P_{Y(x)}$ for all $x$, where $\xi^\star \sim \bb P_\xi^\star$. By contrast, 
using $\mc F$ to model an SCM with \emph{fixed} base distribution $\widehat{\bb P}_\xi$  requires the same property, but the noise must have distribution $\widehat{\bb P}_\xi$. If $\widehat{\bb P}_\xi$ poorly matches the true $\bb P_\xi$, the resulting maps $\{\hat f_x\}_{x \in \bb X}$ may be highly complex or non-bijective and so may lie outside $\mc F$.

In \cref{sec:methodgeneral}, we extend counterfactual cocycle models to a more general setup that can handle confounding, and cover our high-level approach to estimating causal quantities with cocycles. The basic idea is to target a distributional discrepancy of the form
\begin{align}
  \ell(T)
  = \bb E_{x,x'\sim\bb P_X}\,
    D\bigl(\bb P_{Y(x)},\;(T_{x,x'})_{\#}\,\bb P_{Y(x')}\bigr)^2 \;,
  \label{eq:DI}
\end{align}
and use the cocycle to compute counterfactuals $\hat Y^{(i)}(x)$ of observations $\{X^{(i)}, Y^{(i)}\}_{i=1}^n$. From here, one can empirically estimate causal quantities. For instance, for the THR: \vspace{-6pt}

\[\widehat {\text{THR}} = \widehat{\bb P}(Y(1) \preceq Y(0)) = \tfrac1n\sum\nolimits_{i=1}^n \bm 1 \{\hat Y^{(i)}(1) \preceq \hat Y^{(i)}(0)\}\;.\] \vspace{-11pt}

In \cref{sec:estimation}, we derive a tractable cocycle estimation procedure and show that it is asymptotically equivalent to minimizing \eqref{eq:DI} (\cref{thm:cocycle_recovery} and \cref{prop:MMD_prob_bound}). The resulting cocycle estimators are consistent under general conditions (\cref{thm:CMMD_consistency}) and consistency is unaffected by the noise distribution of the (true) SCM. (see \cref{remark:robustness} and \cref{fig:noise_adaptiveness}). We also show $\sqrt{n}$-consistency for parametric classes under regularity conditions (\cref{thm:cmmd-rate}). 

In \cref{sec:cocycle:vs:scm}, we analyze the robustness of modeling counterfactual cocycles versus SCMs in more detail, by studying an invariance under reparameterizations. Specifically, any bijective SCM can be reparameterized using an invertible transformation $g$:\vspace{-6pt}

\[Y = f(X,\xi) = f(X, g\circ g^{-1}(\xi)) = f^{(g)}(X, \xi^{(g)})\]\vspace{-11pt}

Importantly, although the choice of $g$ affects $f^{(g)}$, the associated cocycle is unchanged:\vspace{-6pt}

\[
T_{x,x'}^{(g)}
= f_x^{(g)}\circ\bigl(f_{x'}^{(g)}\bigr)^{-1}
= f_x\circ g\circ g^{-1}\circ f_{x'}^{-1}
= T_{x,x'} \;.
\]\vspace{-11pt}

Thus, while an SCM implicitly commits to a particular $g$ through the noise parameterization, one is free to model the cocycle using the parameterization \(f^{(g^*)}\) that has the lowest functional complexity (e.g., Sobolev norm). 
As we prove in \cref{thm:min_complexity} and show in Examples \ref{ex:simplification} and \ref{ex:dependence}, this property permits using smaller model classes \(\mc F\) for the cocycle than for an SCM with a fixed base distribution, while remaining well-specified. We also show our method is robust to the dependence structure of the noise $\bb P_\xi$, unlike recently proposed quantile methods for SCMs \citep{plevcko2020fair, machado2024sequential}.

In \cref{sec:implementation}, we discuss the implementation details of counterfactual cocycle modeling. In \cref{sec:experiments}, we implement counterfactual cocycles in several simulations and a real experiment, demonstrating state-of-the-art performance and robustness to noise assumptions.

\section{Background and Limitations of Existing Methods} \label{sec:background}

To set the stage, we first present the basic setting of interest and formally introduce counterfactual couplings. We then review existing approaches to modeling couplings and discuss their limitations, which motivate our work.

\paragraph{Causal Inference and Counterfactuals} Let
\(
  \bm V := (V_1,\dots,V_d)
  \in
   \prod_j^d\bb V_j \subset \bb R^d,
\)
 be observed random variables with distribution
\(\bb P_{\bm V}\). For most of this work we assume $V = (Z,X,Y)$, where $X:= (X_1,\dots,X_q)$ are treatment variables we wish to manipulate (e.g., doses of different medications), $Y:= (Y_1,\dots,Y_p)$ are outcomes of interest (e.g., patient blood pressure, resting heart rate), and $Z:= (Z_1,\dots,Z_l)$ are pre-treatment covariates (e.g., age, gender). We denote $Y(x)$ as the {counterfactual} outcomes under a (`do') intervention that fixes the treatment level to $x$ (often denoted $\doInt(X=x)$)  \citep{pearl2000models,causalMLbook2025}. We assume the observed outcomes correspond to the counterfactuals under the received treatment level: $Y=_{\mathrm{a.s.}} Y(X)$ (i.e., the consistency property \citep{rubin1974estimating}). When dealing with an i.i.d.\ dataset of observations, we index the samples as $(\bm V^{(i)})_{i=1}^n \sim_{iid} \bb P_{\bm V}$.

\paragraph{Counterfactual Couplings}  In this work, we focus not only on the problem of identifying and estimating the marginal distribution of each counterfactual $Y(x)$, but on the harder problem of recovering a joint distribution $\pi$ over \emph{collections} of counterfactuals, e.g.,
$$\pi_I := \mc L(\{Y(x)\}_{x \in I}), \quad \text{$I$ is finite}.$$ 
Here $\mc L(\{Y(x)\}_{x \in I})$ denotes the joint law (distribution) of the variables $\{Y(x)\}_{x \in I}$. Such couplings are necessary to identify many distributional effects of treatment. For instance, in medical settings, if $X \in \{0,1\}$ is a treatment and $Y \in \bb R$ a health outcome, we may wish to quantify the extent of adverse effects caused by treatment via the THR as in \eqref{eq:THR}. 
In finance, we may wish to compute the Conditional Value at Risk (CVaR) \citep{rockafellar2002conditional} to assess the risk of a binary investment decision \citep{kallus2023treatment}:
$$\text{CVaR}_{\alpha}(Y(0) - Y(1)) := \bb E[Y(0) - Y(1) \mid Y(0) - Y(1) \geq q_{\alpha}] \;,$$
where $q_\alpha$ is the $\alpha$-quantile (VaR) of the return differential.
In economic policy, one may wish to analyze whether a reform primarily benefits those who would already be well-off under the status quo, by assessing how the policy effect varies with status-quo outcomes:
$$\mu(\alpha) := \bb E[Y(1) - Y(0)|Y(0) = q_{\alpha}] \;.$$
When $X$ is continuous, the same effect measures can be used to assess the dose effects by replacing $Y(1) - Y(0)$ with the contrast $Y(x) - Y(0)$, or the effect of the current treatment policy via the contrast $Y(X) - Y(0)$.  Note when there are multiple outcomes these quantities may be computed per dimension, or in aggregate as in \eqref{eq:THR}. 

Unfortunately, standard causal inference frameworks for identifying the marginal distribution of each $Y(x)$, such as causal graphs \citep{pearl2009causal} and potential outcomes \citep{rubin1974estimating}, cannot identify counterfactual couplings without further assumptions, since there can be infinitely many couplings $\pi$ that admit the same marginals over $\{Y(x)\}_{x \in \bb X}$. Even in the fully randomized (unconfounded) setting, we only ever observe at most one counterfactual outcome per unit, so one can never directly learn the coupling from data. 

Below we review two popular approaches for identifying counterfactual couplings and their limitations: structural causal models, and optimal transport methods.

\subsection{Structural Causal Models}\label{sec:background:scms}

Structural Causal Models (SCMs) \citep{spirtes2000causation,pearl2000models} specify the full causal mechanism on $\bm V = (V_1,\dots,V_d)$ using a set of independent (exogenous)  noise variables $\bm \xi := (\xi_1,\dots,\xi_d) \in \mc E$ with distribution $\bb P_{\bm \xi} := \prod_{j=1}^d\bb P_{\xi_j}$ and a structural map $F: \bb V \times \mc E \to \bb V$,
\begin{align}
    \bm V =_{\mathrm{a.s.}} F(\bm V, \bm \xi) \;.
\end{align}
For identifiability, one usually assumes the variables $\bm V$ admit a known \emph{causal ordering} $V_1 \prec \dots \prec V_d$. This concept, dating back to \citet{simon1953causal}, formalizes the idea that $V_i$ may cause $V_j$ but not vice versa when $i < j$. Here we assume the variables are already ordered, so no permutation is necessary. The ordering is used to restrict the SCM to be \emph{acyclic}:
\begin{align}
    V_j =_{\mathrm{a.s.}} f_j(V_{<j},\xi_{j}), \quad \forall j \in \{1,\dots,d\} \;.
\end{align}
Here $f_j: \bb V_{<j} \times \mc E_{j} \to \bb V_j$ is the $j^{th}$ coordinate of $F$ that determines how $V_j$ depends on its predecessors and the noise $\xi_{j}$. In this case we can write $\bm V = F(\bm \xi)$ with $F$ lower-triangular.

Every acyclic SCM can be associated with a causal directed acyclic 
graph (DAG) $\mc G = (\bm V,E)$, with nodes $\{V_1,\dots,V_d\}$ and edges 
$E = \{ V_i \to V_j : i<j\}$ encoding all possible direct effects consistent 
with the ordering. This \emph{maximal DAG} includes every forward edge allowed 
by the ordering. However, since each $f_j$ may not necessarily depend on all $V_1,\dots,V_{j-1}$, many sparser DAGs may also be consistent with the acyclic SCM. In practice, 
when a specific DAG is known, the SCM can be explicitly restricted to
\[
   V_j =_{\mathrm{a.s.}} f_j(V_{\pa(j)}, \xi_j) \;,
\]
where $V_{\pa(j)} \subseteq V_{<j}$ are the `parents' of $V_j$ in the causal DAG, denoted by the index function $\pa(\argdot)$ which satisfies $j \in \pa(i)$ if and only if $V_j \to V_i$ in $\mc G$
\citep{pearl2000models,peters2017elements}.

The counterfactuals under a do-intervention $\text{do}(X=x)$ for any subset $X \subset \bm V$ are determined by a modified SCM \(\bm V(x) =_{\mathrm{a.s.}} F_x(\bm \xi)\), which sets the coordinate functions for $X$ as $X(x) = x$. \cref{fig:scm-and-intervention} shows an example SCM, corresponding maximal causal DAG, and SCM modification for a four variable example where  $Z \prec X \prec Y_1 \prec Y_2$ and we set $\text{do}(X=x)$.

\usetikzlibrary{positioning}
\begin{figure}[t]
\centering
\begin{tabular}{@{}c@{\quad}c@{\quad}c@{}}

\begin{minipage}{0.28\textwidth}
\[
\begin{aligned}
Z     &= f_1(\xi_1) \\
X     &= f_2(Z,\xi_2)\\
Y_1   &= f_3(Z,X,\xi_3)\\
Y_2   &= f_4(Z,X,Y_1,\xi_4)
\end{aligned}
\]
\end{minipage}

&
\hspace{20pt}

\begin{tikzpicture}[baseline=(current bounding box.center),
>=stealth, every node/.style={draw,circle,inner sep=1pt}, node distance=1cm]
  \node (Z) at (0,0) {$Z$};
  \node (X) at (1,0) {$X$};
  \node (Y1) at (2,0.5) {$Y_1$};
  \node (Y2) at (2,-0.5) {$Y_2$};

  \draw[->] (Z) -- (X);
  \draw[->, bend left=20]  (Z) to (Y1);
  \draw[->, bend right=20] (Z) to (Y2);
  \draw[->] (X) -- (Y1);
  \draw[->] (X) -- (Y2);
  \draw[->] (Y1) -- (Y2);
\end{tikzpicture}

&
\hspace{30pt}

\begin{minipage}{0.28\textwidth}
\[
\begin{aligned}
Z & = f_1(\xi_1) \\
X(x)   &= x\\
Y_1(x) &= f_3\bigl(Z,X(x),\xi_3\bigr)\\
Y_2(x) &= f_4\bigl(Z,X(x),Y_1(x),\xi_4\bigr)
\end{aligned}
\]
\end{minipage}
\end{tabular}
\caption{Left: SCM over $(Z,X,Y_1,Y_2)$. Middle: corresponding maximal DAG. Right: modified SCM after hard intervention \(\text{do}(X=x)\). Equalities hold almost surely.}
\label{fig:scm-and-intervention}

  \vspace{-10pt}

\end{figure}

Given an SCM, statistical functionals of any counterfactuals (e.g., \((\bm V(x),\bm V(x'))\)) conditioned on any subset of observed variables $\bm W \subset \bm V$ can be estimated by the familiar three-step abduct-act-predict recipe \citep{pearl2000models}:
\begin{enumerate}[align = left]
\item[\textbf{Abduct.}]
      {Update} noise distribution to condition on evidence:  
      \(\widehat{\bb P}_{\bm\xi}\gets \bb P_{\bm\xi\mid\bm W=\bm w}\).
\item[\textbf{Act.}]
          {Modify} structural equations to $F_x$, $F_{x'}$.
\item[\textbf{Predict.}]
      {Propagate} noise:  
      \(
  \;\bb E\bigl[h\bigl(\bm V(x),\bm V(x') \bigr)\mid\bm W=\bm w\bigr]
\)
=
\(
  \bb E_{\bm\xi\sim\widehat{\bb P}_{\bm\xi}}
  \bigl[
     h\bigl(F_x(\bm\xi),F_{x'}(\bm\xi)\bigr)
  \bigr].
\)
\end{enumerate}

\paragraph{Identifiability via Bijective Causal Models} Various classes of SCMs have been proposed in recent years, primarily relying on the flexibility of deep neural networks \citep{pawlowski2020deep, sanchez2021vaca, khemakhem2021causal, geffner2022deep, nasr2023counterfactual, javaloy2023causal}. The most flexible models with identifiability guarantees for counterfactual couplings are \emph{bijective causal models (BCMs)}, which constrain each $f_j$ to be bijective on the noise $\xi_j$. If two BCMs $(F,\bb P_{\bm \xi})$, $(\tilde F, \tilde {\bb P}_{ {\bm \xi}})$ induce the same distribution $\bb P_{\bm V}$, and each structural function $f_j$ and $\tilde f_j$ is monotone increasing on $\xi_j$,
they produce the same counterfactual couplings \citep{nasr2023counterfactual}. This functional restriction subsumes many traditional SCM classes such as additive-noise models ($V_i = f(V_{\pa(i)}) + \xi_i$) and location-scale models ($V_i = f(V_{\pa(i)}) + h(V_{\pa(i)})\xi_i$). BCMs are also natural counterparts to our proposed methods, and so we focus on them within the broader SCM framework. 

A popular and state-of-the-art (SOTA) modeling approach for flexible BCMs is to use \emph{normalizing flows} \citep{khemakhem2021causal, javaloy2023causal, nasr2023counterfactual}, which specify $F$ as a sequence of invertible and differentiable transformations 
from a simple base distribution (e.g., $\widehat{\bb P}_{\bm \xi} = \mathsf{N}(0,I)$) \citep{papamakarios2021flows}. As discussed above, the causal order restricts $F$ to be lower-triangular, for which \emph{autoregressive flows} are used (\cref{tab:flow_models}). Such flows can be composed for increased expressiveness while respecting the causal ordering and enabling fast and exact maximum likelihood training. 
We refer to such BCMs as \emph{flow-based SCMs}.

While flow-based SCMs have achieved SOTA performance on causal inference tasks, they suffer from a key practical limitation: if the base distribution $\widehat{\bb P}_{\xi}$ poorly matches the target $\bb P_{\bm V}$, the required flow may be extremely complex or may not even exist. Although these problems have been recognized in the normalizing flows literature, their effects on flow-based SCMs have not been systematically studied.  Alternative estimation approaches for BCMs have been proposed based on quantile methods \citep{plevcko2020fair, machado2024sequential}. However, they too rely on a fixed noise distribution and can therefore suffer from related mis-specification problems, as discussed in \cref{sec:cocyclevsscm:seqot}.
Below we focus on two primary problems that occur in flow-based SCMs: tail and support mis-specification.

\begin{table}[t]
\centering
\caption{Autoregressive normalizing flows, their transforms, and conditions for Lipschitz continuity. $M$ is a rational-quadratic spline \citep{gregory1982piecewise}.}
\label{tab:flows}
\resizebox{\textwidth}{!}{%
\begin{tabular}{l l l}
\toprule
\textbf{Model} & \textbf{Autoregressive transform} $f_{j}(v_{<j},\xi_j)$ & \textbf{Lipschitz Condition} \\
\midrule
{NICE} \citep{dinh2014nice}           
 & $\xi_j + \mu_j(v_{<j}) 1_{k \notin [j]}$              
 & $\mu_j$ Lipschitz \\

{MAF} \citep{papamakarios2017masked} 
 & $\xi_j \mapsto \sigma_j(v_{<j}) \cdot \xi_j + (1-\sigma_j(v_{<j}))\mu_j(v_{<j})$ 
 & $\sigma_j$ bounded \\

{IAF} \citep{kingma2016improved}     
 & $\xi_j \mapsto \exp\bigl(\lambda_j(v_{<j})\bigr)\,\xi_j + \mu_j(v_{<j})$ 
 & $\lambda_j$ bounded, $\mu_j$ Lipschitz \\

{Real-NVP} \citep{dinh2016density}   
 & $\xi_j \mapsto \exp\bigl(\lambda_j(v_{<j}) 1_{k \notin [j]}\bigr)\,\xi_j + \mu_j(v_{<j}) 1_{k \notin [j]}$
 & $\lambda_j$ bounded, $\mu_j$ Lipschitz \\

{Glow} \citep{kingma2018glow}        
 & $\xi_j \mapsto \sigma_j(v_{<j})\xi_j + \mu_j(v_{<j}) 1_{k \notin [j]}$
 & $\sigma_j$ bounded, $\mu_j$ Lipschitz \\ 

{NAF} \citep{huang2018neural}        
 & $\xi_j \mapsto \sigma^{-1}(w(v_{<j}) \cdot \sigma(\sigma_j(v_{<j})\xi_j + \mu_j(v_{<j})))$
 & Always (logistic mixture CDF) \\

{NSF} \citep{durkan2019neural}       
 & $\xi_j \mapsto v_j 1_{v_j \notin [-B,B]}$+$M_j(\xi_j;v_{<j}) 1_{v_j \in [-B,B]}$
 & Always (linear outside $[-B,B]$) \\


\bottomrule
\end{tabular}
}
\label{tab:flow_models}
\end{table}

\paragraph{Tail Mis-specification} It is known that if the base distribution $\widehat{\bb P}_{\bm \xi}$ lies in a different tail class to the observational distribution $\bb P_{\bm V}$ (e.g., exponential vs.\ logarithmic tail decay), then no regular class of bi-Lipschitz flows can match the tails of $\bb P_{\bm V}$ (e.g., see Theorem 3.2 and Corollary 3.3 in \citealp{liang2022fat} and Theorem 3 in \citealp{jaini2020tails}). This is problematic, since most flows are bi-Lipschitz, in some cases by design (see \cref{tab:flow_models}).  Recent work in the normalizing flows literature has mitigated the problem using tail-adaptive base distributions \citep{jaini2020tails, liang2022fat, laszkiewicz2022marginal}. However, those approaches still impose isotropic tails per dimension and so can fail to learn distributions where e.g., some $V_j$ has a heavier right tail than left tail, which occurs frequently in finance and climate applications \citep{verhoeven2004fat, deidda2023asymmetric}. 
This mis-specification can also result in estimation pathologies  due to high-variance log-likelihood gradients \citep{amiri2024practical}. In the extreme case that $\bb E\|\bm V\|_2 = \infty$ and $\widehat{\bb P}_{\xi} = \mathsf{N}(\bm 0,I)$ is chosen, the  gradient may even be infinite and so the minimizer $F_{\theta^*_n}$ can diverge as $n \to \infty$.

\paragraph{Support Mis-specification} If there is no differentiable function that transports $\widehat{\bb P}_{\xi}$ to $\bb P_{\bm V}$, then no normalizing flow can map between them (since normalizing flows are diffeomorphisms). This arises whenever a continuous base is chosen with \(\supp(\widehat{\bb P}_{\xi})=\bb R^d\), while the true BCM noise distribution has disconnected support (e.g., \(\xi_j\sim\alpha \bb P_0+(1-\alpha)\bb P_1\))  or lies on a lower‐dimensional manifold (e.g., discrete \(\xi_j\)).  Such noise structures have been assumed in medical settings, where patients with latent discrete characteristics may respond differently to treatment \citep{yin2018assessing,loh2022evaluating}. In the normalizing flows literature, solutions to this relax the bijectivity of $F$   \citep{tanielian2020learning,cornish2020relaxing}. However, this would sacrifice any identifiability guarantees for counterfactual inference. Additionally, this mis-specification can also lead to estimation identifiability problems. The expected log‐likelihood depends only on the inverse flow and its derivatives over the support of $\bb{P}_{\bm{V}}$. As a result, if the image of $F^{-1}(\bm V)$ does not cover $\supp(\widehat{\bb P}_{\xi})$, the solution may be undetermined and different minimizers can yield different distributions.

\begin{figure}[!t]
    \centering
    \includegraphics[width=0.24\linewidth]{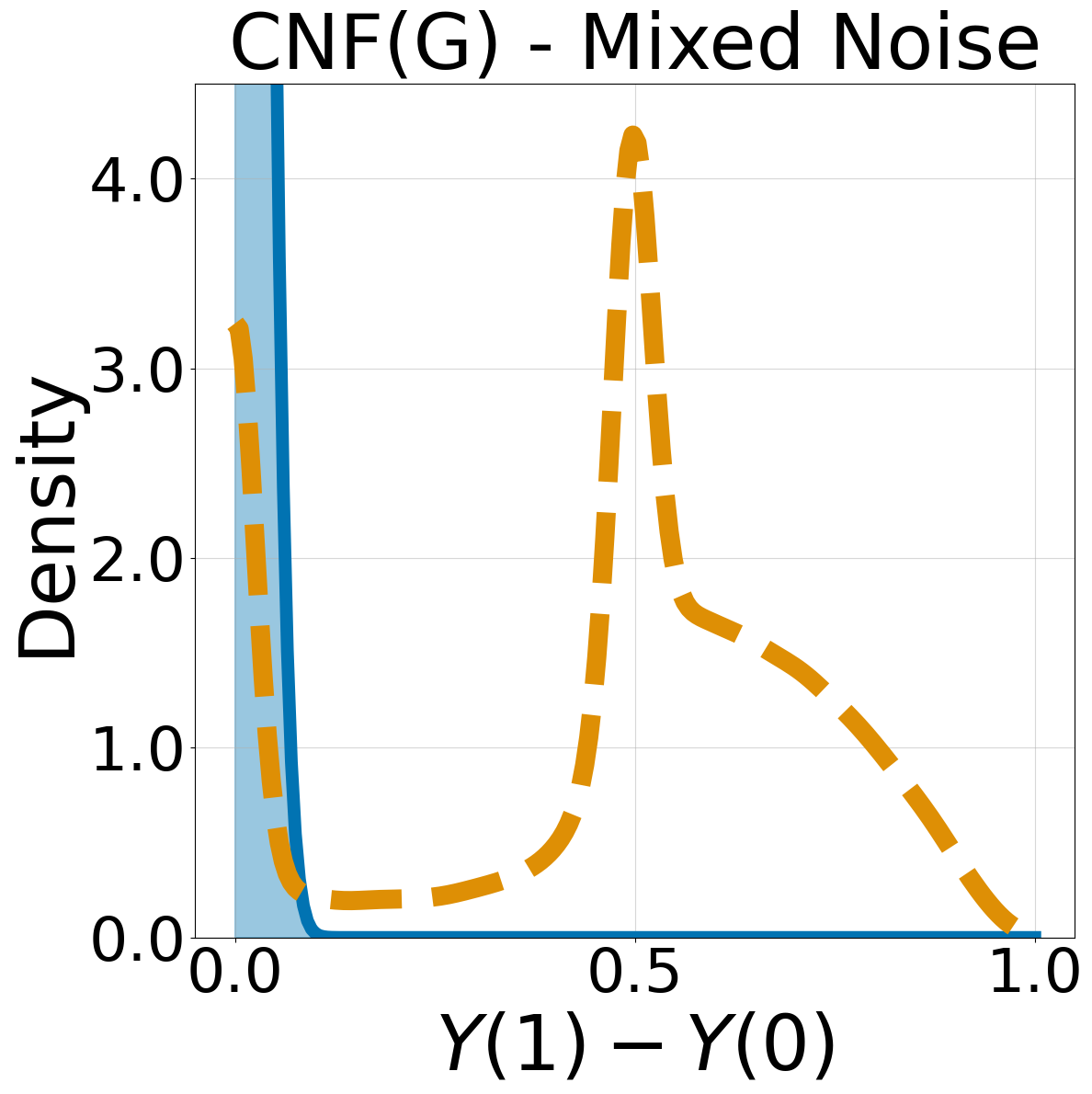}
    \includegraphics[width=0.24\linewidth]{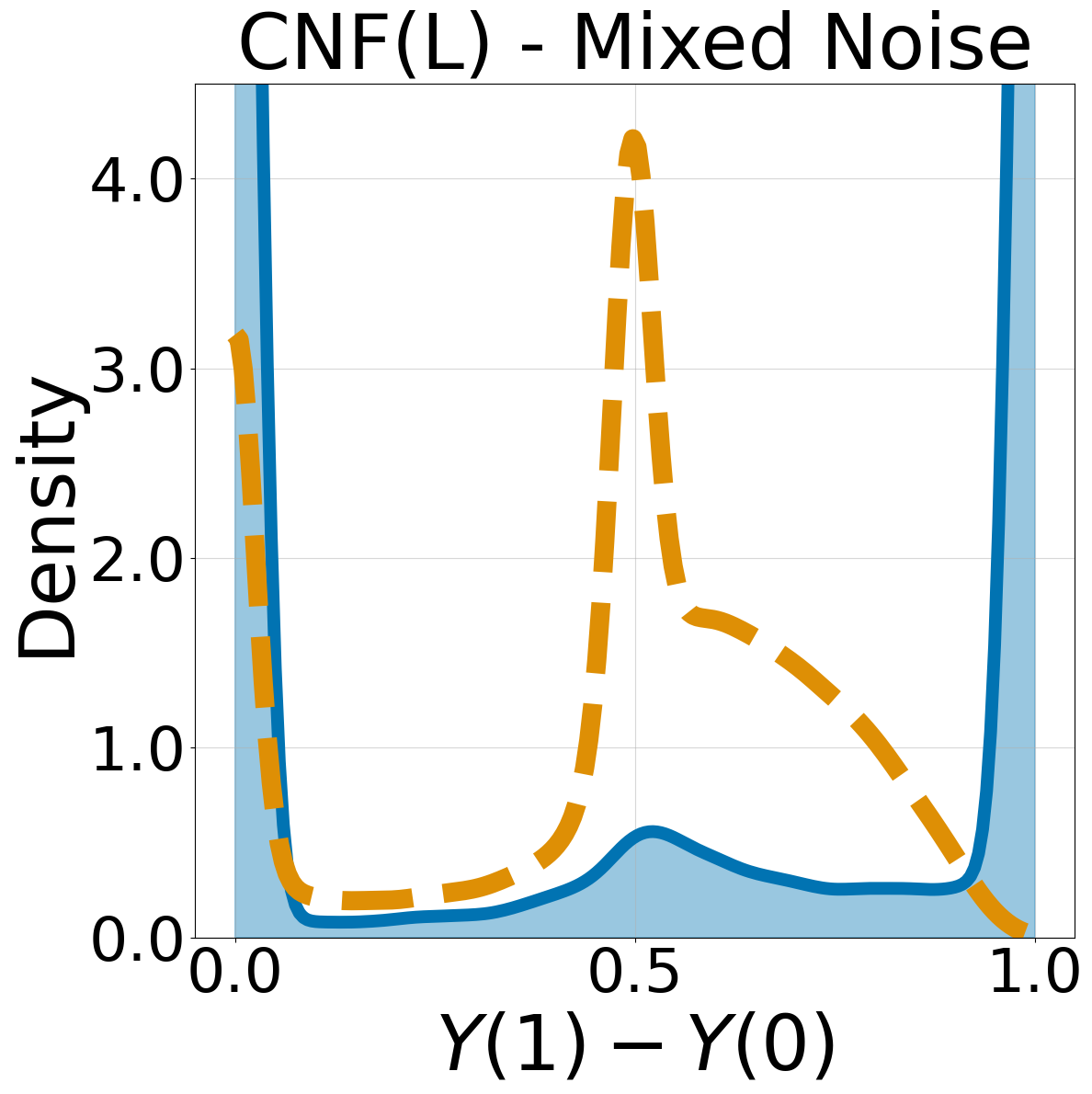}
    \includegraphics[width=0.24\linewidth]{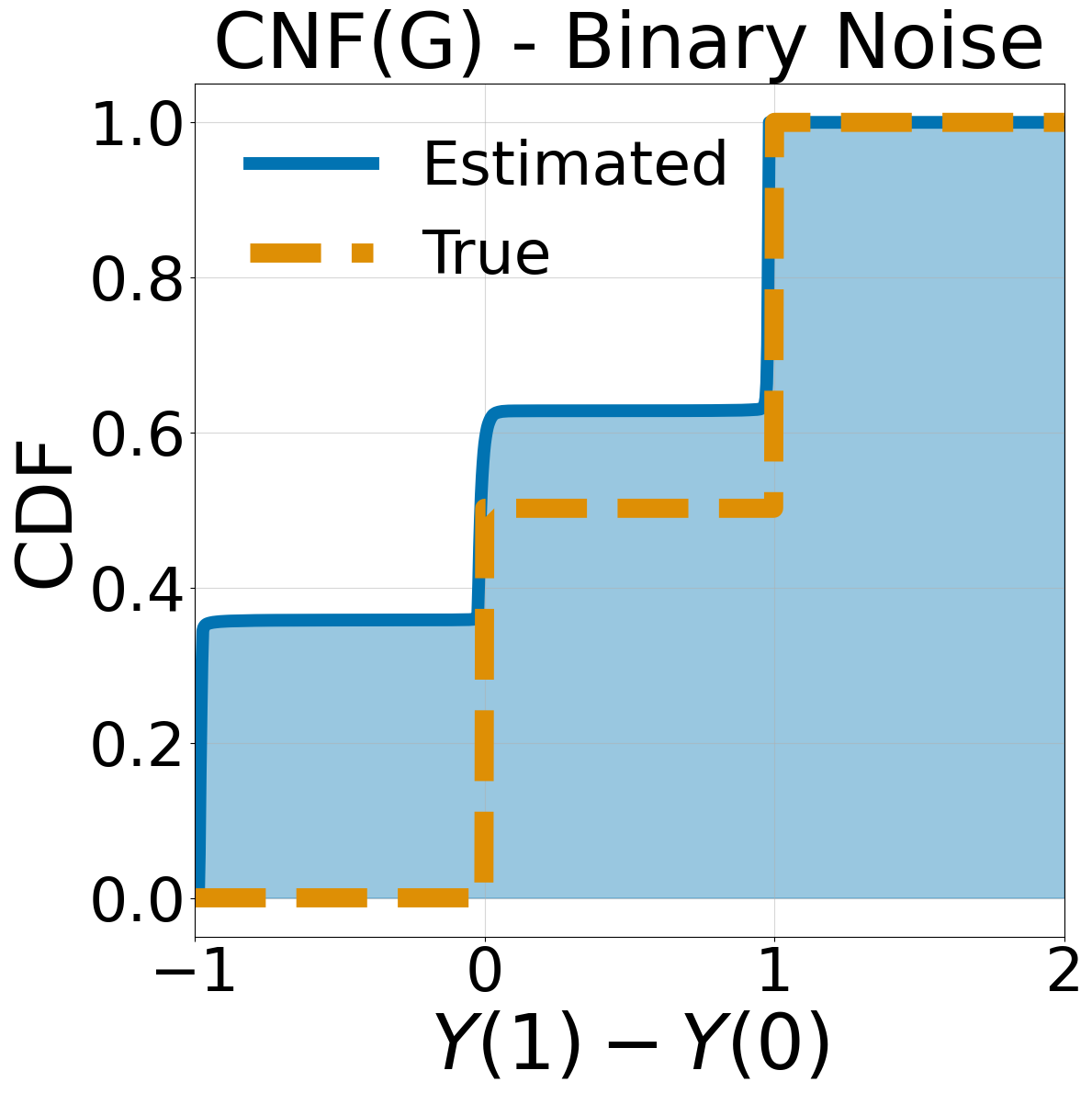}
    \includegraphics[width=0.24\linewidth]{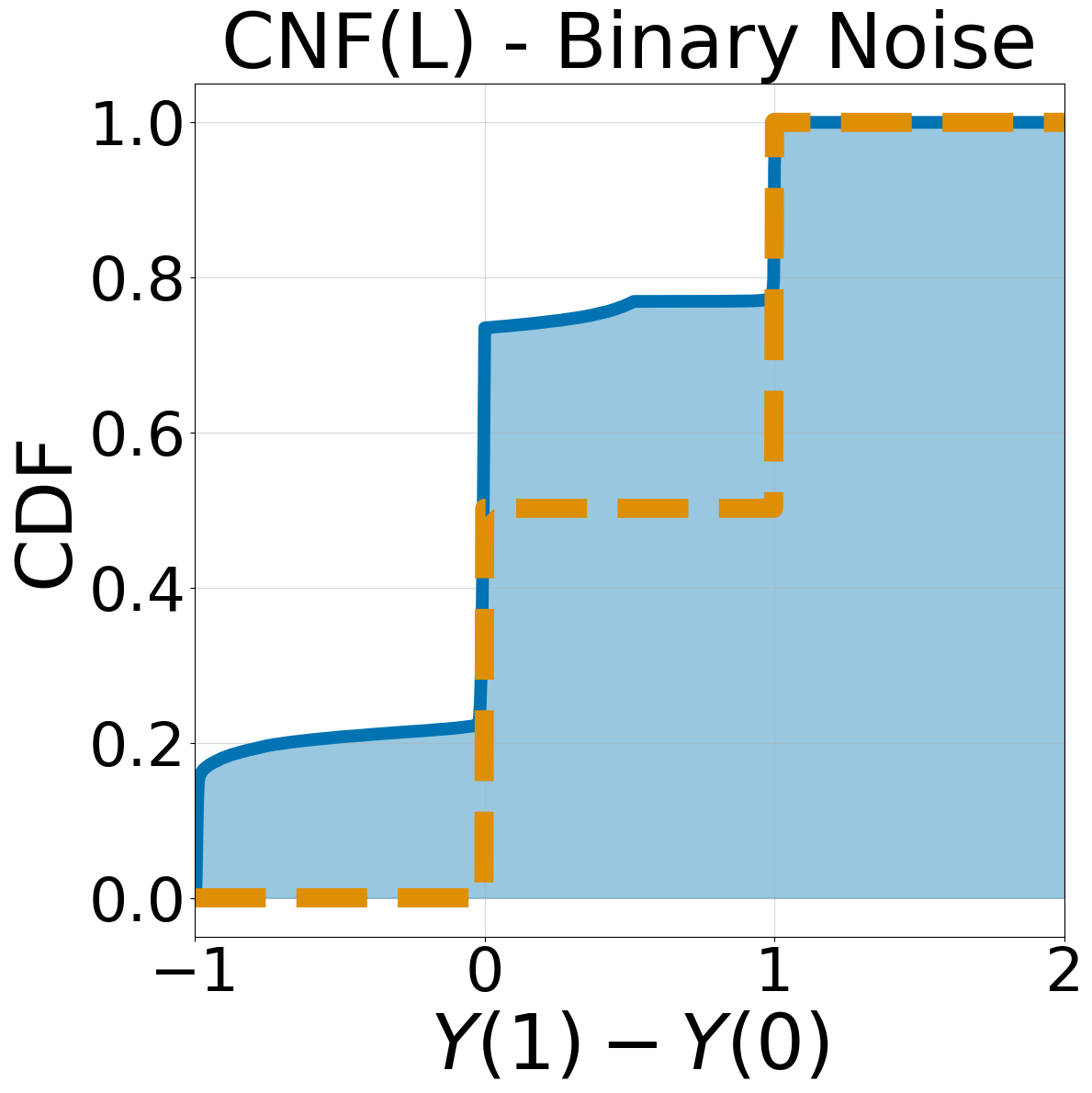}
    \caption{True (orange) vs.\ estimated (blue) distribution of the treatment effect $Y(1)-Y(0)$ for the SCM $Y(x)=(x+1)\xi$, under two different noise laws: (LHS) mixed-tailed $\xi$ (density plots) and (RHS) binary $\xi$ (CDF plots). Estimates are obtained with flow-based SCMs using Gaussian (CNF(G)) and Laplace (CNF(L)) base distributions $\widehat{\bb P}_\xi$, each trained on $n=2000$ samples. For the mixed-tailed case, $Y(1)-Y(0)$ is transformed to $[0,1]$ via $\sigma(y)=1/(1-e^{-y})$ for visualization. Both estimated flows fail to recover the effect distributions.}
    \label{fig:misspec}
    \vspace{-10pt}
\end{figure}

\begin{example} \label{ex:mis-spec}
To demonstrate the severity of these problems, consider the structural model,
\begin{align*}
    Y(x) = (x+1)\xi  \quad , \quad X \sim \mathsf{Bern}(1/2)
\end{align*}
Suppose the estimation target is the distribution of the treatment effect $Y(1) - Y(0) = \xi$ and the noise has a heavy left tail and light right tail $\xi \sim  \frac{1}{2}|\mathsf{N}(0,1)| - \frac{1}{2}|\mathsf{NBP}(0.1,0.1)|$.\footnote{Here, $\mathsf{NBP}(\alpha,\beta)$ is the heavy-tailed Normal-Beta-Prime distribution on $\bb R$ \citep{bai2021beta}.} \cref{fig:misspec} (left + middle left) shows the true density of \(Y(1)-Y(0)\), rescaled to $[0,1]$ via $\sigma(y) = 1/(1+\exp(-y))$ for better visualization, alongside estimated densities via flow-based SCMs with Gaussian (left) and Laplace (middle left) base distributions. Since $x \in \{0,1\}$ is binary, we specify separate flows $f_x$ for each $x \in \{0,1\}$, rather than parameter sharing across $x$. Each flow architecture was specified as a Neural Spline Flow (NSF) \citep{durkan2019neural} with a single (8-bin) spline layer sandwiched between two affine layers. The flow parameters were trained by maximum likelihood on $n=2000$ samples $\{X^{(i)}, Y^{(i)}\}_{i=1}^n$ and the exact training and optimization routine matches that used in \cref{exp:lin_model}. In order to match the heavy left tail, both estimated flows move most probability mass there, but as a consequence fail to recover the rest of the density. Note the large mass \emph{near} zero of the estimated densities under the transformation $\sigma$ does not indicate a learned heavy left tail.

We next re-implement the same flow-based models with binary noise,
$\xi \sim \mathsf{Rad}(1/2)$,
this time showing the learned CDFs (\cref{fig:misspec}, middle right and right). While we acknowledge flows are not
designed for discrete outcomes $Y$, the example serves to highlight the problems that arise when the support of $\bb P_{Y(x)}$ lies on a
lower-dimensional manifold than that of $\bb P_{\xi}$. Since each model only evaluates the likelihood on a set of measure zero under its base distribution, there are infinitely many flows that can achieve the same likelihood. As a consequence, both implemented models fail to learn the size and placement of the modes of the distribution (as indicated by the jumps in the learned vs.\ true CDF). 

 \end{example}

\subsection{Optimal Transport Methods for Counterfactual Inference}
\label{sec:bg:ot}

Several recent works estimate counterfactual couplings with
\emph{optimal transport} (OT)
\citep{charpentier2023optimal,de2024transport,balakrishnan2025conservative}, providing an alternative to SCMs, albeit in a restricted setting. Given discrete treatments $X$ and continuous outcomes $Y$, the idea is to model the coupling $\mc L(\{Y(x)\}_{x \in \bb X})$ via a collection of transports $\{T_{x,x'}\}_{x \in \bb X}$ that model deterministic counterfactuals under treatment changes,
\begin{align}
    Y(x) =_{\mathrm{a.s.}} T_{x,x'}(Y(x')) \;, \quad x, x' \in \bb{X} \label{eq:transport_maps}\;.
\end{align}
Since many sets of transports can give rise to the same marginal counterfactual distributions, the transports are identified via the principle of \citet{lewis1973causation}: out of all couplings, choose that which induces the most ``similar'' counterfactual worlds. This is formalized by choosing $T_{x,x'}$ to minimize the cost of transporting mass from $\bb P_{Y(x')}$ to $\bb P_{Y(x)}$,
\begin{align}
    T_{x,x'}^* = {\arg\min}_{T_{x,x'} : T_{x,x'}(Y(x')) =_d Y(x)} \bb E \left[ \lVert Y(x) - T_{x,x'}(Y(x')) \rVert_2^2 \right] \label{eq:OT}
\end{align}
Since quadratic-cost OT problems between continuous distributions have a unique solution (known as the Brenier map) \citep{villani2021topics}, this guarantees identifiability when $\bb{P}_{Y(x)}$ is absolutely continuous for all $x$. This approach is appealing, as it avoids the need to specify a full generative process (e.g., noise distributions) or causal ordering when there are multiple outcomes $Y := (Y_1,\dots,Y_p)$. Under no confounding, one has
$\bb P_{Y(x)} = \bb P_{Y|X=x}$. In practice, given i.i.d.\ data $\{Y^{(i)}(x)\}_{i=1}^n \sim \bb P_{Y|X=x}$, one can plug the empirical analogues $\widehat{\bb P}_{Y(x)} = \frac{1}{n_x}\sum_{i=1}^{n_x}\delta_{Y^{(i)}(x)}$ of the conditionals into \eqref{eq:OT} and solve the resulting problem (either in closed form or using specialized solvers). When there are measured confounders $Z$, the transports are specified between $\{Y(x,z)\}_{x,z \in \bb X \times \bb Z}$ instead \citep{balakrishnan2025conservative}. 

\begin{figure}[!t]
  \centering
  \includegraphics[width=\textwidth]{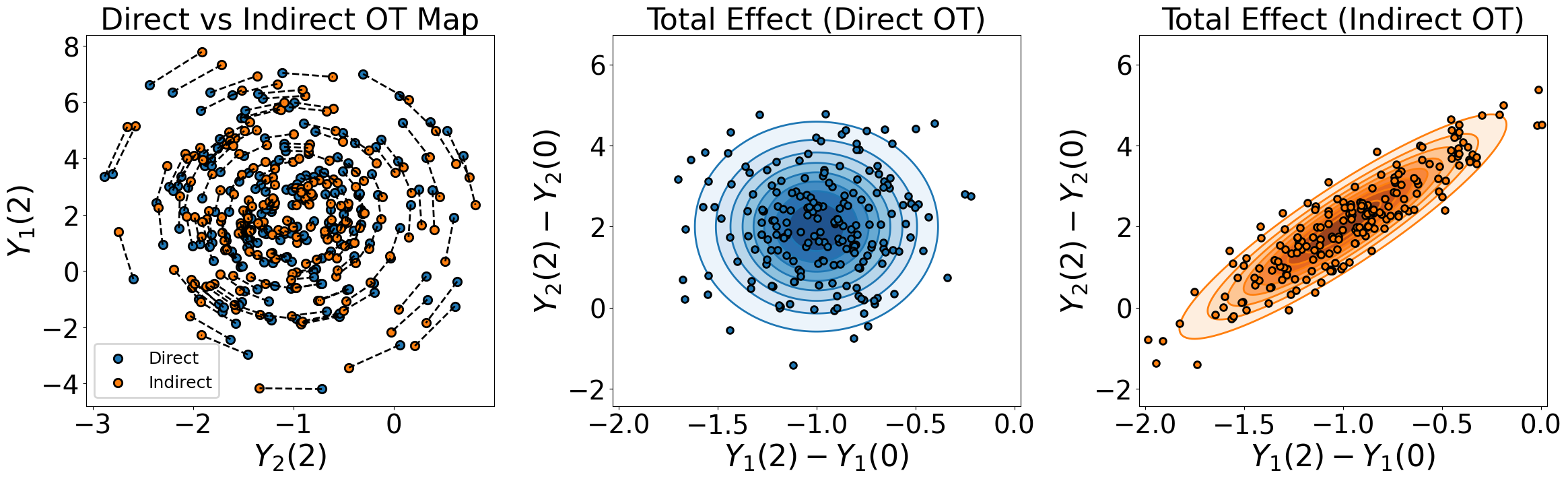}
\caption{Transport of $200$ samples $Y(0) \sim \mathbb P_{0}$ to $\mathbb P_{1}$ and $\mathbb P_{2}$ using OT maps. Blue: direct route $(Y(1), Y(2)) = (T_{1,0}(Y(0)), T_{2,0}(Y(0))$. Orange: indirect route $(T_{1,0}(Y(0)), T_{2,1} \circ T_{1,0}(Y(0))$. Left: Imputed $Y(2)$ under each method; dashed lines highlight the inconsistency. Middle \& right: Density of $Y(2)-Y(0)$ depends on which maps are used (direct vs. indirect), so cannot be identified.}
\label{fig:brenier-inconsistency}
\vspace{-10pt}
\end{figure}

When $Y$ is scalar, the Brenier map is the quantile transform \citep{Santambrogio_2015} and can be estimated via the CDF \citep{balakrishnan2025conservative}. 
However, when $Y$ is multivariate, one needs to solve the OT problem separately for each pair $(x,x')$, typically using a specialized solver, which makes it challenging to handle continuous treatments and covariates.

A more serious issue overlooked in the literature is that, except in special cases, Brenier maps will not yield a valid joint coupling when $\dim(\bb Y)>1$ and $|\bb X|>2$. The solution to \eqref{eq:OT} is the gradient of a convex function, $T_{x,x'}^* = \nabla \varphi_{x,x'}$ \citep{villani2021topics}, but the composition of two such maps is generally not the gradient of a convex function when $\dim(\bb Y)>1$ \citep{Santambrogio_2015}. Consequently, if $T_{x'',x}^* \neq T_{x'',x'}^*\circ T_{x',x}^*$ on a set of positive $\bb P_{Y(x)}$-measure, the transports cannot satisfy \eqref{eq:transport_maps} for any set of counterfactuals with the same marginal distributions. Below we demonstrate this incompatibility and resulting identifiability issues.

\begin{example} \label{ex:ot}
Consider solving the quadratic-cost OT problem for transporting between bivariate Gaussian distributions  $\bb P_0,\bb P_1,\bb P_2$ with the following means and covariances.
\[(\mu_0, \Sigma_0) = \left(\bm 0,I_2\right), \quad (\mu_1, \Sigma_1) = \left(\begin{pmatrix}
    1 \\ 1
\end{pmatrix}, \begin{pmatrix}
    1 & \tfrac{-9}{10} \\ \tfrac{-9}{10}   & 1
\end{pmatrix} \right), \quad (\mu_2, \Sigma_2) = \left(\begin{pmatrix}
    -1 \\ 2
\end{pmatrix}, \begin{pmatrix}
    \tfrac12 & 0 \\ 0  & 5
\end{pmatrix} \right)\]
The Brenier map from
$\bb P_x$ to $\bb P_{x'}$ is 
the affine map
$
  T_{x',x}(y)=\mu_{x'}+A_{x',x}(y-\mu_x)
$, 
with
    \[
      A_{x',x}
      =\Sigma_x^{-1/2}\bigl(\Sigma_x^{1/2}\Sigma_{x'}\Sigma_x^{1/2}\bigr)^{1/2}
        \Sigma_x^{-1/2}\;, \quad x,x' \in \{0,1,2\}.
    \]
However, as $\Sigma_1$ and $\Sigma_2$ do not commute, the maps cannot be valid counterfactual transports in the sense of \eqref{eq:transport_maps}, since using them to impute counterfactuals results in a contradiction: 
\[Y(2) - Y(2) =_{\mathrm{a.s.}}T_{2,0}(Y(0)) - T_{2,1}\circ T_{1,0} (Y(0)) = \underbrace{(\Sigma_2^{\frac 12} - A_{2,1}\Sigma_1^{\frac 12})}_{\neq 0}(Y(0) - \mu_0)\neq_{\mathrm{a.s.}} 0\,.\]
To illustrate the identifiability issues that arise from this, suppose we draw $n=200$ samples $\{Y^{(i)}(0)\}_{i=1}^{n}\sim \bb P_0$ and impute their counterfactuals $\{Y^{(i)}(1),Y^{(i)}(2)\}_{i=1}^n$. There are two natural ways of using the OT maps to do this: (i) \emph{directly}, $y_0 \mapsto (T_{1,0}(y_0), T_{2,0}(y_0))$; and (ii) \emph{indirectly}, $y_0\mapsto (T_{1,0}(y_0), T_{2,1} \circ T_{1,0} (y_0))$. Each approach uses two out of three OT maps to induce a coupling over $\{Y(0),Y(1), Y(2)\}$. \cref{fig:brenier-inconsistency} shows that the direct map
$T_{0,2}$ and the composition $T_{1,2}\circ T_{0,1}$ result in different counterfactual predictions for $Y(2)$ (left) and different distributions of the total effect
$Y(2)-Y(0)$ (middle and right). Thus, the induced coupling depends on the subset of transports used, and is not identifiable from these maps.
\end{example}

\section{Transport-based Counterfactual Couplings in a Simplified Setting}
\label{sec:method}

Our goal is to develop a framework for modeling counterfactual couplings that avoids the limitations of SCMs and OT methods. The basic idea is to directly model counterfactual transports (as in OT), but in a way that guarantees coherence and identifiability under general conditions (as in SCMs). To that end, in this section we analyze the algebraic properties that a set of transports must satisfy to induce a valid coupling between counterfactual outcomes, in a simplified confounding-free setting. We show that these properties are precisely those of a \emph{cocycle}, and so call any system of transports satisfying them a counterfactual cocycle. We then show that any counterfactual cocycle can be represented using a family of injective functions. This lets us characterize conditions for cocycle identifiability and provide a general recipe for parameterizing flexible, identifiable cocycle classes. Finally, we show that every counterfactual cocycle corresponds to an equivalence class of injective SCMs. This suggests that modeling counterfactual cocycles can avoid the mis-specification problems of SCMs, and underpins the estimation approach we later develop.

\subsection{Counterfactual Cocycles as Admissible Transports}
 
 We focus on a simple setting where $X  \in \bb X \subset \bb R$ is a randomized treatment, $Y := (Y_1,\dots,Y_p) \in \bb Y \subset \bb R^p$ are a set of outcomes of interest, and $\{Y(x)\}_{x \in \bb X} \in \bb Y$ are counterfactuals under different treatment levels. Throughout, we assume all spaces (here $\bb X, \bb Y$) are standard Borel and all maps are Borel measurable.  To formalize counterfactuals in this setting, we work under the standard potential outcome assumptions \citep{rubin1974estimating}.
\begin{assumption}\label{ass:OSA_counterfactuals}
\begin{enumerate}
    \item \textbf{Consistency}: $Y =_{\mathrm{a.s.}} Y(X)$ \label{eq:consistency} 
    \item \textbf{Exchangeability}: $\{Y(x)\}_{x \in \bb X} \; \ind \;X$
\end{enumerate}    
\end{assumption}

\cref{ass:OSA_counterfactuals} is consistent with a ``coarse-grained'' causal DAG $X \to Y$ \citep{richardson2013single} and identifies the marginal counterfactual distribution as $Y(x) \sim \bb P_{Y|X=x}$.

To recover a coupling over $\{Y(x)\}_{x \in \bb X}$, we follow previous transport-based approaches and start from the existence of a collection of transport maps between counterfactual outcomes under different treatment levels. 
\begin{assumption}\label{ass:counterfactual_cocycle}
There exist a collection of (Borel measurable) transport maps $\{T_{x,x'} : \bb Y \to \bb Y \mid x,x' \in \bb X\}$ that satisfy the \textbf{Counterfactual Coupling} \eqref{eq:CC} property:
\begin{align} \label{eq:CC}   
 Y(x)&  =_{\mathrm{a.s.}} T_{x,x'}( Y(x'))\;, \quad \text{ for all $x,x' \in \bb X \;.$}  \tag{CC}
\end{align}
\end{assumption}

In contrast to previous transport-based methods, our approach to modeling the transports is motivated by the insight that, as is easily checked, \eqref{eq:CC} can only hold if the transports satisfy the following properties: 
    \begin{enumerate}
        \item \textbf{Identity}: For each $x \in \bb{X}$,
        \begin{align} \label{eq:NI}
            T_{x,x} = \id, \quad \bb P_{Y|X=x}\text{-a.s.}\ \tag{ID}
        \end{align}
        
        \item \textbf{Path Independence}: For each $x,x',x'' \in \bb{X}$,
        \begin{align} \label{eq:CPI}
            T_{x'',x'}\circ T_{x',x} = T_{x'',x},\quad \bb P_{Y|X=x}\text{-a.s.} \tag{PI}\;
        \end{align}
        \item{\textbf{Distribution Adaptedness}}: For each $x,x' \in \bb{X}$,
        \begin{align} \label{eq:DA}
        {(T_{x',x})}_{\#}\bb P_{Y|X=x} = \bb P_{Y|X=x'} \tag{DA}\;
        \end{align}
    \end{enumerate}

\usetikzlibrary{automata,arrows.meta,positioning}
Properties \eqref{eq:NI} and \eqref{eq:CPI} together make the map \[T: (x,x',y) \mapsto T_{x,x'}(y)\;,\] a \textbf{cocycle} 
\citep{arnold1998random}. In classical dynamics, a cocycle describes how a dynamical system state evolves as time flows. In the present context, “time” is replaced by the treatment value \(x\), the ``state'' is the outcome value $y(x)$, and the cocycle \(T\) encodes how $y(x)$ changes as treatment values change $x \to x'$. If a cocycle also satisfies \eqref{eq:DA} we call it a $\bm {\bb P_{Y|X}}$\textbf{-adapted cocycle}. Lastly, if a cocycle $T: \bb X^2 \times \bb Y \to \bb Y$ satisfies \eqref{eq:CC} then we call it a \textbf{counterfactual cocycle}. We thus also refer to \eqref{eq:CC} as the \emph{counterfactual cocycle property}.

\paragraph{Importance of Cocycle Properties}

\begin{figure}[!t]
    \centering
\includegraphics[width = 0.9\textwidth]{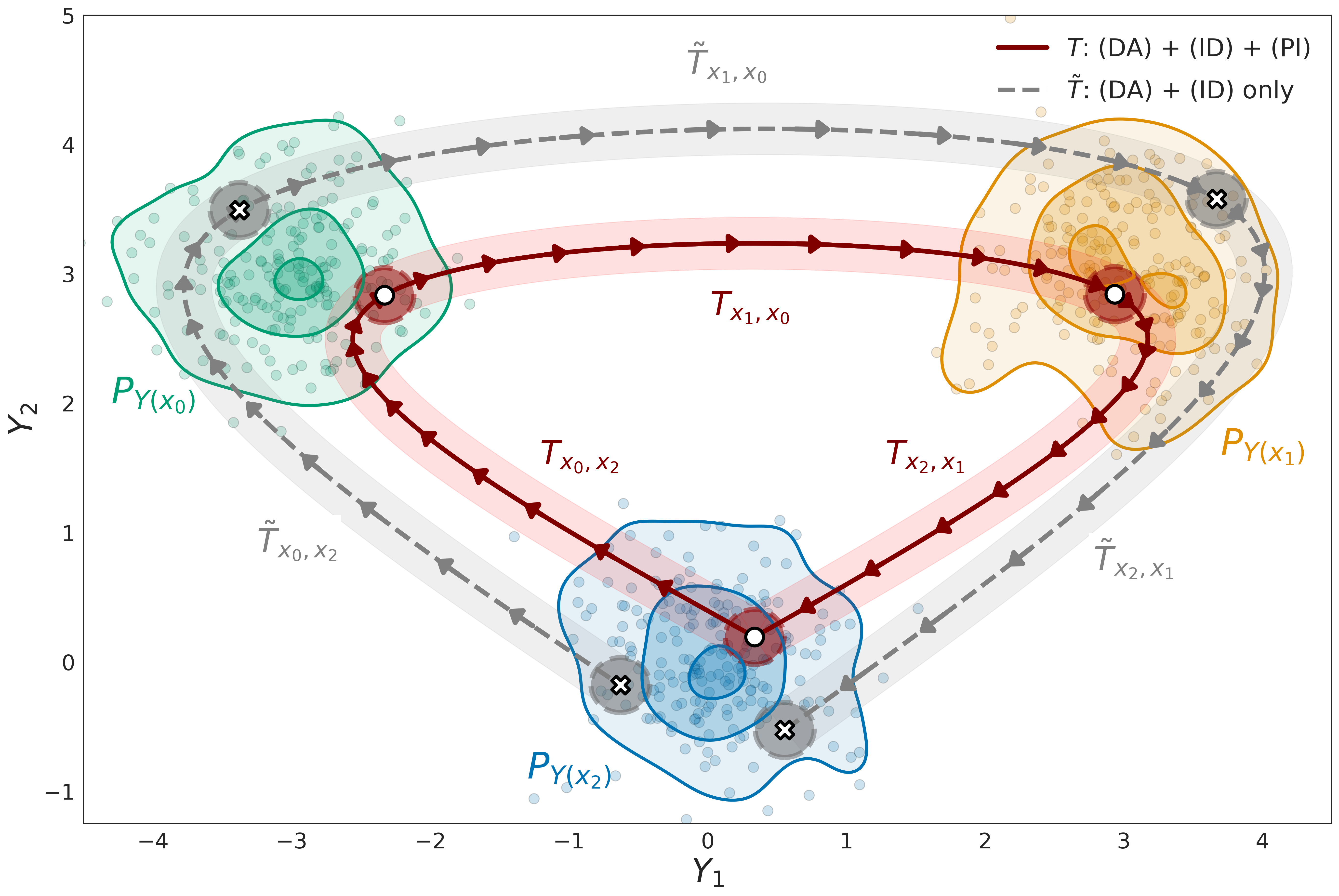}

\caption{Illustration of cocycle property importance for coherent counterfactual transports. 
Shaded regions show trajectories of different values of $Y(x_2) \sim \bb P_{Y(x_2)}$ as we change the treatment level $x_2 \to x_0 \to x_1 \to x_2$, under transport collections $T$ (maroon) and $\tilde T$ (gray). 
Lines indicate trajectory for single value $y(x_2)$ within each region. 
$\{\tilde T_{x,x'}\}_{x,x'\in\bb X}$ satisfies \eqref{eq:DA} and \eqref{eq:NI} but fails \eqref{eq:CPI}, leading to the logical impossibility that $Y(x_2) \neq Y(x_2)$ on a set of strictly positive measure (see mismatch in gray regions of $\bb P_{Y(x_2)})$. 
The transports therefore do not define a valid counterfactual model. 
In contrast, $\{T_{x,x'}\}_{x,x'\in\bb X}$ satisfy all cocycle properties and so cohere with one another, inducing a valid counterfactual model (Theorem \ref{thm:cocycle_sufficiency}).}

\label{fig:cocycle_properties}

\vspace{-20pt}
\end{figure}

\eqref{eq:DA} alone ensures that \eqref{eq:CC} holds in \emph{distribution},
\[
Y(x) =_d T_{x,x'}(Y(x')).
\]
However, without \eqref{eq:NI} and \eqref{eq:CPI}, the transports \(T_{x,x'}\) cannot describe a valid coupling between a set of counterfactuals as in \eqref{eq:CC}. For example, if \eqref{eq:CPI} fails and we try to impose \eqref{eq:CC}, we can arrive at logical impossibilities in the counterfactual outcomes:
\[
Y(x_2) =_{\mathrm{a.s.}} T_{x_2,x_0}(Y(x_0)) \neq_{\mathrm{a.s.}}  T_{x_2,x_1} \circ T_{x_1,x_0}(Y(x_0)) =_{\mathrm{a.s.}} Y(x_2).
\]
\cref{fig:cocycle_properties} illustrates this failure. This issue arises in recent transport-based models \citep{de2024transport,charpentier2023optimal,torous2024optimal,balakrishnan2025conservative}, where \(T_{x,x'}\) is defined as the OT map from \(\bb{P}_{Y|X=x'}\) to \(\bb{P}_{Y|X=x}\). These maps satisfy \eqref{eq:NI}, \eqref{eq:DA} and an invertibility property (\(T_{x,x'} \circ T_{x',x} (Y(x)) =_{\mathrm{a.s.}} Y(x)\)), but, as demonstrated in \cref{sec:bg:ot} can  fail \eqref{eq:CPI} when \(|\bb{X}| > 2\) and \(\dim(\bb{Y}) > 1\), leading to identifiability problems.

The properties \eqref{eq:NI}, \eqref{eq:CPI}, and \eqref{eq:DA} are purely mathematical and so are necessary but not sufficient to guarantee \eqref{eq:CC}---the latter requires that the counterfactual variables \(\{Y(x)\}_{x \in \bb X}\) exist on a common probability space and are actually linked by the transports. However, these properties \emph{are}  necessary and sufficient to ensure \eqref{eq:CC} holds w.r.t. \emph{some} set of variables \(\{\tilde Y(x)\}_{x \in \bb X}\). Thus, these properties characterize all sets of transports which can induce admissible couplings over counterfactuals. This is formalized below.
\begin{definition}
   A set of transports $\{T_{x,x'}: \bb Y \to \bb Y\}_{x,x' \in \bb X}$ are \textbf{admissible} w.r.t.\ $\bb P_{Y|X}$ if there exists a collection of random variables $\{\widetilde Y(x)\}_{x \in \bb X}$ such that $\widetilde Y(x) \sim \bb P_{Y|X=x}$ and $\widetilde Y(x) =_{\mathrm{a.s.}} T_{x,x'}\bigl(\widetilde Y(x')\bigr)$ for every $x, x' \in \bb X$.
\end{definition}

\begin{theorem}[Cocycle Equivalence to Admissible Transports] \label{thm:cocycle_sufficiency}
$\{T_{x,x'}: \bb Y \to \bb Y\}_{x,x' \in \bb X}$ satisfy \eqref{eq:NI}, \eqref{eq:CPI}, and \eqref{eq:DA} w.r.t.\ $\bb P_{Y|X}$ if and only if they are admissible w.r.t.\ $\bb P_{Y|X}$.
\end{theorem}

Our aim is therefore to develop a framework for modeling and estimating counterfactual cocycles. In the rest of this section we focus on the modeling aspects.

\subsection{Structure and Identifiability of Counterfactual Cocycles} \label{sec:method:id}

The previous results tell us that modeling admissible counterfactual transports reduces to targeting the properties \eqref{eq:NI}, \eqref{eq:CPI}, and \eqref{eq:DA}. However, it remains unclear how to construct transports that satisfy these properties in general, or under what conditions the resulting system of transports is uniquely determined.  Both questions must be answered in order to apply these properties in practice to estimate counterfactual transports.

\paragraph{Characterizing Cocycles via Factorization} The following result shows that every cocycle has a special structure: it can always be constructed using a family of injective functions. This provides a clear path to modeling transports that satisfy \eqref{eq:NI} and \eqref{eq:CPI}.

\begin{theorem}[Cocycle Factorization] \label{thm:cocycle_factorization}
    Let $\{T_{x,x'}\}_{x,x' \in \bb X}$ satisfy \eqref{eq:NI}, \eqref{eq:CPI}, and \eqref{eq:DA} w.r.t.\ $\bb P_{Y|X}$. Then there is a set $\bb Y_0 \subseteq \bb Y$ and, for each $x \in \bb X$, a function $f_x : \bb Y_0 \to \bb Y$ with left-inverse $f_x^+: \bb Y \to \bb Y_0$, such that for every $x' \in \bb X$: 
    \begin{align}T_{x,x'} = f_x \circ f_{x'}^+, \quad  \bb P_{Y|X=x'}\text{-a.s.} \label{eq:cocycle_factorization}\end{align}
\end{theorem}

The proof is straightforward: one essentially defines $f_x := T_{x,x_0}$ and $f_x^+ := T_{x_0,x}$ for any $x_0 \in \bb X$ and applies \eqref{eq:NI} and \eqref{eq:CPI} to verify \eqref{eq:cocycle_factorization}. To ensure $f_x^+$ is a left inverse, we restrict the functions to map to and from an appropriate subset $\bb Y_0 \subset \bb Y$ of full $\bb P_{Y|X=x_0}$-measure.

 \cref{thm:cocycle_factorization} implies that specifying a parameterized family of cocycles reduces to specifying a parameterized family of {functions} $f: \bb X \times \bb Y_0 \to \bb Y$  injective in their second argument, 
 \[\mc F \subseteq \mc I  := \bigl\{\,f:\bb X\times\bb Y_0\to\bb Y\;\big|\;f_x := f(x,\argdot)\text{ is injective} \;\forall x \in \bb X\bigr\}\,.\]

Indeed, any element $\hat f \in \mc F$ can be used to construct a candidate cocycle via,
\begin{align}\hat T(x,x',y') := \hat f_x \circ \hat f^+_{x'}(y') \;,\label{eq:coboundary_construction}\end{align}

where $\hat f^+_x$ is a left inverse of $\hat f_x$. To avoid the need to separately model the left inverse (which is only unique on the image of  $\hat f_x$) a convenient approach is to take $\bb Y_0:=\bb Y$ and parameterize
$\hat f_x:\bb Y\to\bb Y$ using \emph{bijective} maps  such as normalizing flows (NFs)
\citep{papamakarios2021flows} or invertible neural networks (INNs) \citep{ishikawa2023universal},
with parameters conditioned on $x$. In this case the left inverse is unique and is an exact inverse $\hat f^+_x=\hat f_x^{-1}$. As a consequence, the construction
\eqref{eq:coboundary_construction} is uniquely determined by $\hat f$ and automatically satisfies \eqref{eq:NI} and \eqref{eq:CPI} on all of $\bb Y$:
\[\hat  T_{x,x} = \hat f_x \circ \hat  f_x^{-1} = \id, \quad \hat  T_{x,x'} \circ \hat  T_{x',x''} = \hat f_{x} \circ \cancel{\hat f_{x'}^{-1} \circ \hat  f_{x'}} \circ \hat f_{x''}^{-1} = \hat  T_{x,x''}\]
We note specializing to bijections is not overly restrictive. The constructed $f_x$ in \cref{thm:cocycle_factorization} is already {bijective} on $\bb Y_0 \to \bb Y_x := f_x(\bb Y_0)$. When the sets \(\{\bb Y_x\}_{x\in\bb X}\) are
  \emph{Borel-isomorphic} (e.g., full-dimensional Borel subsets
  of \(\bb Y\)\footnote{Note, we do \emph{not} require that $\bb Y$ is a full-dimensional subset of $\bb R^p$.}), classical results
  guarantee a measurable bijective extension
  \(\tilde f_x:\bb Y\!\to\!\bb Y\) of $f_x$ \cite[Thm.\ 15.6]{kechris2012classical}. 
 Although the extension is not always continuous, NFs and INNs remain highly expressive, and will be used to specify cocycle models in \cref{sec:implementation}. Cocycles of the form $T(x,x',y') = f_{x} \circ f_{x'}^{-1}(y')$ are known as \textbf{coboundaries} \citep{varadarajan1968geometry} and so we call the function $f$ a \textbf{coboundary map}.

\paragraph{General Conditions for Identifiability} 

\cref{thm:cocycle_factorization} provides a general recipe for parameterizing counterfactual cocycles, but does not tell us how to restrict $\mc F$ so that any cocycle constructed by functions in $\mc F$ that satisfies  \eqref{eq:NI}, \eqref{eq:CPI} and \eqref{eq:DA} is (almost surely) unique. Indeed, many sets of transports may satisfy these properties, but they can only all satisfy \eqref{eq:CC} for the same set of counterfactuals if they are identical up to null sets.

To demonstrate this point, a collection of OT maps $\{T^{(OT)}_{x,x'}\}_{x \in \bb X}$ can in principle be used as coboundary maps to construct a cocycle that satisfies \eqref{eq:NI}, \eqref{eq:CPI} and \eqref{eq:DA}. One can simply define $f_x := T^{(OT)}_{x,x_0}$ and $f_x := T^{(OT)}_{x_0,x}$ for some $x_0$, and define a $\bb P_{Y|X}$-adapted cocycle \(\tilde T(x,x',y') := f_x \circ f_{x'}^+(y')\). However, since OT maps do not necessarily satisfy \eqref{eq:CPI}, the constructed cocycle $\tilde T$ will depend on the reference point $x_0$, and so there are potentially $|\bb X|$-many different construction choices. We already saw this in \cref{ex:ot} and \cref{fig:brenier-inconsistency}: the direct imputation approach used reference $x_0 = 0$, while the indirect approach used $x_0=1$.

The factorization structure of cocycles enables us to provide general conditions for identifiability when working with bijective cocycles (a.k.a. coboundaries) defined by transformation groups.\footnote{For simplicity, we characterize identifiability under the assumption that the cocycles can be characterized using bijections, but a nearly identical version holds in the more general injective case using a more complicated algebraic construction involving monoids instead of groups.} Formally, a \emph{transformation group} $\bb G$ on $\bb Y$  is a set of bijective transformations $g: \bb Y \to \bb Y$ that contains the identity and is closed under compositions and inverses. We say that a cocycle $T$ is \textbf{$\bm{\bb G}$-valued} if its coboundary map $f$ satisfies $f(x,\argdot) := f_x \in \bb G$ for every $x \in \bb X$. For example, any bijective $f_x: \bb Y \to \bb Y$ belongs to the automorphism group of $\bb{Y}$, denoted $\Aut(\bb{Y})$. If $f_x$ is also bi-continuous, it belongs to the group of self-homeomorphisms, $\textrm{Homeo}(\bb{Y})$. If $f_x$ and $f_x^{-1}$ also have continuous derivatives (e.g. if modeling each $f_x$ using typical parameterizations of the flows in \cref{tab:flows}), it lies in the diffeomorphism group, $\textrm{Diff}^1(\bb Y)$.  The set of all $\bb G$-valued coboundary maps is denoted
\begin{align*}
  \mc F_{\bb G} := \{f: \bb X \times \bb Y \to \bb Y \mid f(x,\argdot) \in \bb G\}  \;.
\end{align*}
Using these concepts, we have the following formal notion of cocycle identifiability.

\begin{definition}[Identifiability]\label{def:identifiability}
Let $\mathcal F$ be a set of $\bb G$-valued coboundary maps $f:\mathbb X \times \mathbb Y \to \mathbb Y$ and define for each $f \in \mc F$ the cocycle \(T_f:(x,x',y') \mapsto  f_x \circ f_{x'}^{-1}(y)\). A $\mathbb P_{Y|X}$-adapted cocycle $T$ is {identifiable in ${\mathcal F}$} if 
\begin{enumerate}
    \item $\exists\, f^\star \in \mathcal F$ with $T =T_{f^\star}$ $(\bb P_X \otimes \bb P_{X,Y})$-a.s. \label{eq:id_existence}
    \item For any $f \in \mathcal F$, if $T_f$ is a $\mathbb P_{Y|X}$-adapted cocycle satisfying \eqref{eq:NI}, \eqref{eq:CPI} and \eqref{eq:DA}, then $T_f = T$ $(\bb P_X \otimes \bb P_{X,Y})$-a.s. \label{eq:id_uniqueness}
\end{enumerate}
\end{definition}

Lastly, we denote $\Aut(P) := \{ g \in \Aut(\bb{Y}): g_{\#} P = P \}$ as the set of transformations that leave distribution $ P \in \mc P(\bb Y)$ unchanged, $\Aut(P)|_{\bb{G}} := \Aut(P) \cap \bb{G}$ its restriction to $\bb{G}$, and $[ f ]_{P}$ as the set of functions equal to $f$, $P$-almost surely. With these concepts, we can now state our general identifiability result.

\begin{theorem}[Identifiability of Counterfactual Cocycle]
\label{thm:iden:group:coboundary}
    Let $T$ be a $\bb{P}_{Y|X}$-adapted, $\bb{G}$-valued cocycle. $T$ is identifiable in $\mc F_{\bb G}$
    if and only if \({\Aut(\bb P_{Y|X=x_0})|_{\bb{G}} \subseteq [ \id ]_{\bb P_{Y|X=x_0}}}\) for some $x_0 \in \bb X$.
\end{theorem}

Intuitively, the result is a consequence of the fact that (i) any $\bb G$-valued $\bb P_{Y|X}$-adapted cocycle has an almost sure coboundary representation 
$T_{x,x'} = f_x \circ f_{x'}^{-1}$ with $\{f_x\}_{x \in \bb X} \subseteq \bb G$, and (ii) any other cocycle $\tilde T$ differs only by a twist---i.e., 
$\tilde T_{x,x'} = f_x \circ b_x \circ b_{x'}^{-1} \circ f_{x'}^{-1}$ 
with $b_x \in \Aut(\bb P_{Y|X=x_0})|_{\bb G}$. Thus, identifiability requires that the elements of $\Aut(\bb P_{Y|X=x_0})|_{\bb G}$ behave like the identity map. Any non-trivial elements of $\Aut(\bb P_{Y|X=x_0})|_{\bb G}$ can lead to indeterminacy. This parallels identifiability results in latent variable models and nonlinear ICA \citep{khemakhem20a, xi2023indeterminacy}, where partially identified models can still be useful for some estimands \citep{syrota2025identifying}. 
Note that the choice of $x_0$ in \cref{thm:iden:group:coboundary} is arbitrary: if $g\neq_{\mathrm{a.s.}}\id$ is an automorphism of 
$\bb P_{Y|X=x_0}$ and $g \in \bb G$, then, for any $x\in\bb X$, $\hat g_x := T_{x,x_0} \circ  g \circ T_{x,x_0}^{-1}\neq_{\mathrm{a.s.}}\id$ 
is an automorphism of $\bb P_{Y|X=x}$ and $\hat g_x \in \bb G$. 

\paragraph{Identifiable Parameterizations via TMI Maps} In general, specifying models with a smaller group $\bb G$ preserves identifiability for a larger set of conditional distributions $\bb P_{Y|X}$, but risks violating \eqref{eq:DA}.  A practical way to achieve identifiability without risking \eqref{eq:DA} for continuous outcomes $Y$, is to constrain $\bb  G$ using knowledge of the causal ordering of variables.  In particular, suppose \(Y=(Y_1,\dots,Y_p)\in\bb R^p\) admit a known causal order \(Y_1\prec\cdots\prec Y_p\) (here we assume the variables are already permuted to reflect this order). To preserve the causal order, it is natural to enforce a lower triangular structure in the transports via $f_x$:
\begin{align*}
    f_x(y) =
        \bigl(f_{x,1}(y_1),\,f_{x,2}(y_1,y_2),\dots,f_{x,p}(y_1,\dots,y_p)\bigr)\;. 
\end{align*}
Requiring that each $f_{x,j}$ is strictly increasing in $y_j$ makes $f_x$ a \emph{triangular monotone increasing} (TMI) map. The set of such maps with full support forms a transformation group, $\bb G_{\text{TMI}}$.  It is known that, for any two absolutely continuous distributions $P, Q \in \mc P(\bb R^p)$, there is a (a.s.) unique TMI map $g \in \bb G_{\text{TMI}}$ such that $P = g_\# Q$ \citep{bogachev2005triangular}. Below we use this result to prove that, if $T$ is a  $\bb G_{\text{TMI}}$-valued cocycle adapted to \emph{any} $\bb P_{Y|X}$ (i.e., not necessarily absolutely continuous) and $\bb Y \subset \bb R^p$, then it is identifiable within $\mc F_{\bb G_{\text{TMI}}}$.

\begin{theorem}[Identifiability under TMI maps]\label{thm:tmi_uniqueness}
    Let $T$ be a $\bb P_{Y|X}$-adapted, $\bb G_{\mathrm{TMI}}$-valued cocycle and $\bb Y \subseteq \bb R^p$. Then $T$ is identifiable in $\mc F_{\bb G_{\mathrm{TMI}}}$. 
\end{theorem}

Since any family of \emph{autoregressive} flows (e.g., \cref{tab:flow_models}) lie in $\mc F_{\bb G_{\text{TMI}}}$, these architectures are a natural choice to model the coboundary map $f$ whilst guaranteeing identifiability,  by specifying $f(x,\argdot)$ as a flow with parameters conditioned on $x$. We cover implementation details in \cref{sec:implementation}. Although such architectures lead to mis-specification problems in SCMs in \cref{sec:background:scms}, we will see that using them to model the cocycle avoids those problems.

\subsection{Connection to SCMs}  \label{sec:method:scm}

The factorization structure of cocycles and their viable parameterization using the same function classes as flow-based SCMs suggests a close connection between counterfactual cocycles and SCMs. This connection is formalized below and has important implications.

\begin{theorem}[Cocycle Equivalence to Structural Model] \label{thm:scm_equivalence}
     A collection of counterfactual variables $\{Y(x)\}_{x \in \bb X}$ satisfies \cref{ass:OSA_counterfactuals} and \cref{ass:counterfactual_cocycle} with cocycle $T: \bb X^2 \times \bb Y \to \bb Y$ if and only if there is a function $f: \bb X \times \bb Y_0 \to \bb Y$ injective on its second argument, such that
    \[Y = _{\mathrm{a.s.}}f(X,\xi_Y), \quad \xi_Y \in \bb Y_0 \subseteq \bb Y, \quad  \xi_Y \ind X \,.\]
\end{theorem}

 The proof is straightforward up to some measurability technicalities. To illustrate the first direction, since \eqref{eq:CC} implies \eqref{eq:NI}, \eqref{eq:CPI} and \eqref{eq:DA}, by  \cref{thm:cocycle_factorization} we have $Y(x) =_{\mathrm{a.s.}} f_{x} \circ f_{x'}^+(Y(x'))$ for injective $f_x$ with left inverse $f_x^+$. Thus, one can set $\xi_Y :=  f_{x_0}^+(Y(x_0)) \in \bb Y_{0}$ and apply \eqref{eq:CC} to get $Y(x) =_{\mathrm{a.s.}} f_x(\xi_Y)$. Defining $f: (x,y) \mapsto f_x(y)$, \cref{ass:OSA_counterfactuals} gives $Y =_{\mathrm{a.s.}} f(X,\xi_Y)$ by consistency. $\xi_Y \ind X$ holds since $Y(x_0) \ind X \implies  f_{x_0}^+(Y(x_0)) \ind X$. 
 
 A direct consequence of \cref{thm:scm_equivalence} is that every counterfactual cocycle model \eqref{eq:CC} corresponds to an equivalence class of SCMs of the form
 \begin{align}
     \bm V := \begin{pmatrix}
     X \\ Y 
 \end{pmatrix} =_{\mathrm{a.s.}} \begin{pmatrix}
     \xi_X \\ f(\xi_X,\xi_Y)
 \end{pmatrix} =: F(\bm \xi) \;, \label{eq:scm}
 \end{align} 
 where $\xi_X \ind \xi_Y$ and $F : \bb V \to \bb V$ is injective by the injectivity of $f(x,\argdot)$. Each member of the equivalence class is defined by a different noise distribution $\bb P_{\xi} := \bb P_{\xi_X} \otimes \bb P_{\xi_Y}$. Naturally, when $f(x,\argdot)$ is bijective (or has a bijective extension), the equivalence is to a class of BCMs.
 
 Note that when $Y$ is multivariate, \eqref{eq:scm} shows only a partial factorization of $F$. However, under the TMI restriction in \cref{sec:method:id}, $f$ further decomposes into coordinate maps,
\[
Y_j \;=_{\mathrm{a.s.}}\; f_j\bigl(X,Y_{<j},\xi_{Y,j}\bigr), \quad \xi_{Y,j} \ind X,
\]
where each coordinate function $f_j$ strictly increasing in $\xi_{Y,j}$. This is precisely the identifiability restriction used in BCMs \citep{nasr2023counterfactual, javaloy2023causal} and so is a natural counterpart to the TMI restriction proposed in the previous subsection. However, we do \emph{not} need to assume $\xi_{Y,j} \ind \xi_{Y,i}$ for $i \neq j$ when deriving an SCM from a cocycle.

\definecolor{ok}{RGB}{46,160,67}    

\begin{figure}[t]
    \centering
\includegraphics[width = 0.7\textwidth]{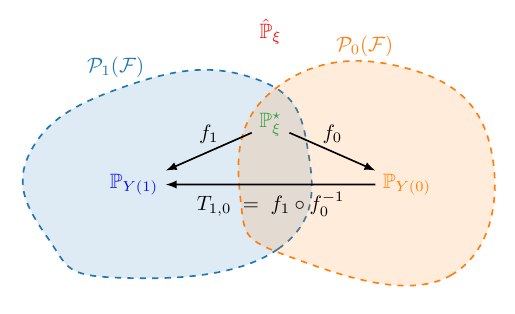}
\vspace{-10pt}
\caption{Illustration of robustness of counterfactual cocycle modeling in a randomized control trial (i.e., $Y(x)\sim \bb P_{Y\mid X=x}$ and $x\in{0,1}$). Each dashed region, ${\color{blue}\mc P_1(\mc F)}$ and ${\color{orange}\mc P_0(\mc F)}$, denotes the set of distributions reachable via pushforwards of $\bb P_{Y(1)}$ and $\bb P_{Y(0)}$ under inverses of elements of $\mc F\subset \{f:\bb Y\to\bb Y\}$. Flow-based SCMs require committing to a fixed base distribution ${\color{red}\widehat{\bb P}_\xi}$; the class $\mc F$ is well-specified if and only if ${\color{red}\widehat{\bb P}_\xi}$ lies in the intersection ${\color{blue}\mc P_1(\mc F)}\cap{\color{orange}\mc P_0(\mc F)}$. By contrast, cocycle models require only that the intersection is non-empty. In this case, there exists some ${\color{ok}\bb P_\xi^\star} \in {\color{blue}\mc P_1(\mc F)}\cap{\color{orange}\mc P_0(\mc F)}$ and $f_0,f_1
\in \mc F$ such that $(f_x)_{\#}{\color{ok}\bb P_\xi^\star}=\bb P_{Y(x)}$ for $x=0,1$, which gives the counterfactual cocycle $T(x,x',y) =f_x\circ f_{x'}^{-1}(y)$. Note the existence of $({\color{ok}\bb P_\xi^\star}, f_0,f_1)$ does not depend on which true noise distribution generated the data.}
\label{fig:overlap}
\vspace{-10pt}
\end{figure}

\paragraph{Noise Invariance of Counterfactual Cocycles}
A key difference between counterfactual cocycles and SCMs is that cocycles are 
\emph{noise-invariant}. To demonstrate, let $Y(x) := f_x(\xi)$ where each $f_x := f(x,\argdot)$ is bijective. Notice that each counterfactual transport
\begin{align}
   T_{x,x'} := f_{x} \circ f_{x'}^{-1} \quad , \quad (T_{x,x'})_{\#} : \bb P_{Y|X=x} \mapsto \bb P_{Y|X=x'} \;,
   \label{eq:cocycle_transport}
\end{align}
remains unchanged if we modify the SCM by changing the noise law $\bb P_\xi$.  This invariance suggests a practical robustness to directly targeting a cocycle that solves \eqref{eq:cocycle_transport}.
In particular, the flow-based SCM approach is to model the system in \eqref{eq:scm} by fixing some base distribution $\widehat{\bb P}_{\xi_Y}$ and class of functions $\mc F$ to model $f$ (note one can always set $\widehat{\bb P}_{\xi_X} = \widehat{\bb P}_X = \tfrac 1n \sum_i \delta_{X^{(i)}}$ (see \cite{nasr2023counterfactual}). As we showed in \cref{ex:mis-spec} in \cref{sec:background:scms}, whether there is a function $\hat f \in \mc F$ that correctly models the data distribution, i.e.,
\begin{align}
    (\hat f_x)_{\#}\widehat{\bb P}_{\xi_Y} = \bb P_{Y|X=x}, \quad \forall x \in \bb X,\label{eq:scm_transport_condition}
\end{align}
depends on whether the properties of the \emph{chosen} noise distribution $\widehat{\bb P}_{\xi_Y}$  (e.g., support and tails) matches the \emph{true} noise distribution $\bb P_{\xi_Y}$.  In contrast, the cocycle-based approach requires there to exist a function $\hat f \in \mc F$ that solves \eqref{eq:cocycle_transport} for every $x,x' \in \bb X$. This is true as long as there exists \emph{some} noise distribution $\bb P_\xi^\star \in \mc P(\bb Y)$ such that \eqref{eq:scm_transport_condition} holds:
\begin{align}
    \exists \,\bb P_\xi^\star \in \mc P(\bb Y) \; : \; (\hat f_x)_{\#} {\bb P}_{\xi}^\star = \bb P_{Y|X=x}, \quad \forall x \in \bb X \;.\label{eq:cocycle_transport_condition}
\end{align}
To see this, note that \eqref{eq:cocycle_transport_condition}  implies $(\hat f_x^{-1})_{\#}\bb P_{Y|x=x} =_d (\hat f_{x'}^{-1})_{\#}\bb P_{Y|x=x'}$ for every $x$ and $x'$, which implies \eqref{eq:cocycle_transport}. \eqref{eq:cocycle_transport_condition} is a much weaker requirement than  \eqref{eq:scm_transport_condition} and, for instance, is satisfied in \cref{ex:mis-spec} for some $\hat f \in \mc F$ when using most of the flow classes in \cref{tab:flow_models}. Moreover, whether  \eqref{eq:cocycle_transport_condition} is satisfied does not depend on the true noise distribution of the corresponding SCM. \cref{fig:overlap} illustrates this point in a simple binary treatment setting. These properties are analyzed in greater detail in \cref{sec:cocycle:vs:scm} and underpin our approach to estimation.

\section{Counterfactual Cocycles: General Framework Under Confounding} \label{sec:methodgeneral}

Having established a viable approach to modeling counterfactual transports in a simple set-up, we now present our framework under a more general setting, cover our approach to model parameterization, and how we propose to estimate causal quantities with cocycles.

\subsection{Counterfactual Cocycle Models Under Partial Orderings and Confounding} \label{sec:method:confounding}

Let $\bm V := (V_1,\dots, V_d) \in \prod_j^d \bb V_j \subset \bb R^d$ now denote the full set of observed variables. We retain focus on a structured setting where \(X:= (X_1,\dots,X_q) \subset \bm V\) are a set of (`treatment') variables we wish to manipulate, $Y := (Y_1,\dots,Y_p) \subset \bm V$ are a set of outcomes of interest that may be affected by the treatments, and \(Z := (Z_1,\dots,Z_l) \subset \bm V\) are a collection of `pre-treatment' covariates which may confound the effect of any $X_j$ on $Y_i$. That is, we assume the variable sets $(Z,X,Y)$ satisfy the following \emph{partial} causal ordering, \vspace{-5pt}
\begin{align}
   Z \prec X \prec Y \;, \label{eq:order} 
\end{align}\vspace{-20pt}

where $Z \prec X \implies Z_i \prec X_j$ for every $(i,j) \in \{1,\dots,l\} \times \{1,\dots, q\}$.
\cref{fig:back_door} shows several causal DAGs consistent with this ordering. Although we are yet to discuss unobserved confounding, the examples in \cref{fig:back_door} are compatible with the framework we develop here. Note that, when $X \in \bm V$ is a single treatment, one can trivially find sets $Z,Y \subset \bm V$ that satisfy \eqref{eq:order} for \emph{any} causal DAG over $\bm V$. The only notable exclusion is the dynamic treatment regime, where intermediate outcomes can affect subsequent treatments. However, we do not envisage any fundamental challenge extending to that setting in future work. 

The partial ordering \eqref{eq:order} lets us straightforwardly extend the framework laid out in \cref{sec:method}. The basic idea is to specify \cref{ass:OSA_counterfactuals} and \cref{ass:counterfactual_cocycle} but on a set of counterfactuals $\{Y(x,z)\}_{x,z \in \bb X \times \bb Z}$ under both levels of treatments $X$ and covariates $Z$. These counterfactuals relate to the counterfactuals of interest $\{Y(x)\}_{x \in \bb X}$ via an additional `nested consistency' property, which is commonly used when combining counterfactuals with graphs \citep{richardson2013single, shpitser2016causal, malinsky2019potential}. In particular, our counterfactual assumptions are as follows.

\begin{figure}[!t]
\centering
\begin{minipage}{0.3\linewidth}
\begin{tikzpicture}[>=stealth,
        every node/.style={draw,circle,inner sep=1pt,minimum size=16pt, node distance = 15pt}]
  \node (Z1)  {$Z_1$};
  \node[below=of Z1] (Z2) {$Z_2$};
  \node[right=of Z1, yshift = -6mm] (X) {$X_1$};
  \node[right=of X, yshift=6mm] (Y1) {$Y_1$};
  \node[right=of X, yshift=-6mm] (Y2) {$Y_2$};
  \node[dashed, gray, above=of X, yshift = 6mm] (U) {$U$};

  \draw[->] (Z1) -- (X);
  \draw[->] (Z1) -- (Z2);
  \draw[->, bend left=20] (Z1) to (Y1);

  \draw[->] (Z2) -- (X);
  \draw[->, bend right=20] (Z2) to (Y2);

  \draw[->] (X) -- (Y1);
  \draw[->] (X) -- (Y2);

  \draw[->] (Y1) -- (Y2);

  \draw[->, gray, dashed] (U) -- (X);
  \draw[->, gray, dashed] (U) -- (Z1);
\end{tikzpicture}
  
\end{minipage}
\begin{minipage}{0.3\linewidth}
    \begin{tikzpicture}[>=stealth,
        every node/.style={draw,circle,inner sep=1pt,minimum size=16pt, node distance = 15pt}]
\node (X)  {$X_1$};
\node[right= of X] (Y2) {$Y_2$};
\node[above= of Y2] (Y1) {$Y_1$};
\node[below= of Y2] (Y3) {$Y_3$};

\node[dashed, right= of Y2, gray] (U) {$U$};

\draw[->] (X)  -- (Y1);
\draw[->] (X)  -- (Y2);
\draw[->] (X)  -- (Y3);

\draw[->, gray, dashed] (U) -- (Y1);
\draw[->, gray, dashed] (U) -- (Y2);
\draw[->, gray, dashed] (U) -- (Y3);
\end{tikzpicture}
\end{minipage}
\begin{minipage}{0.3\linewidth}
    \begin{tikzpicture}[>=stealth,
        every node/.style={draw,circle,inner sep=1pt,minimum size=16pt, node distance = 15pt}]

  \node[dashed, gray] (U) {$U$};
  \node[right =of U] (X2) {$X_2$};
  \node[above= of X2] (X1) {$X_1$};
  \node[below= of X2] (X3) {$X_3$};
  \node[right =of X2] (Y) {$Y_1$};
  \node[right =of Y] (Z2) {$Z_2$};
  \node[above =of Z2] (Z1) {$Z_1$};
v  \node[below =of Z2] (Z3) {$Z_3$};

  \draw [->,dashed, gray] (U) -- (X1);
  \draw [->,dashed, gray] (U) -- (X2);
  \draw [->,dashed, gray] (U) -- (X3);

  \draw  [->] (X1) -- (Y);
  \draw  [->] (X2) -- (Y);
  \draw  [->] (X3) -- (Y);

  \draw[->] (Z3) -- (Y);
  \draw[->] (Z2) -- (Y);
  \draw[->] (Z1) -- (Y);

\end{tikzpicture}

\end{minipage}

\caption{DAGs consistent with ordering in \eqref{eq:order} and \cref{ass:counterfactuals:confounded}. $U$ is unobserved. }
\label{fig:back_door}
\end{figure}

\begin{assumption}[Counterfactual Cocycles with Covariates] \label{ass:counterfactuals:confounded}
Let $\{Y(x,z)\}_{(x,z) \in \bb X \times \bb Z}$ satisfy:
\begin{enumerate}
    \item \textbf{Consistency:} (i) $Y(X,Z) =_{\mathrm{a.s.}} Y$ and (ii)  $Y(x,Z) =_{\mathrm{a.s.}} Y(x)$  \label{eq:consistency_confounding} \vspace{-1pt}
    \item \textbf{Exchangeability:} $\{Y(x,z)\}_{(x,z)} \ind (X,Z)$  \label{eq:indep_confounding} \vspace{-1pt}
    \item \textbf{Counterfactual Cocycle:} $Y(x,z) =_{\mathrm{a.s.}} T_{(x,z),(x',z')}(Y(x',z')),  \forall (x,x',z,z') \in \bb X^2 \times \bb Z^2$ \label{eq:transports_confounding}
\end{enumerate}
\end{assumption} \vspace{-10pt}
We stress that these assumptions do not further restrict the causal structure implied by \eqref{eq:order}. They simply provide a compatible potential outcomes representation which lets us formalize counterfactual transports in this setting. Note that \cref{ass:counterfactuals:confounded}.\ref{eq:consistency_confounding} and \ref{ass:counterfactuals:confounded}.\ref{eq:indep_confounding} imply the well-known `strong ignorability' criteria used in the potential outcomes literature,
\begin{align}
    Y(X) =_{\mathrm{a.s.}} Y  \quad \text{and } \quad  \{Y(x)\}_{x \in \bb X} \;\ind\; X \mid Z \;. \tag{SI} \label{eq:SI}
\end{align}
The assumptions identify $\bb P_{Y(x)}$ via the adjustment formula: $\bb E [h(Y(x))] = \bb E[\bb E[h(Y)\mid X=x,Z]]$ \citep{rubin1974estimating}. However \cref{ass:counterfactuals:confounded}.\ref{eq:consistency_confounding} and \ref{ass:counterfactuals:confounded}.\ref{eq:indep_confounding} are stronger than \eqref{eq:SI} as they additionally identify $\bb P_{Y(x,z)} = \bb P_{Y|X=x,Z=z}$. In terms of the implied causal graphs, this precludes any direct confounding of $Z \leftrightarrow Y$, but does allow confounding within each block $Z,X,Y$ and between $Z \leftrightarrow X$ (see \cref{fig:back_door}). The main benefit of precluding confounding between $Z \leftrightarrow Y$ is it lets us reduce the problem of recovering a coupling over counterfactuals $\{Y(x)\}_{x \in \bb X}$, to the problem of estimating transports $\{T_{(x,z),(x',z')}\}_{(x,z) \in \bb X \times \bb Z}$ between conditional distributions $(\bb P_{Y|X=x,Z=z})_{x,z \in \bb X \times \bb Z}$ (which we show how to do in \cref{sec:estimation}). This is because, by the nested consistency \cref{ass:counterfactuals:confounded}.\ref{eq:consistency_confounding}.(ii) and counterfactual cocycle \cref{ass:counterfactuals:confounded}.\ref{eq:transports_confounding}, the transports $\{T_{(x,z),(x',z')}\}_{(x,z) \in \bb X \times \bb Z}$ determine the (stochastic) coupling over $\{Y(x)\}_{x \in \bb X}$:
\begin{align}
    Y(x) =_{\mathrm{a.s.}} T_{(x,Z),(x',Z)}(Y(x')) \;. \label{eq:stochastic_transports}
\end{align}

\paragraph{Implied Causal Model} Since \cref{ass:counterfactuals:confounded} is equivalent to \cref{ass:OSA_counterfactuals} and \cref{ass:counterfactual_cocycle} but on an augmented set of `treatments' $\tilde X := (X,Z)$, all results in \cref{sec:method} apply equivalently to the set of transports on $\{Y(x,z)\}_{(x,z) \in \bb X \times \bb Z}$. In particular, the set of admissible transports must satisfy \eqref{eq:NI}, \eqref{eq:CPI} and \eqref{eq:DA} w.r.t.\ $\bb P_{Y|X,Z}$. This in turn implies that \(T_{(x,z),(x',z')} =_{\mathrm{a.s.}} f_{x,z} \circ f^+_{x,z'}
\)
for some injective $f_{x,z}: \bb Y_0 \to \bb Y$ and that \(Y =_{\mathrm{a.s.}}  f(X,Z,\xi_Y),\ \xi_Y \ind (X,Z)\). Therefore, any joint BCM over the variable blocks,
\begin{align}
\bm V=\begin{pmatrix}Z\\X\\Y\end{pmatrix}
=_{\mathrm{a.s.}}\begin{pmatrix} \xi_Z\\ h\big(\xi_Z,\xi_X\big)\\ f\big(h(\xi_Z,\xi_X),\,\xi_Z,\,\xi_Y\big)\end{pmatrix} := F(\bm \xi) \;,\label{eq:scm_confounded}
\end{align}
is again consistent with the cocycle $T$. Different choices of $h$ and reparameterizations of the noise $(\xi_Z,\xi_X,\xi_Y)$ all correspond to the \emph{same} cocycle by noise invariance. Thus, counterfactual cocycles provide the \emph{minimal structure} needed to recover the required couplings and counterfactuals, without committing to a full SCM for $(Z,X)$, noise distribution $\bb P_\xi$, or enforcing the noise to factorize within the blocks $(Z,X)$ and $Y$.

The SCM equivalence makes clear that even if $Z \ind X$ (i.e., $Z$ are not confounders, as in \cref{fig:back_door}, right), if $Z \cancel{\ind}Y$ including them explicitly in the model formulation may still be required to satisfy the injectivity assumption implied by the cocycle. That is, we may only have $\dim (\supp (\bb P_{\xi})) \leq \dim( \supp (\bb P_Y))$ after including enough measured causes $Z$ of $Y$. This is an important distinction from estimating marginal causal effects, where conditioning on unnecessary covariates can increase estimation variance \citep{henckel2022graphical}.

\subsection{Cocycle Parameterization and Refinement}

\paragraph{TMI Restrictions} Following the analysis in  \cref{sec:method:id}, for modeling practicality and identifiability we propose to paramterize cocycles using invertible TMI maps, 
\begin{align}
    T_{(x,z),(x',z')}(y') :=  f_{x,z} \circ f_{x',z'}^{-1}(y') \quad \text{ where } \quad  f_{x,z} \in \bb G_{\mathrm{TMI}}\,. \label{eq:tmicocycle}
\end{align}
In the present context, this essentially imposes a known partial causal ordering between the outcomes, i.e., $Y_1\prec \dots \prec Y_p$ and implies an underlying acyclic SCM, 
\begin{align}
Y_j =_{\mathrm{a.s.}} f_j(X,Z,Y_{<j},\xi_{Y,j}), \quad \xi_{Y,j} \ind X,Z\;. \label{eq:tmiscm}
\end{align}
While this is not the only possible restriction that can achieve identifiability, it is commonly used in the literature \citep{javaloy2023causal, nasr2023counterfactual, machado2024sequential} and, as discussed  in \cref{sec:method:id}, guarantees a well-defined transport whenever $Y_1,\dots,Y_p$ are continuous on $\bb R^p$. In practice, we parameterize the coboundary map $f(x,z,y) := f_{x,z}(y)$ using classes of autoregressive flows on $y$ (\cref{tab:flow_models}), with flow parameters conditioned on the values of $x,z$. Implementation details are in \cref{sec:implementation}.

\paragraph{Incorporating Causal DAGs} If $\mc G$ is a known causal DAG over $\bm V$, we can additionally restrict the transports to reflect the sparsity of the direct effects. In particular, if $Y_{\pa(j)} \subset (X,Z,Y_{<j})$ are the parents of $Y_j$ in $\mc G$, then the SCM is of the form
\[Y_j =_{\mathrm{a.s.}} f_j(Y_{\pa(j)},\xi_{Y,j}), \quad \xi_{Y,j} \ind X,Z \;.\]
 In \cref{sec:implementation}, we discuss how we practically constrain the coboundary map to reflect this sparsity. The main idea is to use TMI maps with masks reflecting the structure of the DAG.

\subsection{Estimating Causal Quantities with Cocycles} \label{sec:method:estimation}
Any cocycle parameterized as in \eqref{eq:tmicocycle}  trivially satisfies \eqref{eq:NI} and \eqref{eq:CPI} on all of $\bb Y$. Thus, all that remains is to enforce \eqref{eq:DA} w.r.t.\ $\bb P_{Y|X,Z}$ and use the resulting cocycle to estimate causal quantities. Here we overview our proposed procedure, deferring details on how to estimate the cocycle itself to \cref{sec:estimation}. In short, our estimation procedure is centered entirely around the cocycle and avoids ever specifying or referencing a latent noise distribution. This lets us take advantage of the invariances of counterfactual cocycles discussed in \cref{sec:method:scm} and later in \cref{sec:cocycle:vs:scm}. In what follows, we define $\tilde X := (X,Z)$ for convenience.

\paragraph{Cocycle Estimation} Rather than specifying a base distribution $\widehat{\bb P}_{\xi}$ and learning a flow \emph{to} the conditional as $f_{\tilde x}: \widehat{\bb P}_{\xi} \mapsto \bb P_{Y|\tilde X=\tilde x}$, we directly target the transports \emph{between} the conditionals, as $T_{\tilde x,\tilde x'} = f_{\tilde x} \circ f_{\tilde x'}^{-1}: \bb P_{Y|\tilde X=\tilde x} \mapsto \bb P_{Y|\tilde X=\tilde x'}$. We do so by minimizing a tractable empirical analogue to a distributional discrepancy in the \eqref{eq:DA} property,
\begin{align*}
    \tilde \ell(T) :=  \bb E_{\tilde x,\tilde x' \sim \bb P_{\tilde X}}D(\bb P_{Y|\tilde X=\tilde x}, (T_{\tilde x,\tilde x'})_{\#}\bb P_{Y|\tilde X=\tilde x'})^2   \;.
\end{align*}
In \cref{sec:estimation} we derive this estimator in detail, and show that, under general conditions, the consistency of this estimator does not depend on any distributional assumptions that follow from the (true) latent noise $\xi$ (see \cref{remark:robustness} in \cref{sec:estimation}).

\paragraph{Causal Quantity Estimation} Rather than using the abduct-act-predict procedure in \cref{sec:background:scms}, which may require sampling from a prior or posterior base distribution, we use the cocycles to directly impute the counterfactuals of interest. In particular, given an i.i.d. dataset $\{Z^{(i)}, X^{(i)},Y^{(i)}\}_{i=1}^n \sim \bb P_{Z,X,Y}$, we can impute for each unit the counterfactual outcomes at treatment levels $x_1,\dots,x_m$ using \eqref{eq:stochastic_transports} and the consistency property:
\[
\{\hat Y^{(i)}(x_1),\dots,\hat Y^{(i)}(x_m)\} \;=\; \{\hat T_{(x_1,Z^{(i)}), ( X^{(i)}, Z^{(i)})}\bigl(Y^{(i)}\bigr),.., \hat T_{(x_m,Z^{(i)}), ( X^{(i)}, Z^{(i)})}\bigl(Y^{(i)}\bigr)\}
\]

The imputed counterfactuals can then be used to estimate causal quantities via standard empirical and/or nonparametric techniques.  Below are some examples using empirical averaging and kernel density estimation with smoothing kernel $K_{\lambda}$:

\begin{align}
&\text{Average Effect:}
&&\widehat{\bb E}\bigl[Y(x)-Y(0)\bigr]
=\frac1n\sum_{i=1}^n\bigl(\hat Y^{(i)}(x)-\hat Y^{(i)}(0)\bigr) \label{eq:ATE},\\
&\text{True Harm Rate:}
&&\widehat{\bb P}\bigl(Y(x)-Y(0)\preceq 0\bigr)
=\frac1n\sum_{i=1}^n\bm 1\{\hat Y^{(i)}(x)-\hat Y^{(i)}(0)\preceq 0\} \label{eq:DTE} \\
&\text{Density of Effect:}
&&\widehat{p}_{Y(x)-Y(0)}\bigl(y\bigr)
=\frac1n\sum_{i=1}^nK_{\lambda}(\hat Y^{(i)}(x)-\hat Y^{(i)}(0)-y) \label{eq:TED}
\end{align} 

One can also condition such quantities on covariates $W \subset Z$ to examine effect heterogeneity. For this, one can replace the empirical averages in \eqref{eq:ATE}-\eqref{eq:TED}, with \emph{weighted} averages estimated nonparametrically. For example, the conditional THR can be estimated as
    \begin{align}
        \widehat{\bb P}(Y(x) - Y(0) \preceq 0 \mid W=w) = \sum_{i=1}^n \hat \alpha_i(w) \bm 1\{\hat Y^{(i)}(x) - \hat Y^{(i)}(0) \preceq 0 \}
    \end{align}
    where $\hat \alpha_i(\cdot)$ are smoothing weights estimated via nonparametrically regressing $\bm 1\{\hat Y^{(i)}(x) - \hat Y^{(i)}(0) \preceq 0 \}$ on $W^{(i)}$ using e.g., Nadaraya--Watson or RKHS regression.  While we note that nonparametrics can suffer from slow rates of convergence in high-dimensions, in causal inference $W$ is typically a low-dimensional (e.g., 1d or 2d) set of interest. 

\section{Cocycle Estimation} \label{sec:estimation}

To leverage any benefits of directly modeling counterfactual cocycles, we require
a tractable way to estimate them without making further assumptions on
$\bb P_{Y\mid X,Z}$. This task is non-trivial for general flow classes beyond
simple additive models, where one cannot recover the cocycle merely by regressing
$Y$ on functions of $X$ and $Z$. In what follows, we develop our estimation
procedure from first principles, establish its asymptotic properties, and
demonstrate its performance on the problems in \cref{ex:mis-spec}.

For notational simplicity, we henceforth write $X$ in place of the augmented
variable $(X,Z)$, since including covariates does not affect any of the
results below.

\subsection{Targeting (DA) via Distributional Discrepancy}

Given a parameterized model $\mc F$ for the coboundary map $f: (x,y) \mapsto f_x(y)$ of a cocycle $T$, a natural estimation criterion is to minimize an expected distance that enforces \eqref{eq:DA} across all conditionals. In particular, denoting by $D$ a divergence or metric on $\mc P(\bb Y)$, the set of distributions on $\bb Y$, the criterion evaluated at $f \in \mc{F}$ is
\begin{align}
   \ell_0(f) = \;\bb E_{X,X'\sim \bb P_X}\,
    D\bigl(\bb P_{Y\mid X},\,(f_X\circ f_{X'}^{-1})_\#\bb P_{Y\mid X'}\bigr)^2 \;.\label{eq:opt-f}
\end{align}
where $\bb P_{Y|X} := \bb P(Y \in \argdot|X)$ is treated as a random probability measure on $\bb Y$, and $f_X$ a random function $\bb Y \to \bb Y$.
When the cocycle is identifiable from a model $\mc F$ (e.g., $\mc{F} \subset \mc{F}_{\bb G_\textrm{TMI}}$---see \cref{sec:method:id}), 
then by definition $\ell_0$ admits a minimizer that is almost-everywhere unique. Unfortunately, evaluating $\ell_0$ requires knowledge of $\bb P_{Y|X}$, which defeats the purpose of modeling only the cocycle. Our aim is to modify this objective in a way that bypasses the need to estimate conditional distributions, without harming identifiability. To that end,  we choose $D$ to be the Maximum Mean Discrepancy (MMD) \citep{gretton2012kernel}, which induces the following loss in terms of the cocycle $T$
\begin{align}
    \ell_0(T) =\; \bb E_{X,X'\sim \bb P_X} \lVert \bb E[\psi(Y)|X] - \bb E[\psi(T_{X,X'}(Y'))|X,X']\rVert_{\mc H_k}^2 \;.\label{eq:cmmd_1}
\end{align}
Here $(X,Y) \ind (X',Y')$ are independent copies and $\psi: \bb Y \to \mc H_k$ is a feature map to a reproducing kernel Hilbert space (RKHS) $\mc H_k$ associated with a positive-definite kernel $k: \bb Y^2 \to \bb R$, in the sense that $\psi(y) = k(y,\cdot)$. Note, we assume the kernel $k$ is \emph{characteristic} so that the MMD is indeed a metric on $\mc P(\bb Y)$ \citep{sriperumbudur11a}. Popular characteristic kernels include the Gaussian kernel $k(y,y') = \exp(-\lambda \|y-y'\|^2_2)$, and Laplace kernel $k(y,y') = \exp(-\lambda \|y-y'\|^1_1)$. While one could attempt to estimate $\ell_0$ from data via nonparametric conditional mean embedding estimation \citep{li2022optimal}, we will modify the MMD objective so that it can be estimated using simple empirical averages.

\paragraph{Modifying the Objective} We start by using the standard identity $\bb E[\|W\|^2_{\mc H_k}] = \|\bb E [W]\|^2_{\mc H_k} + \text{Tr}[\text{Cov}_{\mc H_k}[W]]$ for any RKHS-valued random variable $W \in \mc H_k$ with $\bb E \|W\|^2_{\mc H_k} < \infty$ \citep{berlinet2011reproducing}, where
\[
\text{Cov} _{\mc H_k}[W]
:=
\bb E\bigl[(W - \bb E[W]) \otimes (W - \bb E[W])\bigr] 
\;\in\;
B_1(\mc H_k)
\]
is the covariance operator, \(B_1(\mc H_k)\) is the set of trace‐class operators on \(\mc H_k\), and 
\(\text{Tr}\colon B_1(\mc H_k)\to \bb R\)
is the standard trace functional. Applying this identity to the norm inside the outer expectation in \eqref{eq:cmmd_1} with $W = \psi(Y) - \bb E[\psi(T_{X,X'}(Y')|X=x,X'=x']$,  we get, 
\begin{align}
   \hspace{-10pt}\tilde \ell(T) :=  \ell_0(T) - \text{Tr}[\text{Cov}[\psi(Y)]] = \bb{E}_{(X',X,Y) \sim \bb P_{X} \otimes \bb P_{X,Y}}\lVert \psi(Y) -  \bb{E}[\psi(T_{X,X'}(Y'))|X,X'] \rVert ^2_{\mc H_k} \!\!\!\label{eq:cmmd_2}
\end{align}
Note that minimizing $\tilde \ell$ is equivalent to minimizing $\ell_0$, so we can simply work with $\tilde \ell$ instead. This removes one of the two conditional expectations in \eqref{eq:cmmd_1}. Now, to remove the other conditional expectation, we pass the expectation over $X'$ \emph{inside} the norm, yielding
\begin{align}
    \ell(T) = \bb{E}_{(X,Y) \sim \bb P_{X,Y}}\,\lVert \psi(Y) -  \bb E[\psi(T_{X,X'}(Y'))|X] \rVert ^2_{\mc H_k} \;.
\end{align}
Using the reproducing property of $k$ (i.e., $k(y,y') = \langle \psi(y), \psi(y') \rangle_{\mc H_k}$, the resulting loss function can be expressed as an expectation of a real-valued function,
\begin{align}
   \hspace{-10pt} \ell(T) = {\bb E}_{{\scriptscriptstyle (X,Y),(X',Y'), (X'',Y'') \sim \bb P_{X,Y}}}\!\Bigl(\!k(Y,Y) + k(T_{X,X'}(Y'),T_{X,X''}(Y'') ) -2 k(Y, T_{X,X'}(Y'))\!\Bigr)\!\!\!\! \label{eq:CMMD}
\end{align}
We note that exchanging the expectation over $X'$ with the norm does not guarantee to preserve the minimizing set. However, below we prove that any minimizer $T^*$ of $\ell$ in a set of $\bb G$-valued cocycles satisfies \eqref{eq:NI}, \eqref{eq:CPI}, and \eqref{eq:DA} almost surely.

\begin{theorem}[CMMD Identifiability]\label{thm:cocycle_recovery}
    Let $T$ satisfy \eqref{eq:CC} w.r.t.\ $\{Y(x)\}_{x \in \bb X}$ and $T \in \mc T_{\bb G}$, where $\mc T_{\bb G} := \{T_f : f \in \mc F_{\bb G}\}$ is the set of cocycles constructed from coboundary maps in $\mc F_{\bb G}$. Let $\ell$ be defined using characteristic kernel $k: \bb Y^2 \to \bb R$. Then, any $T^* \in \text{arginf}_{\tilde T \in \mc T_{\bb G}}\ell(\tilde T)$ satisfies \eqref{eq:NI}, \eqref{eq:CPI} and \eqref{eq:DA} for $(\bb P_{X} \otimes \bb P_{X})$-almost all $x,x'$.
\end{theorem}
The restriction to $\bb G$-valued cocycles can be relaxed, but as with earlier results lets us avoid certain complexities working with monoids. An immediate consequence of this result is that, whenever $\mc F \subset \mc F_{\bb G_{\text{TMI}}}$ (or in any other situation in which $\mc{F}$ is identifiable), the minimizer of $\ell$ identifies the counterfactual cocycle $T^*$. Going forward, we refer to the objective \eqref{eq:CMMD} as the \textbf{CMMD loss} (short for Cocycle MMD). Intuitively, one can view CMMD as minimizing the (average) error between true and predicted counterfactuals in Hilbert space, $\{\psi(Y(x^{(i)}))\}_{i=1}^n$, where the predicted counterfactual at $x^{(i)}$ is just the average transformed embedding $\widehat {\psi(Y(x^{(i)}))} := \tfrac1n \sum_{j \neq i} \psi(\hat T_{x^{(i)},x^{(j)}}(Y(x^{(j)}))$.

\paragraph{Tractable Empirical Analogues}
The only expectations in \eqref{eq:CMMD} are over $\bb P_{X,Y}$. Therefore, given data $\mc D_n = \{(X^{(i)},Y^{(i)})\} \sim_{iid} \bb P_{X,Y}$, one can replace the population expectations with empirical ones. This gives rise to the following empirical V-statistic and U-statistic estimators for $\ell$ (dropping all terms independent of $T$):\vspace{-3pt}
\begin{align}
   \ell^V_n(T) & = -\frac{2}{n^2}\sum_{i,j}^nk(Y^{(i)}, Y_T^{(i,j)})+ \frac{1}{n^3}\sum_{i,j,k}^nk(Y_T^{(i,j)},Y_T^{(i,k)})\label{eq:CMMD_3_n} \\[-1pt]
 \ell^U_n(T) & = -\frac{2}{n(n-1)}\sum_{i\neq j}^nk(Y^{(i)}, Y_T^{(i,j)}) + \frac{1}{n(n-1)(n-2)}\sum_{i\neq j \neq k}^nk(Y_T^{(i,j)},Y_T^{(i,k)})\label{eq:CMMD_3_n_U} \;,
\end{align}\vspace{-8pt}

where $Y_T^{(i,j)} := T_{X^{(i)}, X^{(j)}}(Y^{(j)})$. Both loss functions $\ell_n^V$ and $\ell_n^U$ can be used to optimize flow-based cocycles via any gradient-based algorithm (e.g., ADAM) and model select between different flow classes. Implementation details are in \cref{sec:implementation}.

\subsection{Properties of CMMD Estimation}
We now analyze the theoretical properties of CMMD estimation. We start off by verifying that both empirical analogues $\ell_n^V, \ell_n^U$ converge to $\ell$ at $\sqrt{n}$-rate under general conditions.

\begin{proposition}
\label{prop:MMD_prob_bound}
   Let $\{(X^{(i)}, Y^{(i)})\}_{i=1}^n \sim_{iid}\bb P_{X,Y}$. For any cocycle $T$ and bounded kernel $k$, we have $\ell^V_n(T) = \ell(T) + c + \mc{O}_P(n^{-\frac{1}{2}})$, where $c \in \bb R$. The same holds for $\ell^U_n$.
\end{proposition}
We next turn to asymptotic analysis of the resulting estimators. For this, we work under standard parametric assumptions on the model class, and assume the cocycle is identifiable and that the kernel is sufficiently regular (as satisfied by typical kernel choices).
\begin{assumption} \emph{(CMMD Consistency)} \label{ass2:consistency} 
\begin{enumerate}
    \item Compactness: $\Theta$ is a compact subset of $\mathbb R^d$.
    \item Continuity:  $\theta \mapsto T_{\theta,x,x'}(y)$ is continuous for every $(x,x',y) \in \bb X^2 \times \bb Y$.
    \item Identifiability: $M = \arg\min_{\theta \in \Theta}\ell(\theta) \neq \emptyset$. $\theta_1, \theta_2 \in M \implies T_{\theta_1} = T_{\theta_2}$ $(\bb P_{X}\otimes \bb P_{X,Y})$-a.s.
    \item Kernel Regularity: The kernel $k$ is continuous and bounded with $\sup_{y,y'}|k(y,y')| \leq 1$. 
\end{enumerate}
\end{assumption}

 Under these conditions, we have the following strong consistency result based on the U-statistic \eqref{eq:CMMD_3_n_U}. By standard theory (e.g., Theorem 5.2.9 in \cite{de2012decoupling}) analogous results hold for the V-statistic \eqref{eq:CMMD_3_n}.

\begin{theorem}[CMMD Strong Consistency] \label{thm:CMMD_consistency} 
Let $\{(X^{(i)}, Y^{(i)})\}_{i=1}^n \sim_{iid} \bb P_{X,Y}$ and $\hat \theta_n \in \text{argmin}_{\theta \in \Theta} \ell^U_n(\theta)$ be a measurable minimizer.
 If \cref{ass2:consistency} holds, then
 \[\inf_{\theta \in M}\lVert\hat \theta_n  - \theta \rVert_2 \xrightarrow[a.s.]{n \to \infty} 0 \text{ a.s.}\] 
 Moreover, $\exists \, \theta_0 \in M$ such that $T_{\theta_n} \overset{n \to \infty}{\to} T_{\theta_0}$ a.s., for $\bb P_{X} \otimes \bb P_{X,Y}$ almost all ($x,x',y')$.
\end{theorem}  

\begin{remark} \label{remark:robustness}
When the data are generated by a BCM $Y = f(X,\xi)$, the only condition in \cref{ass2:consistency} which may be depend on $\bb P_{\xi}$  is the identifiability criterion  \cref{ass2:consistency}.3. However, this condition only requires that the underlying \emph{cocycle} $T$ is almost-everywhere unique, rather than the \emph{parameterization} $\theta_0$. Since there is at most one (a.s. unique) TMI map transporting between two distributions on $\bb R^p$ (\cref{thm:tmi_uniqueness}), as long as $\mc F \subset \mc F_{\bb G_{\text{TMI}}}$ and the model is well-specified, \cref{thm:CMMD_consistency} must hold for any BCM with function $f$. Thus, like the cocycle itself, our CMMD estimator also enjoys an invariance to the noise distribution. In contrast, likelihood-based estimators for the flow from a base distribution can fail to converge for certain noise distributions, as discussed in \cref{sec:background:scms}. Likewise, in \cref{sec:cocyclevsscm:seqot} we will see conditional-quantile based SCM estimators can be biased under dependent noise.
\end{remark}

Under the following additional regularity conditions we obtain $\sqrt{n}$-consistency of the U-statistic estimator to the minimizing set. We expect an analogous result for the V-statistic.
\begin{assumption}\label{ass3:normality} 
\begin{enumerate}\emph{(Additional Regularity for $\sqrt{n}$-rate)}
  \item {Lipschitz Cocycle:}  There exists a measurable function \(L_T: \bb X^2 \times \bb T \to \bb R_{>0}\) with 
    \(\bb E[L_T(X,X',Y')^2]<\infty\), such that for all \(\theta,\theta'\in\Theta\),
    \[
      \bigl\|T_{\theta,x,x'}(y')-T_{\theta,x,x'}(y')\bigr\|_2
      \;\le\;
      L_T(x,x',y')\,\|\theta-\theta'\|_2 \quad (\bb P_{X} \otimes \bb P_{X,Y})\text{-a.s.}
    \]
  \item {Kernel Derivative Regularity:}  $\partial k : (y,y') \mapsto \nabla_yk(y,y')$ is continuous and bounded. 
  \item {Local Strong Convexity:}  There exist \(c,\delta>0\) such that whenever 
    \(\inf_{\theta'\in M}\|\theta-\theta'\|_2\le\delta\),
    \[
      \| \nabla_{\theta}\ell(\theta) \|_2\ge\;
      c\,\inf_{\theta'\in M}\|\theta-\theta'\|_2.
    \]
\end{enumerate}
\end{assumption}

\begin{theorem}[\(\sqrt n\)-Rate of CMMD]\label{thm:cmmd-rate}
Let $\{(X^{(i)}, Y^{(i)})\}_{i=1}^n \sim_{iid} \bb P_{X,Y}$ and \(\hat\theta_n\in\arg\min_{\theta\in\Theta}\ell_n^U(\theta)\) be a measurable minimizer. Then, under Assumption \ref{ass2:consistency} and Assumption \ref{ass3:normality}, we have
\[
  \inf_{\theta\in M}\|\hat\theta_n-\theta\|_2
  = O_p\bigl(n^{-1/2}\bigr).
\]

\end{theorem}

We note the kernel regularity condition is satisfied by popular characteristic kernels such as the Gaussian kernel. The local strong-convexity condition plays an analogous role to the classical positive-definite Hessian assumption used to obtain $\sqrt{n}$-rates in the unique minimizer setting (i.e., that \(
\nabla^{2}\ell(\theta_0)\succeq\lambda_{\min}I_d
\) when $M = \{\theta_0\}$) \citep{van2000asymptotic}. Indeed, in this case a first-order Taylor expansion of $\nabla \ell$ around $\theta_0$ yields the gradient bound
\(
\|\nabla\ell(\theta)\|_2\;\ge\;\lambda_{\min}\,\|\theta-\theta_0\|_2
\) for $\theta$ near $\theta_0$. We assume the bound directly as it requires only first-order derivatives and is more natural when there are multiple minimizers.

\begin{figure}[t]
    \centering
    \includegraphics[width=0.45\linewidth]{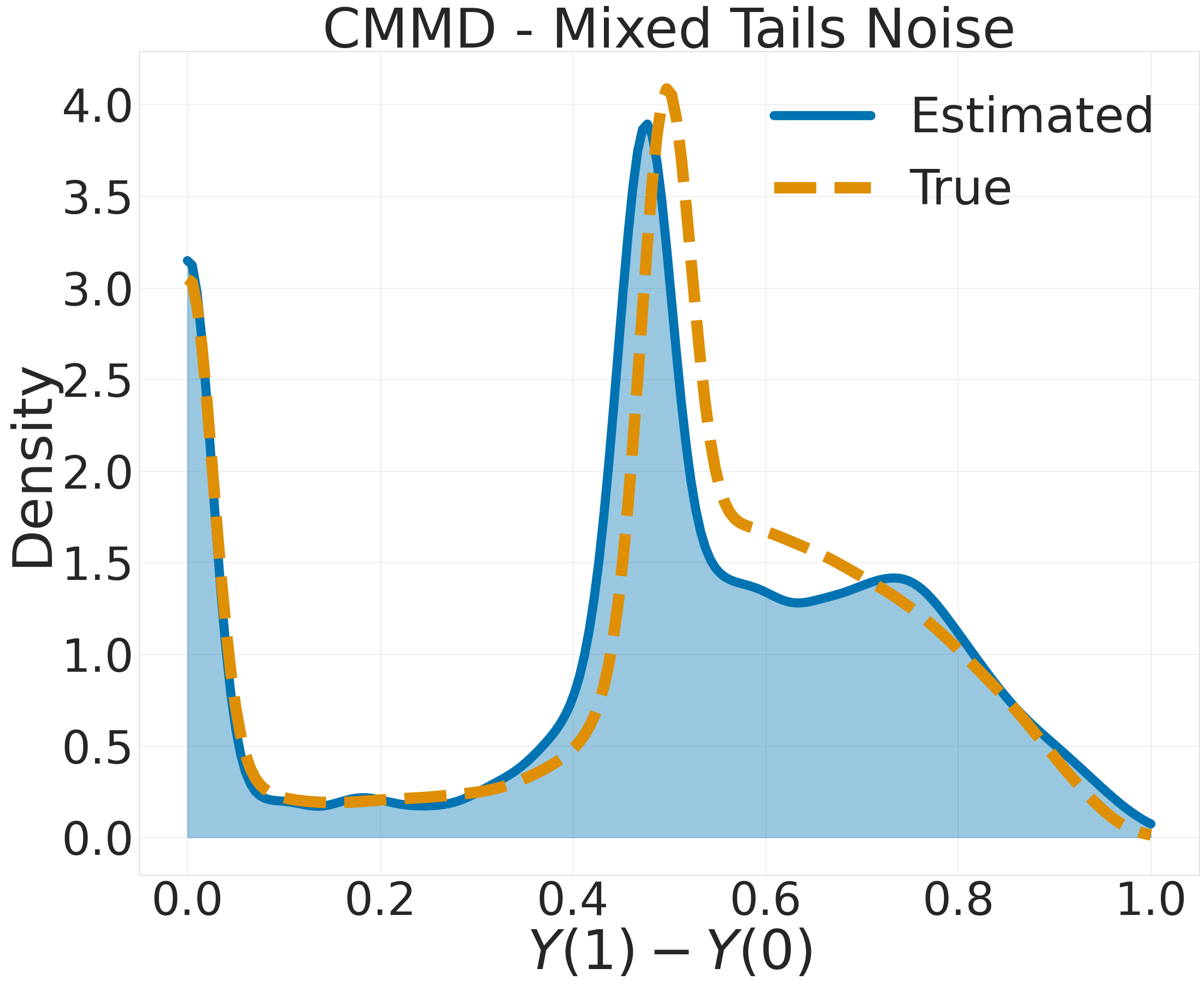}
    \includegraphics[width=0.45\linewidth]{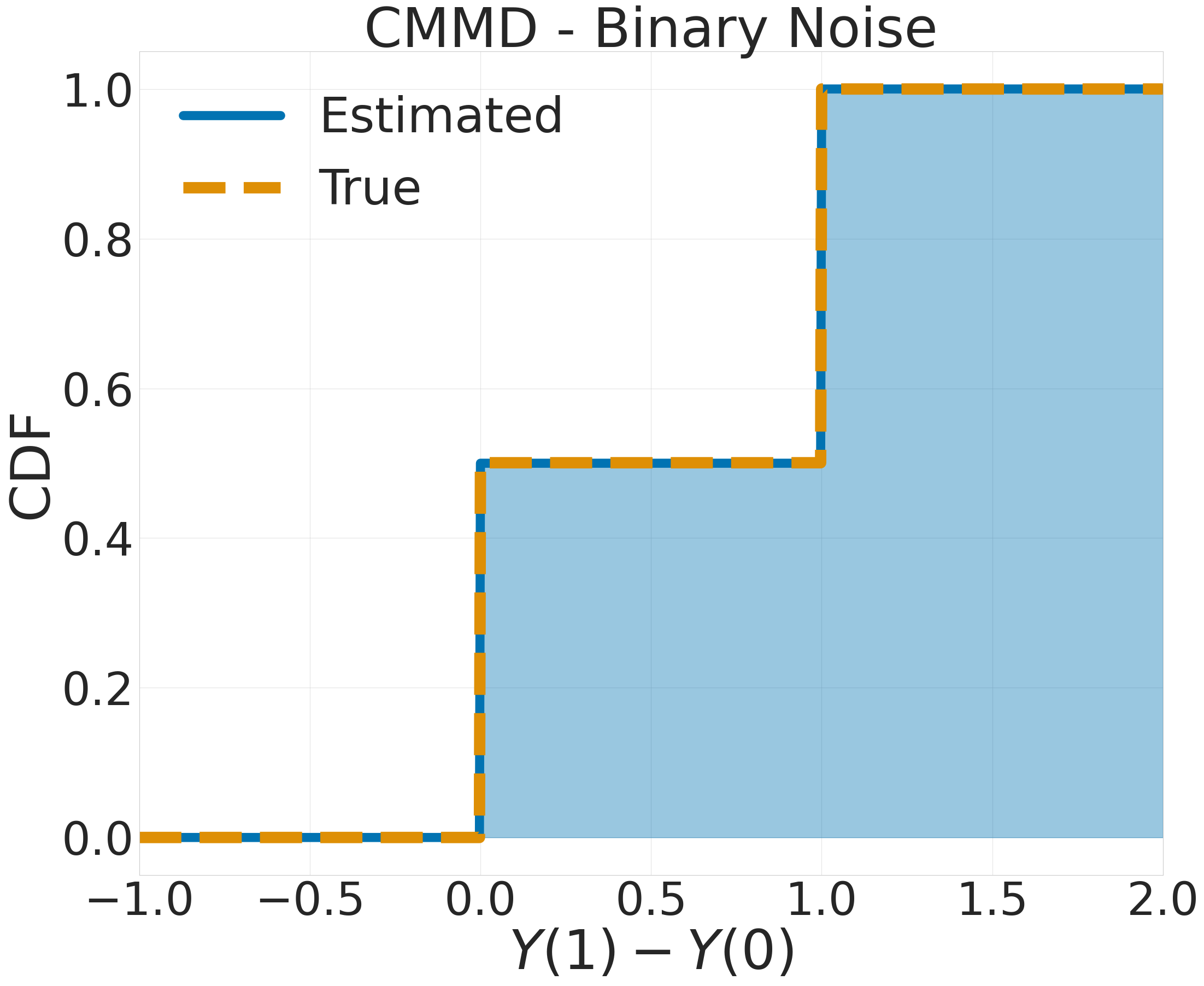}
\caption{Estimated treatment effect distributions for \cref{ex:mis-spec} using flow-based cocycles with CMMD. 
Left: density under mixed-tailed noise. Right: CDF under discrete noise. 
Cocycles accurately recover both, unlike flow-based SCMs in \cref{fig:misspec}.}

    \label{fig:noise_adaptiveness}
      \vspace{-10pt}
\end{figure}

\subsection{Demonstration on a Toy Example} We conclude by demonstrating the performance of counterfactual cocycles estimated with CMMD on \cref{ex:mis-spec}. We assume access to samples $\{X^{(i)}, Y^{(i)}\}_{i=1}^{2000} \sim \bb P_{X,Y}$ and specify cocycle models using the same neural spline flow architecture as the flow-based SCMs in that example. We train each model using the CMMD loss, impute the counterfactuals $\{Y^{(i)}(0),Y^{(i)}(1)\}_{i=1}^n$ using the cocycles, and estimate the treatment effect density and CDF by (Gaussian) kernel smoothing and empirical averaging, as described in \cref{sec:method:estimation} (i.e., see equations \eqref{eq:DTE} and \eqref{eq:TED}). CMMD implementation details are in \cref{sec:implementation} and the exact optimization routine matches that in \cref{exp:lin_model}. \cref{fig:noise_adaptiveness} shows the estimated treatment effect density for the mixed-tailed noise case (left) and CDF for the binary noise case (right). In contrast to the poor performance of flow-based SCMs (see \cref{fig:misspec}), our approach accurately recovers both distributions. This reflects that the CMMD estimator does not require specifying a latent noise distribution, and that its consistency is generally robust to the (true) noise distribution---as established in the previous subsection.

\section{Robustness and Simplicity of Counterfactual Cocycles Versus SCMs}
\label{sec:cocycle:vs:scm}

In this section we analyze in detail the advantages of modeling counterfactual cocycles in contrast to SCMs, by studying a formal invariance of cocycles under model reparameterizations. This invariance means that centering the modeling and estimation process around the cocycle is more robust to mis-specification, can greatly simplify estimation, and in many cases lets us sidestep the problems with flow-based SCMs identified in \cref{sec:background:scms}. We also discuss advantages over recently proposed SCM-based transport methods using conditional quantile estimation \citep{plevcko2020fair, machado2024sequential}.  

In what follows, we restrict attention to the simplified setting  
\[
Y =_{\mathrm{a.s.}} f(X,\xi), \quad \xi \ind X, \quad f(x,\argdot) \text{ is bijective } \forall x \in \bb X
\]  
and compare modeling a counterfactual cocycle \(T\) versus a \emph{bijective generating mechanism} (BGM) \((f,\bb P_{\xi})\). While a full SCM specifies the entire joint system (cf. \eqref{eq:scm} and \eqref{eq:scm_confounded}), when the treatments \(X \in \bm V\) are pre‐defined and $X \prec Y$ one can always set \(X = \xi_X\) and estimate the noise distribution via the empirical distribution of \(X\). In this case, the substantive difference between the two approaches lies in how the conditional law \(\bb P_{Y|X}\) is modeled. The same arguments apply with \((X,Z)\) in place of \(X\) under the more general framework in \cref{sec:methodgeneral}, so we need not explicitly distinguish between treatments and covariates. 

\subsection{Noise Invariance of Counterfactual Cocycles}

It is known that BGMs can only be identified up to noise automorphisms \citep{nasr2023counterfactual}. In particular, let $Y =_{\mathrm{a.s.}} f(X,\xi)$, where $\xi \in \mc E$, $\xi \ind X$, and $f_x := f(x,\argdot): \mc E \to \bb Y$ is bijective. Then, letting $\Aut(\mc E)$ be the group of bi-measurable bijections on $\mc E$, we have

\begin{align*}
    f(x,\xi) = f(x,g \circ g^{-1}(\xi )) \quad \forall g \in \Aut(\mc E) \;.
\end{align*} 
Defining $f^{(g)}: (x,\xi) \mapsto  f(x ,g (\xi))$, it is clear that $\{(f^{(g)},g^{-1}_{\#}\bb P_{\xi})\}_{g \in \Aut(\mc{E})}$ is an equivalence class of BGMs that all generate the same conditional distribution $\bb P_{Y|X}$---each one with a different noise parameterization $\bb P_{g^{-1}(\xi)}$. However, while the structural function $f^{(g)}$ depends on the choice of $g$, the corresponding cocycle that satisfies \eqref{eq:CC} does not:
\begin{align} \label{eq:cocycle:reparam:invar}
    T_{x,x'} := f_{x} \circ f_{x'}^{-1} = f_x \circ g \circ g^{-1} \circ f^{-1}_{x'} \;, \quad \text{ for any $g \in \Aut(\mc{E})$, $x,x'\in\bb{X}$.}
\end{align}
This has implications for whether a model is well-specified. To illustrate, fix an underlying generating mechanism $(f^*,\bb{P}^*_{\xi})$. 
With a flow-based SCM, one specifies a fixed base distribution $\widehat{\bb P}_{\xi}$ and a class of functions, $\mc{F} := \{ f\colon \bb{X}\times \mc E \to \bb{Y} \}$. In this case, the model is well-specified if there is some $f \in \mc{F}$ such that $f_{x\#} \widehat{\bb P}_{\xi} = f^*_{x\#} \bb{P}^*_{\xi}$ for each $x \in \bb{X}$. Denoting $\mc{T}(\bb{P}^*_{\xi}, \widehat{\bb P}_{\xi})$ as the set of invertible transports from $\bb{P}^*_{\xi}$ to $\widehat{\bb P}_{\xi}$, this is true if and only if: 
\begin{enumerate}[label=(\Roman*)]
    \item There exists $h\in \mc{T}(\bb{P}^*_\xi, \widehat{\bb P}_\xi)$ such that $f^{*(h)} \in \mc F$.
\end{enumerate} 
On the other hand, the cocycle approach specifies a function class of the same type, $\mc{F} := \{ f\colon \bb{X}\times \mc E \to \bb{Y} \}$, whose elements will be used to construct candidate cocycles via $T_{x,x'} = f_x\circ f_{x'}^{-1}$. The model $\mc{F}$ is well-specified if there is some $f \in \mc{F}$ such that $f_x\circ f_{x'}^{-1} = T^*_{x,x'} = f^*_x \circ f^{* -1}_{x'}$. By \eqref{eq:cocycle:reparam:invar}, it is easy to see that this is true if and only if:
\begin{enumerate}[label=(\Roman*)]
    \setcounter{enumi}{1}  
    \item There exists $h \in \Aut(\mc E)$ such that $f^{*(h)} \in \mc F$.
\end{enumerate}
Since $\mc{T}(\bb{P}^*_{\xi}, \widehat{\bb P}_{\xi})$ is a strict subset of $\Aut(\mc E)$, flow-based models with fixed $\widehat{\bb P}_{\xi}$ and function class $\mc{F}$ are well-specified for a strictly smaller set of generating mechanisms than cocycle models that use the same function class $\mc{F}$. The following example demonstrates how cocycle modelling can avoid the tail and support mis-specification problems outlined in \cref{sec:background:scms}.

\begin{figure}[t]
 \centering
 \includegraphics[width=\linewidth]{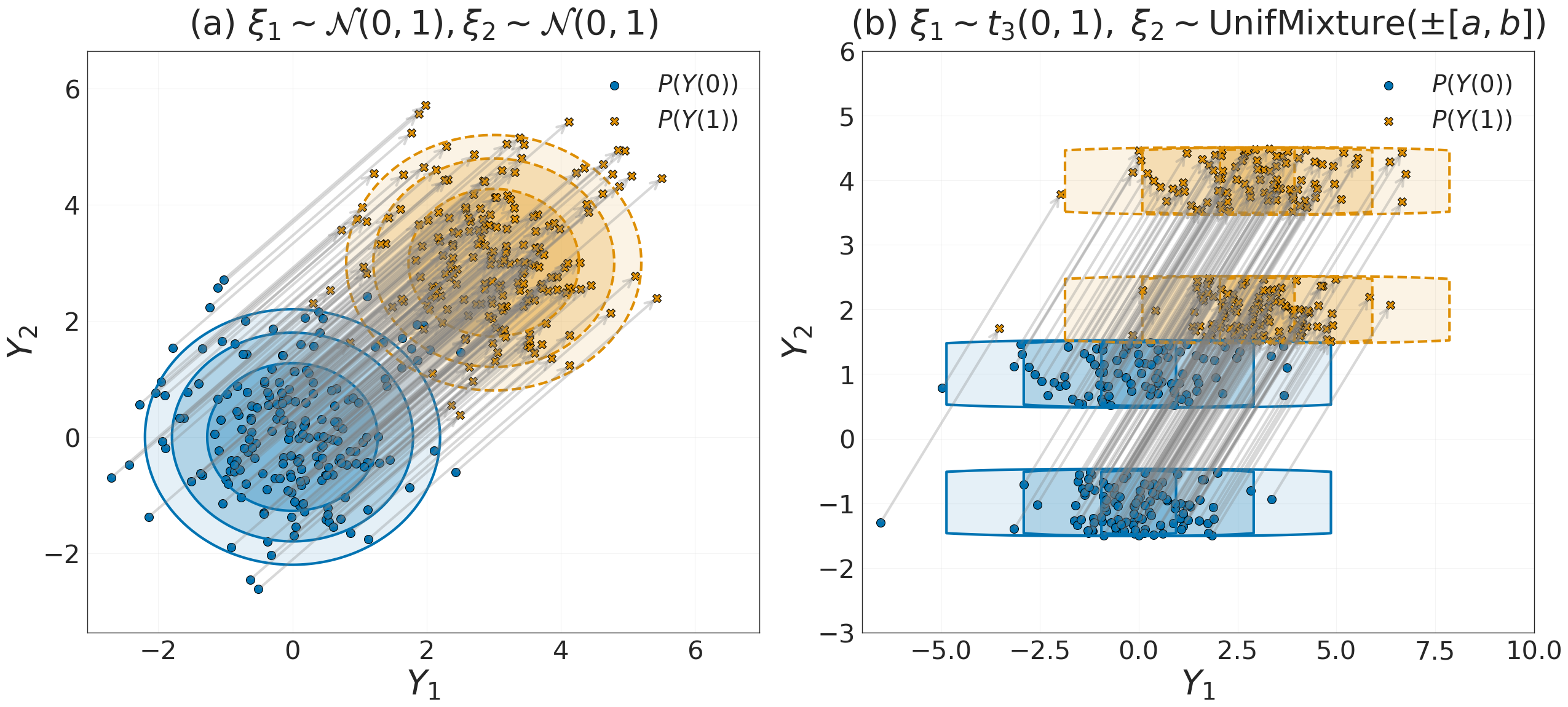}
 \caption{Illustration of invariance of cocycle complexity to noise distribution. Counterfactual transport (gray arrows) $T_{1,0}$ from samples of $\bb P_{Y(0)}$ (blue) to  $\bb P_{Y(1)}$ (orange) for structural model $Y(x) = Ax + \xi$ where (i) $\xi \sim \mathsf{N}(0,I)$ (left) and (ii) $\xi \sim  \mathsf{t}_3(0,1) \otimes \left(\frac{1}{2}\mathsf{U}\left(-\tfrac{3}{2},-\tfrac{1}{2}\right) +\frac{1}{2}\mathsf{U}\left(\tfrac{3}{2},\tfrac{1}{2}\right) \right)$ (right). If $\widehat{\bb P}_{\xi} = \mathsf{N}(0,I)$ is used as a base distribution to learn the BGM $(f,\widehat{\bb P}_{\xi}$), there is no continuous bijection $\widehat{\bb P}_{\xi} \mapsto \bb P_{Y(x)}$ in case (ii). In contrast, the direct transport $T_{1,0}(y) = A + y$ is a simple linear map, regardless of the true noise distribution $ {\bb P}_{\xi}$. Intuitively, one can view $T_{x,0}$ as a structural map from `base' distribution $\bb P_{Y(0)}$ to $\bb P_{Y(x)}$.}
 \label{fig:noise_invariance_2d}

 \vspace{-10pt}
\end{figure}

\begin{example} \label{ex:simplification}
    Suppose that $Y = f^*(X,\xi^*)$, where  $\xi^* \in \mc E \subset \bb R^2$, $X \in \bb R$ and 
    \begin{align*}
        {f^*}(X,{\xi^*}) = A X + {\xi^*}, \quad \xi^* \sim \mathsf{t}_3(0,1) \otimes \left(\frac{1}{2}\mathsf{U}\left(-\tfrac{3}{2},-\tfrac{1}{2}\right) +\frac{1}{2}\mathsf{U}\left(\tfrac{1}{2},\tfrac{3}{2}\right) \right) \;.
    \end{align*}
    Suppose we model this mechanism using base distribution $\widehat{\bb P}_\xi = \mathsf{N}(0,I)$ and a class of autoregressive Lipschitz diffeomorphisms $\mc F \subset \mc F_{\bb G_{\text{TMI}}}$, as in SOTA flow-based SCMs. In this case, both the tails and support of $\bb P_{\xi}^*$ are mis-specified. By the theory of TMI maps \citep{bogachev2005triangular}, the only $h \in \mc T(\bb P^*_\xi, \widehat{\bb P}_\xi))$ that results in $f^{*(h)} \in \mc F_{\text{TMI}}$ is the map $h(\hat \xi) = (h_1(\hat \xi_1), h_2(\hat \xi_2))$ where $h_1$ and $h_2$ are the quantile transforms from $\widehat{\bb P}_{\xi_1}$ to $\bb P_{\xi_1}^*$ and $\widehat{\bb P}_{\xi_2}$ to $\bb P_{\xi_2}^*$ respectively:
\[
h_1(\hat\xi)
=\operatorname{sign}(\hat\xi_2)\sqrt{\frac{3\bigl(1-I^{-1}(2\Phi(\hat\xi_2);\tfrac32,\tfrac12)\bigr)}
                                     {I^{-1}(2\Phi(\hat\xi_2);\tfrac32,\tfrac12)}},\qquad
h_2(\hat\xi_2)
=\begin{cases}
-\tfrac32 + 2\,\Phi(\hat\xi_2), & \Phi(\hat\xi_2)<\tfrac12,\\[3pt]
2\,\Phi(\hat\xi_2) - \tfrac12,   & \Phi(\hat\xi_2)\ge\tfrac12
\end{cases}
\]
 Here $I$ is the regularized incomplete Beta function and $\Phi$ is the CDF of $\mathsf{N}(0,1)$. $h_1$ is not Lipschitz and $h_2$ has a discontinuous jump. Therefore, $f^{*(h)}$ does not lie in $\mc F$. In contrast, the cocycle  $T_{x,x'}(y) = A (x - x') + y$ does not depend on $\bb P_{\xi}^*$ or $\widehat{\bb P}_{\xi}$ and can be modeled using the set of linear maps $\mc F_{\text{LIN}} = \{f(x,\xi)  = A x + \xi \mid A \in \bb R^2\} \subset \mc F$, since $f^{*(h)} \in \mc F_{\text{LIN}}$ for $h = \id \in \Aut(\mc E)$. This is illustrated on \cref{fig:noise_invariance_2d}.
\end{example}

The same concept applies to \cref{ex:mis-spec}. In that example, one has $f^* \in \mc F_{\text{LIN-SCALE}} = \{f(x,\xi) = \sigma(x)\epsilon \mid \sigma(x) = \beta x + \alpha , (\alpha,\beta) \in [-1,1]^2\}$, a set of linearly-parameterized scale transforms, but using fixed $\widehat{\bb P}_{\xi} \in \{\mathsf{N}(0,1), \mathsf{Lap}(0,1)\}$ results in a non-affine, non-Lipschitz flow for the mixed-tailed noise design, and no well-defined flow for the binary noise design.

For our last example, we show the \emph{dependence} structure between the coordinates of $\xi$ can also induce mis-specification problems for SCMs, but again has no effect on the cocycle.

\begin{example}\label{ex:dependence}
    Consider again the set-up of \cref{ex:simplification}, but now where there is a single latent cause of both outcomes: \[\xi^*_1 =_{\mathrm{a.s.}} \xi^*_2 \sim \mathsf{N}(0,1)\;.\]
    In this case, using a factored Gaussian base distribution $\widehat{\bb P}_{\xi} = \mathsf{N}(0,I_2)$ for the SCM admits the correct marginals for the noise, but since $\supp(\bb{P}^*_{\xi})$ is a lower-dimensional set than $\supp(\widehat{\bb P}_{\xi})$ we have $\mc{T}(\bb{P}^*_{\xi},\widehat{\bb P}_{\xi}) = \varnothing$. In contrast, the true cocycle $T$ is unaffected by the particular dependence structure in $\xi$ and can still be well-specified using the class $\mc F_{\text{LIN}} = \{f(x,\xi)  = A x + \xi \mid A \in \bb R^2\} \subset \mc F$, since we still have $f^{*(\id)} \in \mc F_{\mathrm{LIN}}$.
\end{example}

The above examples all serve to demonstrate an underlying point: any properties of the conditionals $(\bb P_{Y|X=x})_{x \in \bb X}$ inherited from the (true) noise distribution $\bb P_{\xi}$ need not be learned by the counterfactual cocycle, since it is invariant to this distribution. As a practical implication, we can employ popular Lipschitz flow classes (e.g., \cref{tab:flow_models}) to model counterfactual cocycles while often avoiding the mis-specification issues that arise in flow-based SCMs.

\subsection{Counterfactual Cocycles as Minimum Complexity SCMs}

Another way to view the mis-specification robustness of cocycle modeling is as a `minimal complexity' property of counterfactual cocycles. In particular, the cocycle can be constructed using \emph{any} coboundary map from the equivalence class of SCMs $\{f^{(g)}\}_{g \in \Aut(\mc E)}$. We are therefore free to choose a representative $f^{\star}$ that minimizes a given notion of functional complexity (e.g., Sobolev norm). Since each map $f^{(g)}$ corresponds to a base distribution $\bb P^{(g)}$ such that $\bb P_{Y|X=x} = (f^{(g)}_x)_{\#}\bb P^{(g)}$, counterfactual cocycles are therefore no more complicated than the \emph{minimal complexity} SCM, induced by an `optimal' noise distribution $\bb P^{\star}$.
 
This means we may be able to use  simpler flow-based model classes $\mc F$ than otherwise for the coboundary map of a cocycle, while remaining well-specified. Using a simpler class of models can improve finite sample performance via reduced estimation variance. In practice, we will cross-validate over a hierarchy of flows of increasing expressivity to adapt to the underlying complexity (see \cref{sec:implementation}). 

\paragraph{Conditions for Stricter Cocycle Simplicity} It is natural to ask under what conditions a counterfactual cocycle can be constructed using a strictly simpler model class than for an SCM with a fixed base distribution $\bb{\hat P}_\xi$. Below we provide an exact characterization when the simplicity of a model class is measured by its size. In what follows we choose the noise space to be $\mc E = \bb Y$ without loss of generality\footnote{Choosing $\mc E=\bb Y$ just corresponds to a reparameterization of the BGM, since by construction there exists a bijection $h:\bb Y \to \mc E$. 
Moreover, the automorphism groups are conjugate: $\Aut(\mc E) \cong \Aut(\bb Y)$, with $g \in \Aut(\mc E)$ if and only if $h \circ g \circ h^{-1} \in \Aut(\bb Y)$. 
Thus, working with $\mc E=\bb Y$ entails no loss of generality.
}, so that each $f_x$ is bijective $\bb Y \to \bb Y$.

 Recall from \cref{sec:method:id} that whenever the maps $(f_x)_{x \in \bb X}$ are exactly invertible on $\bb Y$, they lie in a transformation group $\bb G$ on $\bb Y$. In this case, we call the coboundary map (i.e., $f$) $\bb G$-valued. The smallest group containing $(f_x)_{x \in \bb X}$ is the subgroup generated by them, which we denote $\bb G_f := \langle f_x : x \in \bb X \rangle$. The set of all $\bb G_f$-valued coboundary maps is denoted $\mc{F}_{\bb{G}_f}$. This model class naturally reflects the smallest model class that is guaranteed to be well-specified for $f$, when the dependence of the function $f$ on $x$ is not known. Note that size of this model class $\mc F_{\bb G_f}$ is controlled by the size of $\bb G_f$: if two coboundary maps $f_1$ and $f_2$ generate groups $\bb G_{f_1}$ and $\bb G_{f_2}$ with $\bb G_{f_1} \subsetneq \bb G_{f_2}$, then $\mc F_{\bb G_{f_1}} \subsetneq \mc F_{\bb G_{f_2}}$. Altogether, this implies that if $\bb G_{f_1} \subsetneq \bb G_{f_2}$, then in a certain sense $f_1$ can be modeled with a smaller (i.e., simpler) model class than $f_2$. For an intuitive example, let $\bb G_{f_1} := \text{Diff}^k(\bb Y)$, the set of diffeomorphisms on $\bb Y$ with $k$ continuous derivatives, and $\bb G_{f_2} := \text{Diff}^1(\bb Y)$. Then, $\mc F_{\bb G_{f_1}}$ contains only those functions in $\mc F_{\bb G_{f_2}}$ that are at least $k$-smooth on $\bb Y$.

Below we show that, out of all equivalence class members $(f^{(g)})_{g \in \Aut(\bb Y)}$ that can be used to construct a given counterfactual cocycle, the construction $f^\star_x := T_{x,x_0}$ from \cref{thm:cocycle_factorization} induces the smallest possible set $\mc F_{\bb G_{f^\star}}$ and so can be modeled using the smallest class.

\begin{figure}[t]
\centering
\includegraphics[width = 0.65\textwidth]{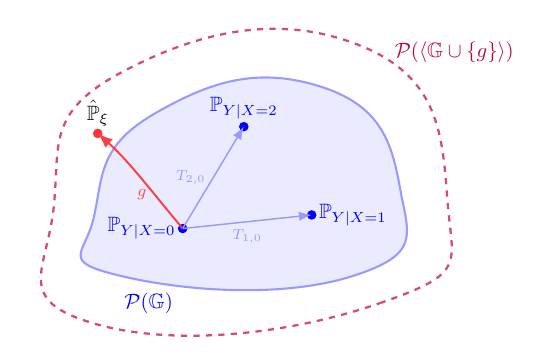}
\caption{
Illustration of \cref{thm:min_complexity} (minimum cocycle complexity). 
The blue region ${\color{blue}\mc P(\bb G)}$ shows the set of distributions reachable from the base distribution $\bb P_{Y|X=0}$ under the group $\bb G = \langle T_{x,0} : x \in \{0,1,2\}\rangle$. 
The family $(T_{x,0})_{x\in\{0,1,2\}}$ defines a coboundary map $f \in \mc F_{\bb G}$ for the cocycle $T$, via $f(x,y) = T_{x,0}(y)$. 
If each $T_{x,0}$ is composed with some ${\color{red} g}\notin \bb G$, the base distribution shifts to $\widehat{\bb P}_\xi = ({\color{red}g})_{\#}\bb P_{Y|X=0}$, which lies outside ${\color{blue}\mc P(\bb G)}$. 
The smallest group transporting to all $(\bb P_{Y|X=x})_{x \in \{0,1,2\}}$ is then $\langle\bb G\cup \{g\}\rangle$, which is strictly larger than $\bb G$ and has reachable set ${\color{purple}\mc P(\langle\bb G\cup \{g\}\rangle)}$. 
Since enlarging the group enlarges the associated function class of coboundary maps (i.e., $\mc F_{\bb G} \subsetneq\mc F_{\langle\bb G\cup \{g\}\rangle}$), starting from $\bb P_{Y|X=0}$ (or any $\bb P_{Y|X=x}$) yields the smallest possible transformation group $\bb G$ and hence the smallest associated model class $\mc F_{\bb G}$ in which the cocycle can be represented. }
\label{fig:min_complexity}
\vspace{-10pt}
\end{figure}

\begin{theorem}[Minimum Cocycle Complexity]\label{thm:min_complexity}
    Let $\{T_{x,x'}\}_{x,x' \in \bb X}$ satisfy \eqref{eq:DA}, \eqref{eq:NI} and \eqref{eq:CPI} with respect to $\bb P_{Y|X}$. For each $x \in \bb X$, define $f_x := T_{x,x_0}$ and suppose each $f_x: \bb Y \to \bb Y$ is bijective. Let $\bb G_{f} = \langle  f_x : x \in \bb X\rangle$ 
    and $\bb G_{f^{(g)}} =  \langle  f_x \circ g : x \in \bb X\rangle$, for any $g \in \Aut(\bb Y)$. 
    Then, 
    \[\text{(i)} \;\mc F_{\bb G_{f}} \subseteq \mc F_{\bb G_{f^{(g)}}}\quad \text{ and } \quad \text{(ii)}\;\mc F_{\bb{G}_{f}} \subsetneq \mc F_{\bb{G}_{f^{(g)}}}\; \forall g \notin \bb{G}_{f}\;.\]
\end{theorem}

Since choosing $f^\star_x := T_{x,x_0}$ corresponds to the base distribution $\widehat{\bb P}^{\star}_{\xi} := \bb P_{Y|X=x_0}$ (i.e., $(f_x^\star)_{\#}\bb P_{Y|X=x_0} = \bb P_{Y|X=x}$) and $x_0$ is arbitrary above, \cref{thm:min_complexity} implies $\bb P_{Y|X=x}$, for any $x \in \bb X$ is always an `optimal' base distribution. Moreover, any choice of base distribution $\tilde {\bb P}_{\xi}$ such that $\tilde {\bb P}_{\xi} \neq g_\# \bb  P_{Y|X=x}$ for all $g \in \bb G_{f^\star}$ must require a more complex coboundary map than can be used to construct the cocycle. This result may seem strange, since \cref{fig:overlap} illustrated an example where the valid set of noise distributions under a function class $\mc F$ did not include either conditional. However, when the class $\mc F$ is specified via a transformation group, the sets of reachable distributions from any two conditionals $\bb P_{Y|X=x_0},\bb P_{Y|X=x_1}$ either do not intersect (if the group is too small), or are identical (if the group if large enough to map between them). \cref{fig:min_complexity} illustrates the minimum complexity result in more detail.

As a concrete application of this result, in \cref{ex:simplification}, we saw that $f^*_x \in (\bb R^2,+)$ , the group of shifts on $\bb R^2$. Since there $\bb P_{\xi}^* = \bb P_{Y|X=0}$, we know that $f^*(x,y) = Ax+y$ is a minimal complexity coboundary map for the cocycle. Moreover, if $\widehat{\bb P}_{\xi}$ is not a translation of $\bb P_{Y|X=0}$ then the structural map from this base distribution will necessarily be more complex. Note that the transport in \cref{fig:noise_invariance_2d} corresponds to this map for $x=1$, since $T_{1,0}(y) = f^*(1,y)$.

\subsection{Robustness of Cocycle-based Approach to Causal Quantity Estimation}

Since counterfactual cocycles are well‐specified under strictly milder conditions than flow‐based BCMs, any procedure for estimating causal quantities that depends only on the cocycle is naturally more robust.  For instance, consider any estimand of the form
\[
\gamma(x)\;=\;\bb E\bigl[\rho\bigl(Y(x),Y(0)\bigr)\bigr], \quad \text{for some \(\rho:\bb R^{2p}\to\bb R\)}\,,
\]
 (e.g., those in \eqref{eq:ATE}-\eqref{eq:TED}). Suppose the true cocycle $T$ can be constructed by a flow-based coboundary map parameterized in $\bb R^d$, $f_{\theta_0} \in \mc F = \{f_\theta : \theta \in \Theta \subset \bb R^d\}$.  Now, let \({\hat\theta}\) be \emph{any} estimator of $\theta_0$ from $\{X^{(i)},Y^{(i)}\}_{i=1}^n$. Our approach uses the cocycle to empirically estimate the causal estimand via the imputed counterfactuals
\[
\hat\gamma(x)
:=  \frac{1}{n}\sum_{i=1}^n 
\rho\bigl(T_{\hat\theta,\,x,X^{(i)}}(Y^{(i)}),\,T_{\hat\theta,\,x_0,X^{(i)}}(Y^{(i)})\bigr).
\]

Under standard empirical process theory arguments on $\theta \mapsto \bb E \rho (T_{\theta,\,x,X}(Y),\,T_{\theta,\,x_0,X}(Y)\bigr)$ (e.g., see \cite{kennedy2016semiparametric}), if \(
\hat\theta\;\xrightarrow{p}\;\theta^*\), then we have
\(
\hat\gamma(x)\;\xrightarrow{p}\;\gamma(x)
\). By contrast, the abduct–act–predict (AAP) estimator by Monte Carlo sampling from a flow‐based BCM is
\[
\hat\gamma^{\rm AAP}(x)
=\frac{1}{m}\sum_{i=1}^m
\rho\Bigl(f_{\hat\theta,\,x}(\xi^{(i)}),\,f_{\hat\theta,\,0}(\xi^{(i)})\Bigr),
\quad
\{\xi^{(i)}\}_{i=1}^m\sim_{iid}\widehat{\bb P}_\xi.
\]
Although one can appeal to equivalent empirical process conditions for this estimator, one can still have $\hat \gamma^{AAP}(x) \not\to_p \gamma(x)$ even if the estimator for the cocycle is consistent, $\hat \theta \to_p \theta_0$. This is because, while $f_{\theta_0} \in \mc F$, we may have $\hat f \notin \mc F$, where $\hat f$ is the conditional flow from $\widehat{\bb P}_{\xi} \mapsto \bb P_{Y|X}$ (e.g., if $\widehat{\bb P}_{\xi}$ has a mis-specified tail and support---as in \cref{ex:mis-spec} and \ref{ex:simplification}).

As we showed in \cref{sec:estimation}, our proposed cocycle estimator actually ensures $\hat \theta \to_p \theta_0$ under \emph{more general} conditions than existing estimators which use a base distribution to estimate the flow parameters, giving our approach an additional source of robustness.

\subsection{Comparison to SCM-based Transport Methods via Conditional Quantiles} \label{sec:cocyclevsscm:seqot}

SCMs combining conditional-quantile estimation techniques with causal DAGs have recently been used to construct counterfactual transports in fairness applications \citep{plevcko2020fair,machado2024sequential}. These are situated in a longer line of work in causality that uses the quantile transform \citep[e.g.,][]{Chernozhukov_Hansen_2005_IV_quantile,athey_imbens_2006_DiD,Vansteelandt_2014}. The methods in \cite{plevcko2020fair} and \cite{machado2024sequential} start from an SCM with independent, uniform noise, 
\[
   Y_j =_{\mathrm{a.s.}} f_j(Y_{\pa(j)},\xi_j), \quad \xi_j \sim \mathsf{U}[0,1] \quad \forall j \in \{1,\dots,p\}\;.
\]
In their context, $X := Y_0$ is a “sensitive’’ source attribute (e.g., sex, race) to be manipulated, and $Y_1,\dots,Y_p$ are downstream outcomes that follow a known causal DAG. 
The main idea is that, under the independent, uniform noise assumption, $f_j$ coincides with the conditional quantile function, $f_j(y_{\pa(j)},\xi_j) = Q_{y_{\pa(j)}}(\xi_j)$, of the conditional law $\mathbb{P}_{Y_j \mid Y_{\pa(j)}=y_{\pa(j)}}$. It can therefore be estimated using standard quantile regression techniques \citep{meinshausen2006quantile}. To generate counterfactuals under the intervention $\doInt(X=x)$ one can recursively compute for each node in the DAG,
\[
   \hat y_j(x) =  Q_{\hat y_{\pa(j)}(x)} \circ F_{y_{\pa(j)}}(y_j), \quad \forall \;j = 1,\dots,p
\]
where $F_{y_{\pa(j)}}$ is the conditional CDF given the \emph{factual} parent values $y_{\pa(j)}$, and $\hat y_{\pa(j)}(x)$ are the counterfactual values of the parents obtained in previous steps. Thus, each node is updated by a map of the form $T_{\hat y_{\pa(j)}(x), y_{\pa(j)}} = Q_{\hat y_{\pa(j)}(x)} \circ F_{y_{\pa(j)}}$, which is precisely the conditional OT map between $\mathbb{P}_{Y_j \mid Y_{\pa(j)}=y_{\pa(j)}}$ and $\mathbb{P}_{Y_j \mid Y_{\pa(j)}=\hat y_{\pa(j)}(x)}$ \citep{Carlier_2016,Hosseini_etal:2025:conditonalOT}. 
The resulting procedure recently has been branded sequential-OT \citep{machado2024sequential}, since it decomposes the transport into a set of conditional OT maps.

When the true noise variables $(\xi_j)_{j=1}^p$ are independent, the sequential-OT transports coincide with the (TMI) counterfactual cocycle. However, the conditional-quantile approach requires estimating the mapping $(Q_x)_{\#}: \mathsf{U}[0,1]^p \mapsto \prod_{j=1}^p\bb P_{Y_j|Y_{\pa(j)}}$, and so still implicitly commits to a particular noise distribution for the purposes of estimation. Thus, in principle, this approach can suffer from similar mis-specification problems as flow-based SCMs, depending on the exact function class used to learn the quantile function. This is somewhat mitigated by using nonparametric CDF estimation techniques, but in either case fails to exploit the minimal complexity of counterfactual cocycles.

A potentially more serious limitation is that the resulting transports generally do \emph{not} coincide with the true counterfactual cocycle when the noise variables are dependent, thus resulting in biased, inconsistent estimators. This is demonstrated in the following example.

\begin{example}\label{ex:seq-ot}
Consider the three-variable SCM with treatment variable $X$,
\[
X\sim\mathsf{N}(0,1),\quad Y_1=X+\xi_1,\quad Y_2=Y_1+\xi_2,
\]
We assume $(\xi_1,\xi_2)\sim\mathsf{N}(0,\Sigma_\rho)$, $\text{Var}(\xi_j)=1$, $\text{Corr}(\xi_1,\xi_2)=\rho$, and $(\xi_1,\xi_2) \ind X$.  
In this case, the true counterfactual cocycle reduces to a joint shift
\[
T_{x',x}^\star(y_1,y_2)=(y_1+\Delta,\;y_2+\Delta),\qquad \Delta=x'-x.
\]

By contrast, the sequential-OT procedure applies conditional quantile maps node by node.  
For $Y_1$, since $(Y_1|X=x)\sim\mathsf{N}(x,1)$, the quantile transform gives $\hat y_1(x')=y_1+\Delta$.  
For $Y_2$, since $(Y_2|Y_1=y_1)\sim\mathsf{N}((1+\tfrac{\rho}{2})y_1,\,1-\rho^2/2)$, the conditional quantile transform gives 
\[
\hat y_2(x')=(1+\tfrac{\rho}{2})\hat y_1(x')+\sqrt{1-\tfrac{\rho^2}{2}}\;\Phi^{-1}(\hat \xi_2)
= y_2+(1+\tfrac{\rho}{2})\Delta.
\]

Combining these transports together gives the sequential-OT map
\[
T^{\mathrm{seq}}_{x',x}(y_1,y_2)=\bigl(y_1+\Delta,\; y_2+(1+\tfrac{\rho}{2})\Delta\bigr),
\]
which coincides with $T^\star_{x',x}$ only when $\rho=0$.  
The discrepancy arises because the TMI map between joint laws $\bb P_{Y_1,Y_2|X}$ does not factorize into the TMI maps of the conditionals $\bb P_{Y_1|X}\otimes \bb P_{Y_2|Y_1}$ when $\xi_1,\xi_2$ are dependent. By targeting the joint cocycle directly, our method avoids this issue and can recover the correct transport regardless of $\rho$, even when using the DAG structure to sparsify the cocycle architecture.
\end{example}

\section{Implementation Details}\label{sec:implementation}

In this section, we discuss implementation and optimization details for counterfactual cocycle models parameterized by autoregressive flows. Formal algorithms are in \cref{app:algs}.

\subsection{Flow-based Cocycle Parameterizations}\label{subsec:flows}

As discussed in \cref{sec:method}, a natural modeling choice for the coboundary map $(x,y) \mapsto f_x(y)$ of a cocycle is to use {conditional normalizing flows}~\citep{kobyzev2020normalizing,papamakarios2021flows}.  Each flow is the composition of
\emph{(i)~a conditioner} $\tau_{\theta}\colon\bb{X}\to\Lambda$, mapping inputs $x$ to a vector of flow parameters $\lambda$, and
\emph{(ii)~a \emph{bijector}} $g_{\lambda}\in \bb G$ that transforms $y$:
\[
  \hat f_{\theta,x}(y)=g_{\tau_{\theta}(x)}(y) \quad \Leftrightarrow \quad  
  T_{\theta, x,x'}(y)=g_{\tau_{\theta}(x)} \circ g_{\tau_{\theta}(x)}^{-1}(y)
\]

The conditioner $\tau_{\theta}$ can be \emph{any} learnable
function class---linear model, MLP, convolutional network, or
transformer---so long as it maps $x$ to valid flow parameters
$\lambda$.  The bijector $g_{\lambda}$ can be a single transform or a multi-layer composition
of such transforms that lie in $\bb G_{\text{TMI}}$.  Popular examples of such transforms are given by the autoregressive flows in \cref{tab:flow_models}. \cref{tab:coboundary:param} presents several cocycles constructed using simple transforms and existing autoregressive flows, together with the lower triangular restrictions under a known partial ordering of the variables in $Y$ used to preserve identifiability. When further constraining these flows using a known causal DAG, we follow \cite{javaloy2023causal} and specify the inverse map $f_x^{-1}$ as the (forward) autoregressive flow with a mask on the adjacency matrix reflecting the sparsity of the DAG. When using a single layer this prevents spurious correlations from being induced by the architecture (see \cite{javaloy2023causal} for details).

\begin{table}[t]
  \centering
  \caption{Example cocycle parameterizations with classes of TMI maps.
  Here $\bb G$ denotes the transformation group of the cocycle.
  MAF = Masked Autoregressive Flow \citep{papamakarios2017masked};
  NSF = Neural Spline Flow \citep{durkan2019neural}.}
  \label{tab:coboundary:param}
  \small
  \resizebox{\textwidth}{!}{%
    \begin{tabular}{@{}lllll@{}}
      \toprule
      Transformation group $\bb G$ & Conditioner output & Cocycle $T_{x',x}(y)$ & Restriction for identifiability \\
      \midrule
      Shifts $(\mathbb{R}^d,+)$
        & $a_x \in \mathbb{R}^d$
        & $y + a_{x'} - a_x$
        & None \\
      $\mathrm{GL}_{+}(\mathbb{R}^d)$
        & $A_x \in \mathrm{GL}_{+}(\mathbb{R}^d)$
        & $A_{x'} A_x^{-1} y$
        & $A_x$ lower-triangular, $[A_x]_{jj}\geq 0$ \\
      $\mathrm{GA}_{+}(\mathbb{R}^d)$
        & $(A_x, a_x)$
        & $a_{x'} + A_{x'} A_x^{-1} (y - a_x)$
        & $A_x$ lower-triangular, $[A_x]_{jj}\geq 0$ \\
      $\mathrm{Diff}(\mathbb{R}^d) \cap \bb G_{\mathrm{TMI}} $
        & MAF parameters $\theta_x$
        & $\operatorname{MAF}^{-1}[\theta_{x'}] \circ \operatorname{MAF}[\theta_x](y)$
        & None \\
      $\mathrm{Diff}(\mathbb{R}^d) \cap \bb G_{\mathrm{TMI}}$
        & NSF parameters $\theta_x$
        & $\operatorname{NSF}^{-1}[\theta_{x'}] \circ \operatorname{NSF}[\theta_x](y)$
        & None \\
      \bottomrule
    \end{tabular}
  }
\end{table}

\subsection{Optimization and Model Selection}\label{sec:implementation:opt}

 \paragraph{CMMD Implementation} We optimize all flow-based cocycles using gradient descent on our empirical CMMD losses (V/U-statistic) introduced in \cref{sec:method}. Our default kernel choice for CMMD is the Gaussian kernel $k(y,y') = \exp(-\lambda \lVert y-y'\rVert^2)$, where the bandwidth $\lambda$ is chosen using the median heuristic on the observations $\{Y^{(i)}\}_{i=1}^n$ \citep{garreau2017large}. For gradient-based optimization, since both evaluation and gradients have a computational complexity of $\mc{O}(n^3)$, at each iteration we subsample $B \ll n$ datapoints, and then approximate $\nabla_{\theta}\ell_n^U$ (resp.\ $\nabla_{\theta}\ell_n^V$) 
with $\nabla_{\theta}\ell_{B}^U$ (resp.\ $\nabla_{\theta}\ell^V_{B}$). This stochastic optimization approach has been used for kernel-based estimators in several works \citep{greenfeld2020robust,jankowiak2021scalable,dance2022fast} and, in our case, estimates $\ell^U_n$ without bias and $\ell^V_n$ with a bias of order $1/B$. We use the ADAM optimizer \citep{kingma2014adam} with a default batch size is $B = \min (n,128)$. Algorithm \ref{alg:cmmd} in \cref{app:algs} presents pseudo-code for this procedure, for the V-statistic.

\paragraph{Model Selection} As established in earlier sections,  a key advantage to modeling counterfactual cocycles rather than an SCM with a base distribution, is that we maximize the `chance' of remaining well-specified using a simpler conditional flow (i.e., one contained in a smaller transformation group $\bb G$). We therefore advocate training a \emph{hierarchy} of flow-based cocycle classes with increasing transformation group expressivity, such as those presented in \cref{tab:coboundary:param}---coupled with conditioners of matched or increasing capacity. In practice we use $K$-fold cross-validation to do this. Algorithm \ref{alg:CV} in \cref{app:algs} demonstrates the procedure.

\section{Experiments}\label{sec:experiments}

We now implement counterfactual cocycles in a range of simulations and a real application, and compare against state-of-the-art SCM and OT methods for counterfactuals. Code for our method and experiments can be found at \url{https://github.com/HWDance/Cocycles}.

\subsection{Noise Ablation in Flow-based SCMs}\label{exp:lin_model}
We begin by comparing counterfactual cocycles to equivalent flow-based SCMs in a simple linear causal model. The goal of this experiment is to assess the robustness of our estimation approach to the underlying noise distribution $\bb P_{\xi}$, and to illustrate how our method can leverage the simplicity of the underlying cocycle, in contrast to the flow that arises from pushing forward a fixed base distribution $\mathbb{P}_0$.

\paragraph{Experimental Set-up} We generate $n=1000$ points from a linear structural model $Y = \beta X + \xi$ under different settings of $\bb P_{\xi}$. We implement cocycles trained using both CMMD-V and CMMD-U losses implemented as in \cref{sec:implementation:opt} (see Algorithm \ref{alg:cmmd}), and benchmark them against flow-based SCMs which estimate the bijective generative mechanism $(f, \mathbb{P}_\xi)$ via maximum likelihood, using different base distributions ($\mathsf{N}(\mu,\sigma^2)$, $\mathsf{Lap}(\mu,\sigma^2)$, $\mathsf{t}_\nu(\mu,\sigma^2)$) with learnable parameters. For all methods, we perform 2-fold cross-validation across flow architectures of increasing complexity: (i) linear $f_x(\xi)=\theta x+\xi$; 
(ii) additive $f_x(\xi)=m_\theta(x)+\xi$ with $m_\theta$ an MLP; 
(iii) a one-layer MAF \citep{papamakarios2017masked}; 
(iv) a one-layer NSF \citep{durkan2019neural} with 8 spline bins, flanked by affine transforms. In (iii)--(iv), we use the standard autoregressive architectures implemented in the \href{https://github.com/probabilists/zuko}{\texttt{zuko}} library, extended to also condition on $x$ via neural networks. All neural networks are set as MLPs with $L=2$ layers and $W=32$ nodes per layer. All models are trained in PyTorch using gradient descent on the respective objective functions (i.e., CMMD, ML) with the Adam optimizer with learning rate=$0.01$ (except NSF with $0.001$), batch size=$128$ and $1000$ epochs. 

\begin{table}[t]
\centering
\small
\resizebox{\textwidth}{!}{%
\begin{tabular}{lccccc}
\toprule
\textbf{Method}
  & $\bb P_\xi = \mathsf N(0,1)$
  & $\bb P_\xi = \mathsf{Ga}(1,1)$
  & $\bb P_\xi =\mathsf{t}_1(0,1)$
  & $\bb P_\xi =\mathsf{IG}(1,1)$
  & $\bb P_\xi =\mathsf{Rad}(1/2)$ \\
\midrule
\multicolumn{6}{c}{\textbf{Interventional KS}} \\
\cmidrule(lr){1-6}
ML-G  
  & $\bm{0.015 \pm 0.007}$  
  & $0.081 \pm 0.041$  
  & $0.121 \pm 0.079$  
  & $0.208 \pm 0.162$  
  & $0.405 \pm 0.069$ \\
ML-L  
  & $0.055 \pm 0.010$  
  & $0.074 \pm 0.036$  
  & $0.110 \pm 0.069$  
  & $0.120 \pm 0.090$  
  & $0.415 \pm 0.067$ \\
ML-T  
  & $0.018 \pm 0.008$  
  & $0.079 \pm 0.040$  
  & $0.029 \pm 0.007$  
  & $0.075 \pm 0.031$  
  & $0.412 \pm 0.062$ \\
CMMD-V  
  & $0.028 \pm 0.009$  
  & $\bm{0.027 \pm 0.008}$  
  & $\bm{0.027 \pm 0.009}$  
  & $\bm{0.027 \pm 0.008}$  
  & $\bm{0.271 \pm 0.031}$ \\
CMMD-U  
  & $\bm{0.010 \pm 0.004}$  
  & $\bm{0.012 \pm 0.005}$  
  & $\bm{0.008 \pm 0.003}$  
  & $\bm{0.009 \pm 0.004}$  
  & $\bm{0.268 \pm 0.008}$ \\
\midrule
\multicolumn{6}{c}{\textbf{Counterfactual RMSE}} \\
\cmidrule(lr){1-6}
ML-G  
  &  $\bm{0.035 \pm 0.035}$  
  &  $0.277 \pm 0.118$  
  &  $113.667 \pm 341.875$  
  &  $114.946 \pm 147.841$  
  &  $0.326 \pm 0.258$ \\
ML-L  
  &  $0.036 \pm 0.028$  
  &  $0.258 \pm 0.073$  
  &  $97.745  \pm 351.815$  
  &  $112.300 \pm 165.014$  
  &  $0.480 \pm 0.294$ \\
ML-T  
  & $\bm{0.033 \pm 0.031}$  
  &  $0.270 \pm 0.097$  
  &  $0.044   \pm 0.053 $  
  &  $29.872  \pm 41.628$  
  &  $0.391 \pm 0.307$ \\
CMMD-V  
  &  $\bm{0.035 \pm 0.026}$  
  & $\bm{0.020 \pm 0.015}$  
  & $\bm{0.040   \pm 0.031}$  
  & $\bm{0.028   \pm 0.024}$  
  & $\bm{0.017 \pm 0.019}$ \\
CMMD-U  
  &  $0.040 \pm 0.027$  
  & $\bm{0.022 \pm 0.016}$  
  & $\bm{0.033   \pm 0.027}$  
  & $\bm{0.027   \pm 0.023}$  
  & $\bm{0.014 \pm 0.011}$ \\
\midrule
\multicolumn{6}{c}{\textbf{True Architecture Selection \%}} \\
\cmidrule(lr){1-6}
ML-G  
  & 96\%  & 14\%  & 0\%   & 2\%   & 2\%   \\
ML-L  
  & $\bm{100\%}$ & 2\%   & 4\%   & 0\%   & 0\%   \\
ML-T  
  & 98\%  & 8\%   & 94\%  & 0\%   & 0\%   \\
CMMD-V  
  & $\bm{100\%}$ & $\bm{100\%}$ & $\bm{100\%}$ & $\bm{100\%}$ & $\bm{98\%}$  \\
CMMD-U  
  & $\bm{100\%}$ & $\bm{100\%}$ & $\bm{100\%}$ & $\bm{100\%}$ & $\bm{100\%}$ \\
\bottomrule
\end{tabular}
}
\caption{Mean ± SE of the interventional Kolmogorov–-Smirnov distance (top block) and counterfactual RMSE (middle block), averaged over 50 trials, plus the percent of correct architecture selections (bottom block), for SCM $Y=X+\xi$ under $\doInt(X=0)$ across different noise laws.  
“ML-” denotes ML flows with Gaussian (G), Laplace (L) or Student-$t$ (T) bases; “CMMD-” denotes our cocycle estimators (CMMD-V/U).  
Boldface marks the top two performers per column.}

\label{tab:linear_flows}

  \vspace{-10pt}

\end{table}

\paragraph{Results} 

\cref{tab:linear_flows} shows performance results for different true noise distributions under a hard intervention $\doInt(X=0)$, averaged over 50 seeds. The top block shows the Kolmogorov--Smirnov (KS) distance between the true and estimated marginal distribution of $Y(0)$; the middle block shows the RMSE between the true and estimated counterfactuals $Y(0)$ for the units in the dataset $\{X^{(i)},Y^{(i)}\}_{i=1}^n$, and the bottom block shows the fraction of trials on which the true linear architecture was selected by each method. When the true noise is Gaussian, all methods performed similarly well for both metrics, and selected the true linear architecture in almost all cases. This reflects the fact that the base distributions are either well-specified (i.e., Normal and Student's t) or close enough to well-specified (i.e., Laplace). The performance of the cocycle-based approach is roughly invariant to the true noise distribution, reflecting the robustness of our estimation approach to this aspect of the data generating process. The one exception is interventional KS for $\bb P_{\xi} = \mathsf{Rad}(1/2)$, which is naturally higher since the true interventional distribution lies on two points and so is very sensitive to imperfect matching. In contrast, the performance of all other methods deteriorates drastically when the noise distribution does not match the specified base distribution, and a more complex flow is chosen in an attempt to correct for this mis-specification. As a result, our cocycle estimators performed best for all non-Gaussian noise distributions. For counterfactual RMSE, the performance gain is in some cases more than two orders of magnitude. 

For the interventional KS, the performance gap was greatest when $\bb P_\xi \in \{\mathsf{t}_1(0,1)$, $\mathsf{IG}(1,1)$, $\mathsf{Rad}(1/2)\}$, reflecting that the base distributions have mis-specified tails or support in these cases (except the Student's-t base in the $\mathsf{t}_1(0,1)$ noise case). For $\bb P_\xi = \mathsf{Ga}(1,1)$, there is still a substantial performance gap, since the flow-based methods need to use a very complex neural spline flow in order to effectively learn the base distribution, which results in worse finite sample performance. For counterfactual RMSE, all methods compute counterfactuals using the same formula: $Y(0) = f_0 \circ f_X^{-1}(Y)$. Thus, the performance gain using cocycle targeting here purely reflects the robustness of our CMMD estimator and that we do not need a more complex flow to compensate for a poorly matched base distribution.

\begin{figure}[t]
    \centering
    \includegraphics[width=\linewidth]{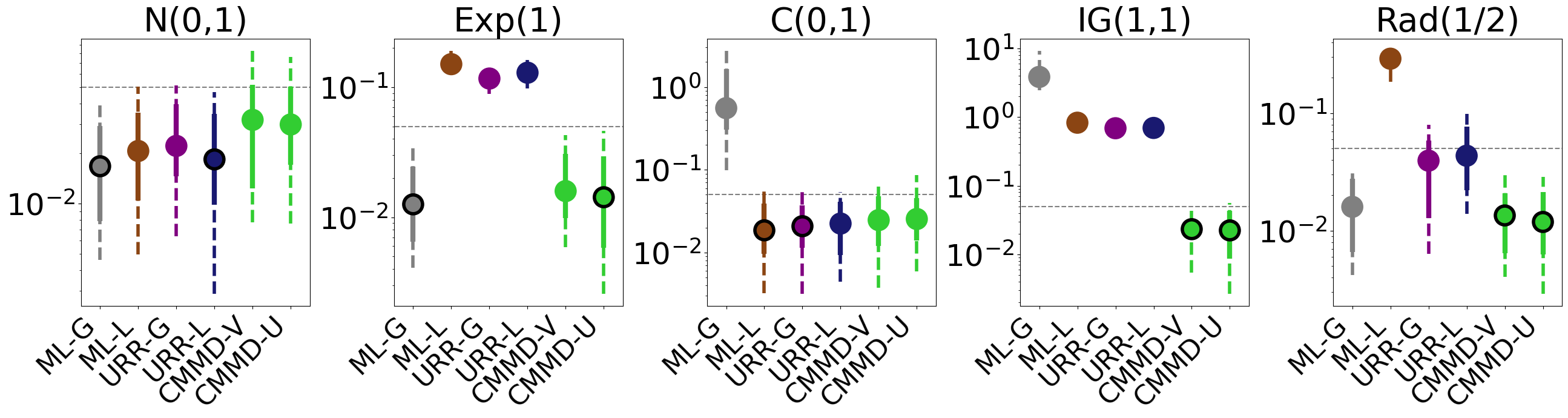}
    \caption{Mean (circle), (25-75) percentiles (solid line) and (10-90) percentiles (dashed line) of absolute error in treatment effect $Y(X+1) - Y(X) = 1$ for different estimation methods in a linear model $Y = X+\xi$ under different noise designs, when fixing the flow to be the true architecture $f_x(\xi) = \beta x + \xi$. Dashed horizontal line = $5\%$ error. Black edges = best two methods on average.}
    \label{fig:lin_model_plots}

      \vspace{-10pt}

\end{figure}

\paragraph{Extension} To analyze how the CMMD estimator performs when all architectures are fixed, in Figure \ref{fig:lin_model_plots} we also produce counterfactual RMSE results under a shift intervention $X \mapsto X+1$, when restricting all flow architectures to be the true (linear) architecture $f_x(\xi) = \beta x + \xi$. Note that in this case, $\hat Y(X+1) - Y(X+1) = \hat \beta - \beta$, so counterfactual error is isometric to cocycle estimation error. For this we compare the CMMD estimator against ML estimators with the Gaussian and Laplace base distributions (i.e., $\ell_2$ and $\ell_1$ regression), as well as a recently proposed MMD-based estimator for conditional generative models: Universal Robust Regression (URR) \citep{alquier2023universal}. The latter bears similarities with the CMMD estimator, with the exception that it requires specifying the full generative model (i.e., a base distribution). For URR we optimize the conditional estimator Eq.\ (5) from \citep{alquier2023universal}, using $m=n$ Monte Carlo samples from $\widehat{\bb P}_{Y|X}^{\theta}$. The kernel is chosen identically to CMMD.  Both CMMD estimators estimate $\beta$ within $5\%$ error on average across all noise designs, reflecting its noise-robustness. By contrast, all other estimators perform poorly (i.e., $\gg 50\%$ error) on at least two noise distributions.

\subsection{Confounding and Path-Consistency Ablation in Transport-based Models} \label{sec:experiments:OT}

\begin{figure}[t]
    \centering
\begin{minipage}{0.48\linewidth}\centering
\begin{tikzpicture}[>=stealth,
        every node/.style={draw,circle,inner sep=1pt,minimum size=16pt}]
  \node (X)  at (0,0)   {$X$};
  \node (Y1) at (1,0)   {$Y_1$};
  \node (Y2) at (2,0)   {$Y_2$};
  \node[dashed, color = gray] (E1) at (1,1) {$\varepsilon_1$};
  \node[dashed, color = gray] (E2) at (2,1) {$\varepsilon_2$};
  \node[draw=none] (titleL) at (-2.5,0.45) {\textbf{Confounded Chain}};

  \draw[->] (X)  to (Y1);
  \draw[->] (Y1) to (Y2);

  \draw[->, dashed, color = gray] (E1) -- (Y1);
  \draw[->, dashed, color = gray] (E2) -- (Y2);

  \draw[-, color = gray] (E1) -- (E2);
\end{tikzpicture}\end{minipage}\hfill
\begin{minipage}{0.48\linewidth}\centering
\begin{tikzpicture}[>=stealth,
        every node/.style={draw,circle,inner sep=1pt,minimum size=16pt}]
  \node (X)  at (0,0)   {$X$};
  \node (Y1) at (1,0)   {$Y_1$};
  \node (Y2) at (2,0)   {$Y_2$};
  \node[dashed, color = gray] (E1) at (1,1) {$\varepsilon_1$};
  \node[dashed, color = gray] (E2) at (2,1) {$\varepsilon_2$};
  \node[draw=none] (titleR) at (-2.5,0.45) {\textbf{Non-additive Triangle}};

  \draw[->] (X)  to (Y1);
  \draw[->] (Y1) to (Y2);
  \draw[->, bend right=30] (X)  to (Y2); 

  \draw[->, dashed, color = gray] (E1) -- (Y1);
  \draw[->, dashed, color = gray] (E2) -- (Y2);
\end{tikzpicture}
\end{minipage}
\vspace{-30pt}

    \includegraphics[width=0.49\linewidth]{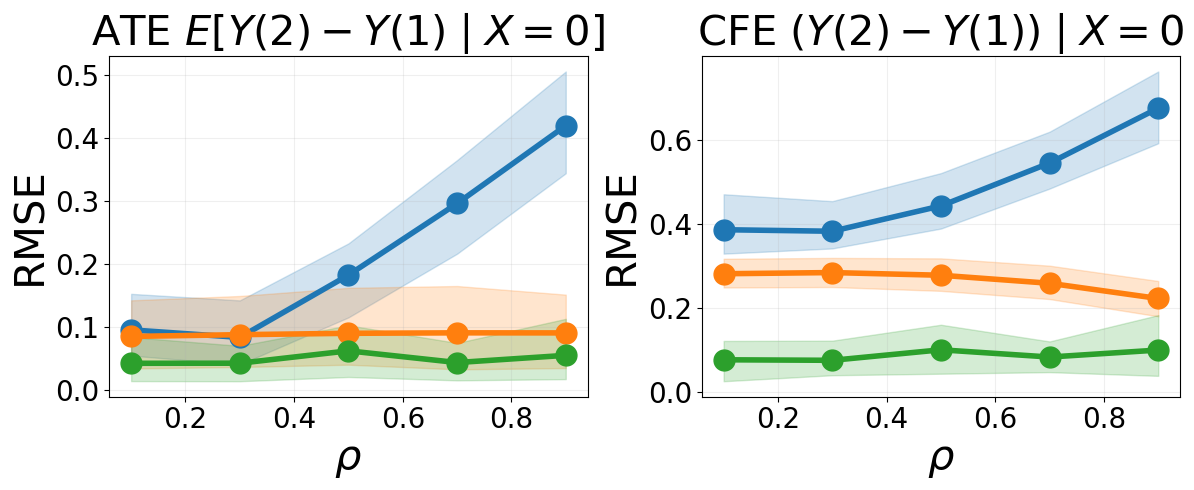}
    \includegraphics[width=0.49\linewidth]{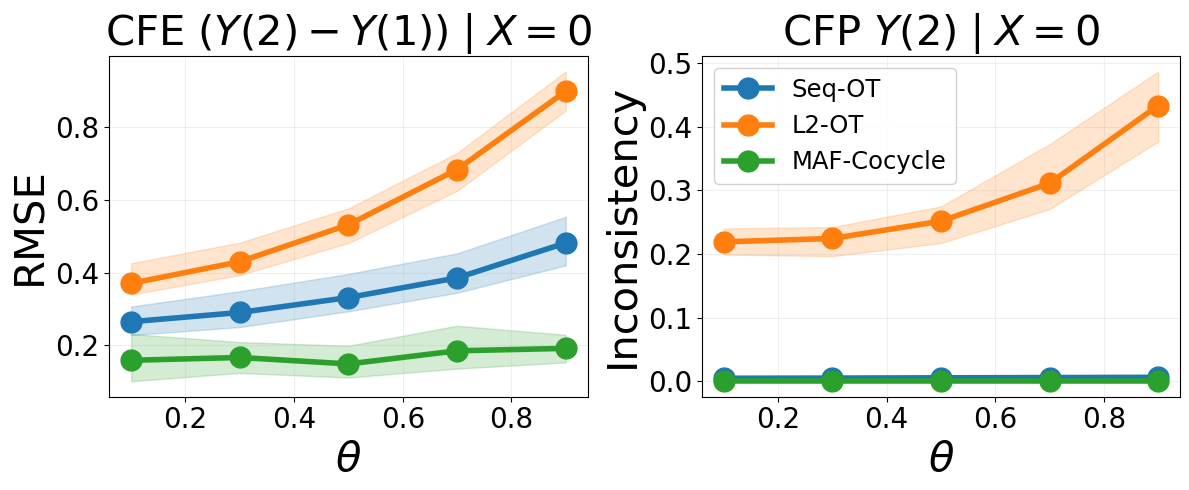}
   \caption{
Left and middle left (Confounded Chain): Mean $\pm$ SD (20 trials) of error in estimated average incremental effect (left) and incremental counterfactual effect (middle left) of second treatment, for observed units $\{Y^{(i)}(0)\}^n_{i=1}$ in control group. {\color{blue}blue} = sequential optimal transport \citep{machado2024sequential}, {\color{orange}orange} = optimal transport (Brenier maps) \citep{de2024transport}, {\color{green} green} = Masked Autoregressive Flow cocycle. Under the chain DAG, Sequential-OT becomes biased as correlation strength $\rho := \text{Corr}(\xi_1,\xi_2)$ increases, whereas the MAF-cocycle does not, due to the robustness of our estimation procedure. Right and middle right (Non-additive Triangle): Mean $\pm$ SD (20 trials) error in incremental counterfactual effect (middle right) and inconsistency in counterfactual predictions $Y(2)$ (right) when imputing for observed units $\{Y^{(i)}(0)\}^n_{i=1}$ in control group, via (a) indirect $T_{2 ,1 } \circ T_{1 , 0}$ and (b) direct $T_{2,0}$ transports. $\theta$ controls degree of non-additivity in true transport maps. As $\theta$ increases, OT performance degrades and transport inconsistency increases, reflecting underlying model incoherence.
}
    \label{fig:ot_inconsistency}

  \vspace{-10pt}

\end{figure}

In this experiment, we compare counterfactual cocycles against OT-based approaches and assess how estimation accuracy and path-consistency varies under different assumptions.
 
\paragraph{Experimental Set-up}
Suppose we have collected data under a randomized controlled trial which tests two treatments and a control, i.e., \(X\in\{0,1,2\}\). We observe two outcomes $Y:= (Y_1,Y_2)$ for $n=500$ patients under control, \(\{Y^{(i)}(0)\}_{i=1}^{n}\), under treatment \(X=1\), \(\{Y^{(i)}(1)\}_{i=n}^{2n}\), and under alternative treatment $X=2$, \(\{Y^{(i)}(2)\}_{i=2n}^{3n}\)). We aim to estimate the incremental effectiveness of treatment $X=2$ by estimating transports \(T_{0,1},T_{0,2},T_{1,2}\) between each state, and using them to compute the contrast $Y(2) - Y(1)$ for each unit. For ease of demonstrating path-consistency issues, we only do this for the control group units.

We consider two generating designs to isolate different weaknesses of competing methods:  
(i) a chain SCM $Y_1 = X + \xi_1,\; Y_2 = Y_1 + \xi_2$ with $(\xi_1,\xi_2) \sim \mathsf{Lap}(\bm 0,\rho\, \bm 1\bm 1^\mathsf{T} + (1-\rho)I)$; and 
(ii) a non-additive SCM $\bm Y = \bm 1 X + L(X)\bm \xi$ with $\bm \xi \sim \mathsf{Lap}(0,I_2)$ and
\[L(0) = \begin{pmatrix} 1 & 0 \\ 0 & 1\end{pmatrix}, \quad L(1) = \begin{pmatrix} 1 & -\theta \\ -\theta & 1\end{pmatrix}, \quad L(2) = \begin{pmatrix} \theta + 1 & 1 \\ 1 & (1+\theta)^{-1}\end{pmatrix}\]
In design (i) we vary the noise correlation $\rho  \in [0,1]$; in design (ii) we vary $\theta \in [0,1]$, which intuitively controls the degree of non-additivity in the true transport maps. The former lets us analyze how noise dependence affects estimation performance of sequential-OT methods, while the latter lets us analyze how non-additivity affects path-consistency of OT methods (in general, OT maps are coherent between distributions which are shifts of one another).

For the cocycle, we use a single layer MAF \citep{papamakarios2017masked} per treatment level $x\in\{0,1,2\}$, i.e., $f_x := \mathrm{MAF}[\theta_x]$ with $\theta_x = (\theta_0,\theta_1,\theta_2)[x]$. The network that parameterizes each conditional affine transform of $y_2$ given $y_1$ is set as a two‐layer MLP with $L=2$ layers and $W=32$ nodes per layer. We optimize the cocycle to target \eqref{eq:DA} w.r.t.\ $\bb P_{Y_1,Y_2|X}$ via CMMD using Algorithm \ref{alg:cmmd}. We use the same optimization settings as \cref{exp:lin_model}.

We compare our flow-based cocycle against OT with quadratic cost \citep{de2024transport, charpentier2023optimal, balakrishnan2025conservative}, implemented using the network‐simplex solver and barycentric projection via the \texttt{POT} package, following \citet{de2024transport}. These methods directly estimate $T_{x,x'}$ between each pair of treatment levels without enforcing path-consistency \eqref{eq:CPI}. When the true transport maps are not additive (as in design (ii)), the resulting pairwise maps may be mutually inconsistent. We also compare against the sequential-OT approach discussed in \cref{sec:cocyclevsscm:seqot} \citep{plevcko2020fair, machado2024sequential}. This approach factorizes the joint transport into a set of (1D) conditional transports of each node given its parents in the DAG, with each transport estimated via the conditional quantile transform. We estimate the conditional CDFs using (Gaussian) kernel smoothing, with bandwidths chosen via the median heuristic \citep{garreau2017large}. Sequential-OT is expected to guarantee \eqref{eq:CPI} in the large sample limit, since each coordinate map satisfies \eqref{eq:CPI} in this limit. However, given our analysis in \cref{sec:cocyclevsscm:seqot}, we expect performance to degrade as the noise correlation increases (as in design (i)).

\paragraph{Results}
For design (i), \cref{fig:ot_inconsistency} reports the RMSE in the estimated average contrast $\bb E[Y(2) - Y(1)|X=0]$ (ATE) (left), and the RMSE in the counterfactual effect $Y(2) - Y(1)$ (CFE) for {control group units} (middle left), over different noise correlation levels (averaged over 20 trials). Increasing the noise correlation $\rho$ leaves cocycle and OT performance essentially unchanged, while sequential OT shows steadily increasing bias in both ATE and CFE. This is to be expected given the analysis in \cref{sec:cocyclevsscm:seqot}. Note, by default we impute counterfactuals using $T_{1,0}$ and $T_{2,0}$, i.e., $(\hat Y(1), \hat Y(2)) = \bigl(T_{1,0}(Y(0)), T_{2,0}(Y(0))\bigr)$.

For design (ii), \cref{fig:ot_inconsistency} reports the RMSE of the CFE again (middle right) as well as the estimated RMSE between the predicted counterfactuals $\hat Y(2)$, when imputing them from the control group units via the direct transport $T_{2,0}$, and the indirect composition $T_{2 ,1} \circ T_{1 ,0}$ (right). Note that each approach is equally valid here, as to impute both $Y(1)$ and $Y(2)$ for control-group units two out of three maps are always needed. However, each pair may induce different couplings over all three states. As expected, increasing the `non-additivity' parameter $\theta$ leads to large increases in both counterfactual RMSE and path‐inconsistency for global OT. This shows that the incoherence of OT transports and resultant non-identifiability problem generalizes beyond the 2D Gaussian example in \cref{sec:bg:ot}.  Sequential OT is less affected here, but its performance remains inferior to the MAF-cocycle.

\subsection{Performance on SCM Benchmarks} \label{exp:bd}

We now assess how our method performs on causal benchmarks used in the SCM literature, against state-of-the-art flow-based SCMs.

\paragraph{Experimental Setup}
We consider linear and non-linear variants of the following benchmark SCMs used in previous work \citep{geffner2022deep, javaloy2023causal}: (i) \verb|Triangle|, a 3-node SCM with a dense causal graph; (ii) \verb|Fork| a 4-node SCM with a sparse causal graph; and (iii) \verb|5-chain|, a 5-node SCM with a chain structure. The linear and non-linear mechanisms for \verb|Triangle|, \verb|Fork| and \verb|5-chain| can be found in \cite{javaloy2023causal}. We also implement a two-variable SCM (\verb|2var (lin)|: $Y = X + \xi$, \verb|2var (nonlin)|: $Y = \sin(X) + \xi$). Unlike in previous implementations where all noise distributions were Gaussian, we set each node in the SCM with a different noise distribution, enabling us to assess performance in a more challenging and realistic setting. In particular, we set $\xi_1 = \mathsf{N}(0,1)$ and $\xi_{d:2} =  \{\mathsf{IG}(1,1), \mathsf{Rad}(1/2), \frac{1}{2}\mathsf{N}(-\sqrt{3}/2,1/2)+\frac{1}{2}\mathsf{N}(\sqrt{3}/2,1/2), \mathsf{Ga}(1,1)\}$. 

We implement flow-based cocycles on $(X,Y) = (V_1,V_{>1})$ with the CMMD-V loss, against several state-of-the-art flow-based SCMs: (i) CAREFL \citep{khemakhem2021causal}, which uses affine autoregressive flows to learn $\bb P_{X,Y}$ (ii) CAUSALNF \citep{javaloy2023causal}, which extends CAREFL to arbitrary flows but enforces a single (abductive) flow layer to prevent the flow from enforcing spurious edges in the adjacency matrix, and (iii) BGM \citep{nasr2023counterfactual}, which trains a conditional flow to match $\bb P_{Y|X}$ and uses the empirical $\widehat{\bb P}_X$. For all methods (including ours) we assume the causal ordering is known, but the DAG is unknown. Hence, all autoregressive network architectures are dense. We use the same architectures, training and cross-validation procedure for all methods as in \cref{exp:lin_model}. However, note CAREFL is restricted to affine architectures, and we additionally cross-validate over Gaussian and Laplace base distributions for the baselines. 

\begin{table}[ht]
  \centering
  \caption{Mean \(\pm\) SD of KS$_\mathrm{int}$ and KS$_\mathrm{CF}$ on the \emph{linear} SCMs.}
  \label{tab:ks_linear}
  \resizebox{\textwidth}{!}{
  \begin{tabular}{l  cc  cc  cc  cc}
    \toprule
    Method
      & \multicolumn{2}{c}{2var (lin)}
      & \multicolumn{2}{c}{triangle (lin)}
      & \multicolumn{2}{c}{fork (lin)}
      & \multicolumn{2}{c}{5chain (lin)} \\
    \cmidrule(lr){2-3}\cmidrule(lr){4-5}\cmidrule(lr){6-7}\cmidrule(lr){8-9}
      & KS$_\mathrm{int}$ & KS$_\mathrm{CF}$
      & KS$_\mathrm{int}$ & KS$_\mathrm{CF}$
      & KS$_\mathrm{int}$ & KS$_\mathrm{CF}$
      & KS$_\mathrm{int}$ & KS$_\mathrm{CF}$ \\
    \midrule
    BGM  
      & $0.13 \pm 0.07$ & $0.19 \pm 0.12$
      & $0.37 \pm 0.05$ & $0.06 \pm 0.01$
      & $0.05 \pm 0.07$ & $0.61 \pm 0.07$
      & $0.14 \pm 0.05$ & $0.06 \pm 0.01$ \\
    CausalNF 
      & $0.31 \pm 0.11$ & $0.24 \pm 0.10$
      & $0.44 \pm 0.05$ & $0.06 \pm 0.02$
      & $0.04 \pm 0.02$ & $0.66 \pm 0.09$
      & $0.14 \pm 0.02$ & $0.06 \pm 0.01$ \\
    CAREFL  
      & $0.40 \pm 0.04$ & $0.15 \pm 0.14$
      & $0.43 \pm 0.05$ & $0.06 \pm 0.01$
      & $0.19 \pm 0.01$ & $0.58 \pm 0.10$
      & $0.13 \pm 0.02$ & $0.06 \pm 0.01$ \\
    CocycleNF
      & $\bm{0.03 \pm 0.02}$ & $\bm{0.04 \pm 0.04}$
      & $\bm{0.23 \pm 0.19}$ & $\bm{0.02 \pm 0.01}$
      & $\bm{0.02 \pm 0.01}$ & $\bm{0.19 \pm 0.23}$
      & $\bm{0.02 \pm 0.01}$ & $\bm{0.03 \pm 0.01}$ \\
    \bottomrule
  \end{tabular}
  }
\end{table}

\begin{table}[ht]
  \centering
  \caption{Mean \(\pm\) SD of KS$_\mathrm{int}$ and KS$_\mathrm{CF}$ on the \emph{nonlinear} SCMs.}
  \label{tab:ks_nonlinear}
  \resizebox{\textwidth}{!}{
  \begin{tabular}{l  cc  cc  cc  cc}
    \toprule
    Method
      & \multicolumn{2}{c}{2var (nonlin)}
      & \multicolumn{2}{c}{triangle (nonlin)}
      & \multicolumn{2}{c}{fork (nonlin)}
      & \multicolumn{2}{c}{5chain (nonlin)} \\
    \cmidrule(lr){2-3}\cmidrule(lr){4-5}\cmidrule(lr){6-7}\cmidrule(lr){8-9}
      & KS$_\mathrm{int}$ & KS$_\mathrm{CF}$
      & KS$_\mathrm{int}$ & KS$_\mathrm{CF}$
      & KS$_\mathrm{int}$ & KS$_\mathrm{CF}$
      & KS$_\mathrm{int}$ & KS$_\mathrm{CF}$ \\
    \midrule
    BGM  
      & $0.12 \pm 0.07$ & $0.27 \pm 0.13$
      & $0.41 \pm 0.07$ & $\bm{0.09 \pm 0.05}$
      & $\bm{0.04 \pm 0.01}$ & $0.59 \pm 0.08$
      & $0.07 \pm 0.03$ & $0.41 \pm 0.12$ \\
    CausalNF  
      & $0.30 \pm 0.11$ & $0.26 \pm 0.08$
      & $0.47 \pm 0.04$ & $0.23 \pm 0.08$
      & $0.07 \pm 0.06$ & $0.61 \pm 0.09$
      & $0.06 \pm 0.06$ & $0.24 \pm 0.16$ \\
    CAREFL  
      & $0.44 \pm 0.04$ & $0.22 \pm 0.15$
      & $0.47 \pm 0.06$ & $0.22 \pm 0.09$
      & $0.19 \pm 0.01$ & $0.51 \pm 0.21$
      & $0.17 \pm 0.04$ & $0.60 \pm 0.25$ \\
    CocycleNF
      & $\bm{0.04 \pm 0.02}$ & $\bm{0.13 \pm 0.05}$
      & $\bm{0.28 \pm 0.13}$ & $0.20 \pm 0.16$
      & $0.08 \pm 0.05$ & $\bm{0.20 \pm 0.24}$
      & $\bm{0.05 \pm 0.02}$ & $\bm{0.20 \pm 0.09}$ \\
    \bottomrule
  \end{tabular}
  }
\end{table}

\paragraph{Results}
For each method we evaluate the learned distribution of $V_{>1}(0)$ and the distribution of the current policy‐effect $V_{>1} - V_{>1}(0)$ under the intervention $\doInt(V_1 = 0)$, by computing the \emph{average marginal Kolmogorov--Smirnov (KS) distance}:
\[
\mathrm{KS}_\mathrm{int}
= \frac{1}{d-1}\sum_{j=2}^d D_{\mathrm{KS}}\bigl(\widehat{F}_{V_j(0)},F_{V_j(0)}\bigr), \quad \mathrm{KS}_\mathrm{CF}
= \frac{1}{d-1}\sum_{j=2}^d D_{\mathrm{KS}}\bigl(\widehat{F}_{V_j - V_j(0)},F_{V_j - V_j(0)}\bigr),
\]
Here $\hat F$ is the CDF, which for flow-based cocycles is estimated empirically using the imputed counterfactuals $\hat V_{>1}^{(i)}(0) = \hat T_{0,V_1^{(i)}}(V_{>1}^{(i)})$, and for baselines is estimated via the abduct-act-predict procedure in \cref{sec:background:scms}, with $m=10^5$ Monte Carlo samples.
Tables~\ref{tab:ks_linear} and~\ref{tab:ks_nonlinear} report the mean \(\pm\) SD results from 10 trials. Cocycles achieves the best KS$_\mathrm{int}$ and KS$_\mathrm{CF}$ in all SCMs except one (\verb|Triangle-nonlin|  for  KS$_\mathrm{CF}$ and \verb|Fork-nonlin|  for  KS$_\mathrm{int}$). The performance gap is generally greatest when the true DGP is linear, reflecting our method’s ability to be well‐specified with simpler architectures. Out of the baselines, CAREFL performed worst, likely reflecting its restriction to affine flows, while BGM generally performed best. The latter is to be expected, as under the intervention $\doInt(V_1=0)$, the only part of the flow used at test time is the component which is directly optimized by BGM, $f:(\xi_1,\xi_{>1}) \mapsto f_{\xi_1}(\xi_{>1})$.

\subsection{Application: Counterfactual Effects of 401(k) Pension Plan Eligibility} \label{sec:application}

 As an application to real data, we use counterfactual cocycles to estimate the impact of 401(k) eligibility on net financial assets, using the well-known economic dataset studied in \cite{chernozhukov2004effects}. The dataset contains $n=9915$ households with variables $(Y^{(i)},D^{(i)},Z^{(i)})_{i=1}^n$. $Y^{(i)} \in \mathbb R_+$ is net financial assets, $D^{(i)} \in \{0,1\}$ is a binary indicator for eligibility to enroll in a 401(k) savings plan, and $Z^{(i)} \in \bb Z \subseteq \mathbb R^9$ are covariates measuring demographics and earnings, as described in \cite{chernozhukov2004effects}.

  \begin{figure}[t]
    \centering
\includegraphics[width = 0.49\textwidth]{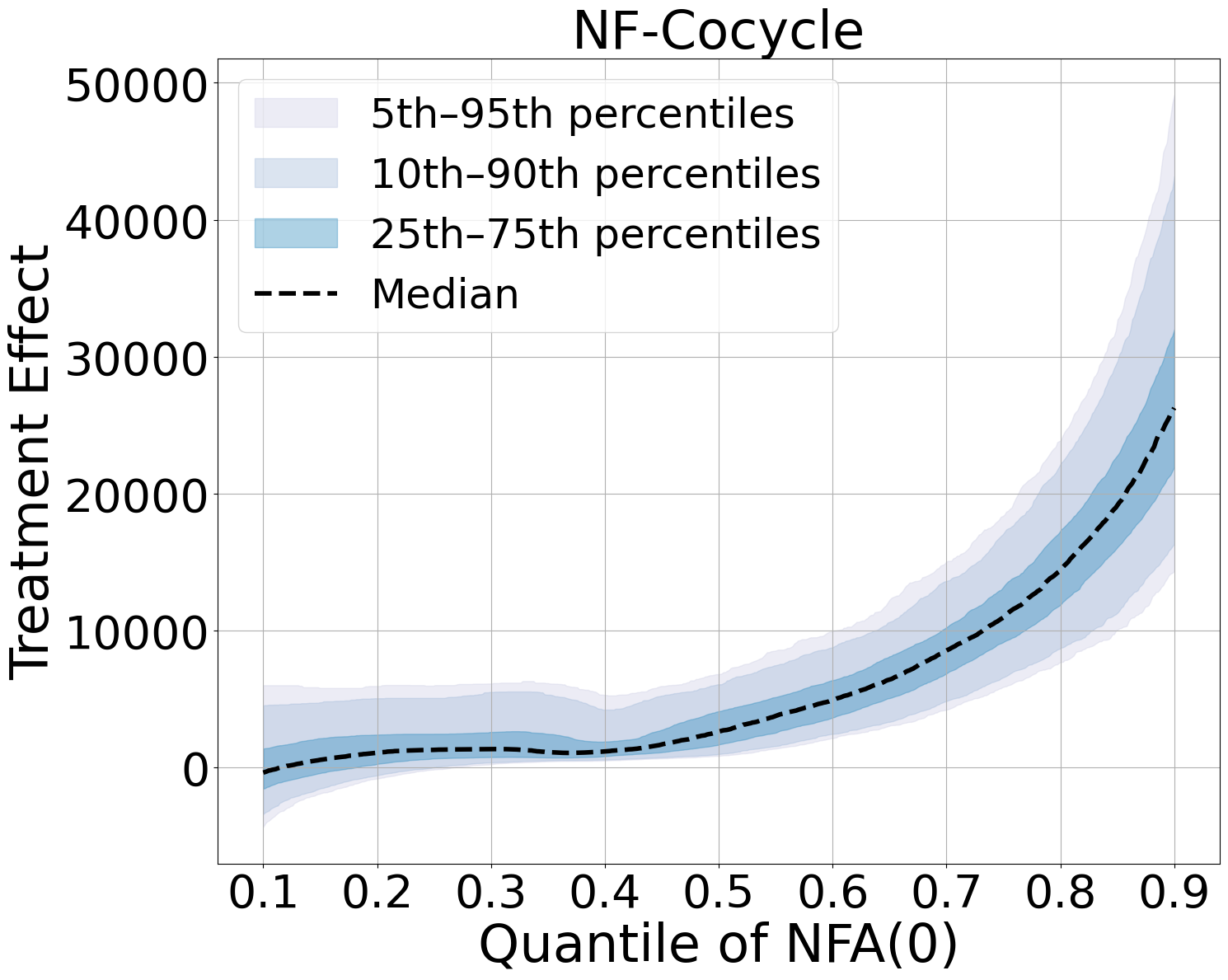}
\includegraphics[width = 0.49\textwidth]{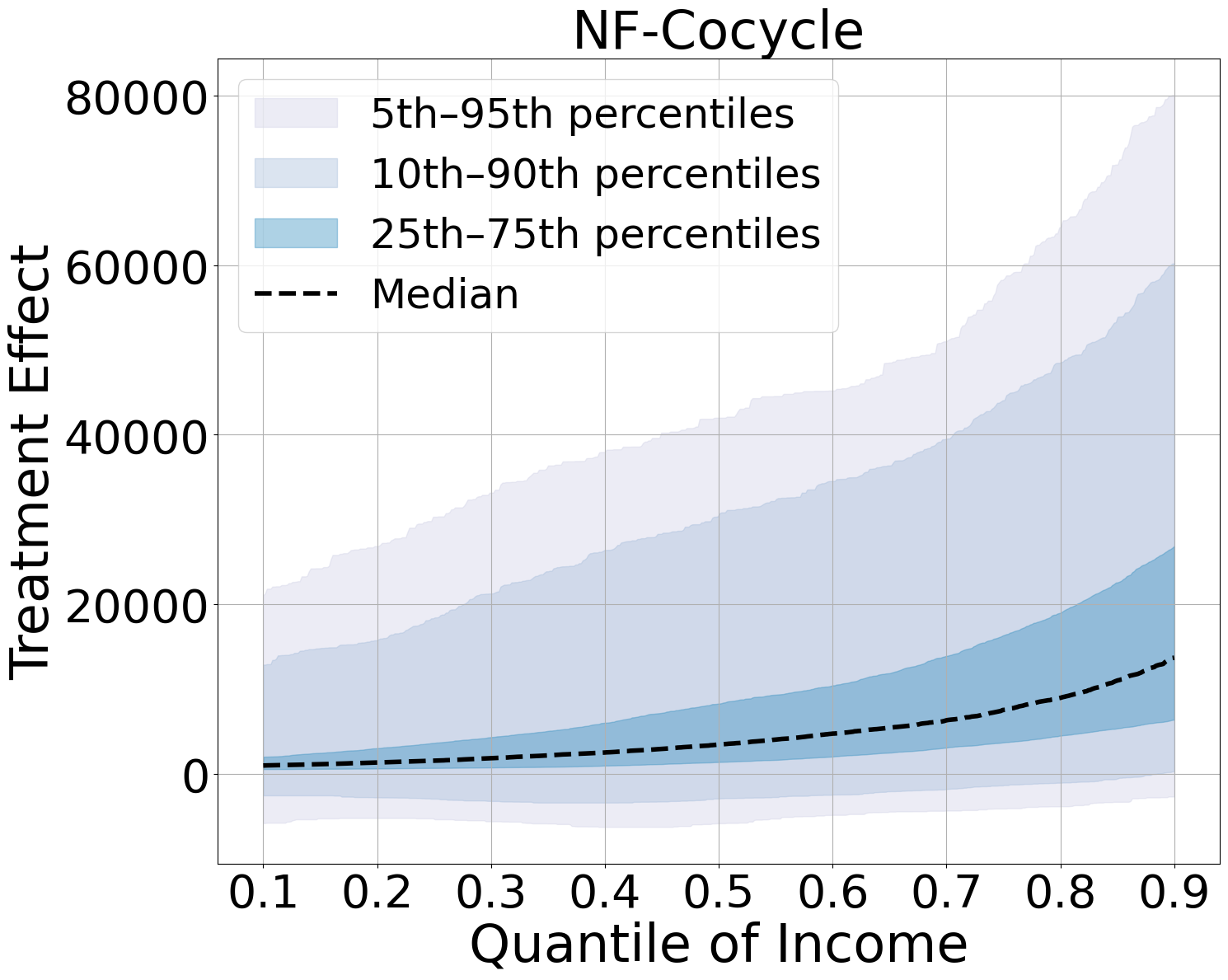}
    \caption{Estimated quantiles of the treatment effect $Y(1) - Y(0)$ (i.e., change in net financial assets) given quantiles of (i) $Y(0)$ (i.e., net financial assets under the no treatment scenario) and (ii) income, using normalizing flow-based cocycle.
    }
    \label{fig:e401(k)}
    \vspace{-10pt}
\end{figure}

We investigate the hypothesis that the effect of the 401(k) pension scheme on wealth accumulation follows a ``rich get richer'' phenomenon---i.e., whether those that benefited most from the 401(k) scheme are those who would otherwise have most wealth in the first place. To answer this question, we  estimate how quantiles of the treatment effect $Y(1) - Y(0)$ varies across quantiles of (i) income levels $I \in Z$, and (ii) net financial assets under the no-treatment scenario $Y(0)$. Estimating (i) will help us to determine whether individuals who are richer in income are able to take better advantage of the 401(k) pension scheme distributions, through their ability to save. Estimating (ii) will help us to determine the extent to which those who benefit most are those who were better off regardless.

We work under the assumptions laid out in \cref{sec:method:confounding} on counterfactuals $\{Y(d,z): (d,z) \in \{0,1\} \times \bb R^9\}$.  We estimate flow-based counterfactual cocycles on these counterfactuals using the same architectures and procedure as in \cref{exp:lin_model}. We use the estimated cocycle to impute counterfactuals 
$\hat Y(1) := \hat Y(1,Z)$ and $\hat Y(0) := \hat Y(0,Z)$, 
and estimate the quantile of the treatment effect 
$\tau = Y(1)-Y(0)$ at level $q$ given covariate $V \in \{I,Y(0)\}$ by
\[
\hat Q_{\tau \mid V=v}(q) 
:= \inf \Bigl\{ t \in \mathbb{R} : 
{\sum\nolimits_{i=1}^n \hat w_i(v)\,\mathbf{1}\{\hat \tau_i \le t\}}
      \;\ge\; q \Bigr\},
\]
i.e., the empirical quantile of $\{\hat \tau_i\}_{i=1}^n$ reweighted via 
Nadaraya--Watson (Gaussian) kernel weights 
$\hat w_i(v) \propto K_\lambda(v-V_i)$. 
The bandwidth $\lambda$ is selected by $5$-fold cross-validation, 
minimizing the squared error of predicting 
$\mathbf{1}\{\hat \tau \le t_j\}$ from $V$, across different thresholds $\{t_j\}_{j=1}^{100}$.

The estimated conditional quantiles of the treatment effect are displayed in \cref{fig:e401(k)}. Both distributions are consistent with the ``rich get richer'' phenomenon. In particular, the effects are generally largest for those who would have had greater net financial assets without access to a 401(k) pension scheme in the first place. While the majority of the 10th percentile of the distribution barely see a positive effect, the majority of the 90th percentile of the distribution see increases of $\geq \$20,000$. The story is somewhat similar when conditioning the effect on income levels (i.e., those with larger incomes see greater increases in net financial assets), albeit with larger treatment effect variance per quantile.

The fact that $Y(1) - Y(0)$ typically increases with $Y(0)$ implies that the effect of treatment on the quantile, $ETQ(\tau) = Q_{Y(1)}(\tau) - Q_{Y(0)}(\tau)$ is also increasing over $\tau$, which has been reported in previous studies \citep{belloni2017program,chernozhukov2004effects}.

 \section{Conclusion}  \label{sec:conclusion}

In this work, we introduced a general framework for modeling 
transport-based couplings over counterfactuals. Such couplings 
are essential for estimating measures of treatment risk and heterogeneity. 
To overcome the incoherence and identifiability problems of previous transport approaches, our 
key idea was to model transports between counterfactuals in a way that guarantees the necessary and sufficient algebraic properties to induce valid couplings. 
We called the resulting set of transports a \emph{counterfactual cocycle}, given the 
connection to cocycles in dynamical systems \citep{arnold1998random}.  

To achieve this, we characterized the structure of all counterfactual cocycles, and showed that each cocycle is equivalent to a 
class of injective SCMs. This equivalence enables parameterization via 
the same autoregressive flows used in flow-based SCMs, and 
identifiability under a known causal ordering. Crucially, however, cocycles are \emph{noise invariant}: 
they depend only on the transports, not on the choice of latent noise 
distribution. This allows estimation to be centered directly on the 
transports, eliminating the need to model the noise law. Moreover, the 
flows required to represent a cocycle can be significantly simpler than 
those needed for the corresponding SCM, yielding models that are both 
well-specified under milder conditions and less prone to mis-specification.

To estimate cocycles efficiently, we proposed a new estimator based on 
minimizing the maximum mean discrepancy (MMD) between the true and predicted counterfactual 
marginals under the transports. In contrast to maximum likelihood approaches used in traditional 
flow-based SCMs, the estimator is \emph{noise-robust}: its consistency does 
not rely on the properties of the underlying noise distribution. These 
advantages translate into strong empirical performance: across synthetic 
benchmarks and a 401(k) eligibility study, cocycle models outperformed both 
OT-based methods and flow-based SCMs.

One interesting direction for future research is how to construct identifiable classes of counterfactual cocycles without knowledge of the causal ordering. A promising avenue could be to combine the algebraic structure of cocycles with the optimal transport criteria, yielding valid transports that satisfy the causal 
principle of counterfactual similarity \citep{lewis1973causation}. It also remains to be seen how one could extend the framework to settings with more severe forms of unobserved confounding, where counterfactual marginals are not conditionals. 

\paragraph{Acknowledgements} The authors are grateful to Peter Orbanz for introducing them to cocycles, and for helpful conversations about them. HD is supported by the Gatsby Charitable Foundation. BBR acknowledges the support of the Natural Sciences and Engineering Research Council of Canada (NSERC): RGPIN2020-04995, RGPAS-2020-00095, DGECR2020-00343.

\appendix

\section{Proofs}

\subsection{Proofs for \cref{sec:method}}

\begin{proof}[\cref{thm:cocycle_sufficiency}]
For the first direction, suppose $\{T_{x,x'}\}_{x,x' \in \bb X}$ satisfies \eqref{eq:NI}, \eqref{eq:CPI} and \eqref{eq:DA} for a given $\mathbb P_{Y\mid X}$.  Now, fix the (standard Borel) probability space $(\bb Y, \mc B(\bb Y), \bb P_{Y|X=x_0})$, where $\mc B(\bb Y)$ is the Borel $\sigma$-algebra on $\bb Y$. We can therefore define the random variable $Y(x_0): \bb Y \to \bb Y$ as the identity map, which is Borel measurable. By construction, $Y(x_0) \sim \bb P_{Y|X=x_0}$. Now, define the random variables
\[\widetilde Y(x) := T_{x,x_0}(Y(x_0)) \;, \quad \forall x \in \bb X\]
Note by measurability of each $T_{x,x_0}$, they are well-defined. By \eqref{eq:DA} we have $\widetilde Y(x) \sim \bb P_{Y|X=x}$ for each $x \in \bb X$, which is the marginal distribution requirement of admissibility. To show \eqref{eq:CC} holds almost surely, we note by  \eqref{eq:NI}  and \eqref{eq:CPI} that $T_{x_0,x} \circ T_{x,x_0} = T_{x_0,x_0} = \text{id}$, $\bb P_{Y|X=x_0}$-a.s. This means that 
\[T_{x_0,x}\bigl(\widetilde Y(x)\bigr) :=T_{x_0,x} \circ T_{x,x_0}\bigl( Y(x_0)\bigr) =_{\mathrm{a.s.}}Y(x_0) ,\; \forall x \in \bb X\,.\] Applying $T_{x',x_0}$ to both sides and noting that $x$ and $x'$ are arbitrary gives 
\[ T_{x',x_0} \circ T_{x_0,x}\bigl(\widetilde Y(x)\bigr) =_{\mathrm{a.s.}}T_{x',x_0}\bigl(Y(x_0)\bigr) := \widetilde Y(x') \;, \quad \forall x,x' \in \bb X\,.\]
By \eqref{eq:CPI}, the LHS = $T_{x',x}(\widetilde Y(x)\bigr)$ a.s. Thus, the transports are admissible. 

For the reverse direction, let $\{T_{x,x'}\}_{x,x'\in\bb X}$ be admissible w.r.t.\ $\bb P_{Y\mid X}$. In this case, there exist random variables $\{\widetilde Y(x)\}_{x\in\bb X}$ with $\widetilde Y(x)\sim\bb P_{Y\mid X=x}$ and
$\widetilde Y(x)=_{\mathrm{a.s.}}T_{x,x'}(\widetilde Y(x'))$ for all $x,x' \in \bb X$. Below we show that this implies \eqref{eq:DA}, \eqref{eq:NI}, and \eqref{eq:CPI} hold:
\begin{enumerate}
\item[(DA)] For any $x,x' \in \bb X$, the fact that $\widetilde Y(x)=_{\mathrm{a.s.}}T_{x,x'}(\widetilde Y(x'))$ implies $\widetilde Y(x)=_{\mathrm{d}}T_{x,x'}(\widetilde Y(x'))$. Since $Y(x) \sim \bb P_{Y|X=x}$, this means $(T_{x,x'})_{\#}\,\bb P_{Y\mid X=x'}=\bb P_{Y\mid X=x}$.
\item[(NI)] For any $x \in \bb X$ we have $\widetilde Y(x)=_{\mathrm{a.s.}}T_{x,x}(\widetilde Y(x))$, so $T_{x,x}=\mathrm{id}\;$ $\bb P_{Y\mid X=x}$-a.s.
\item[(CPI)] For any $x,x',x'' \in \bb X$, we have 
\[T_{x'',x'}\circ T_{x',x}(\widetilde Y(x))
=_{\mathrm{a.s.}} T_{x'',x'}(\widetilde Y(x')) =_{\mathrm{a.s.}} \widetilde Y(x'')
=_{\mathrm{a.s.}} T_{x'',x}(\widetilde Y(x))\]
So,
$T_{x'',x'}\circ T_{x',x}=T_{x'',x}\;$ $\bb P_{Y\mid X=x}$-a.s.
\end{enumerate}
\end{proof}

\begin{proof} [\cref{thm:cocycle_factorization}]
For each $x \in \bb X$ set $\tilde f_x := T_{x,x_0}: \bb Y \to \bb Y$, where $x_0 \in \bb X$ is arbitrary. Now, let $\bb Y_{x_0}$ be the set of full $\bb P_{Y|X=x_0}$-measure on which $T_{x_0,x_0} = \id$. Since $(\bb Y, \mc B(\bb Y))$ is a Polish space and $\bb P_{Y|X=x_0}$ a probability measure on this space, it is a standard fact that any measurable subset $A \subseteq \bb Y$ with $\bb P_{Y|X=x_0}(A) = 1$ contains a Borel subset $B \subseteq A$ also with $\bb P_{Y|X=x_0}(B) = 1$. Thus, we choose $\bb Y_0$ as the Borel subset of $\bb Y_{x_0}$ of full $\bb P_{Y|X=x_0}$-measure and note that $\bb Y_0 \subseteq \bb Y$, as required. Let $f_x := \tilde f_x\!\!\restriction_{\bb Y_0} : \bb Y_0 \to \bb Y$ be the restriction of $f_x$ to $\bb Y_0$. It is Borel measurable since $\bb Y_0$ is a Borel subset of $\bb Y$. Furthermore, by \eqref{eq:NI} we have $T_{x_0,x} \circ f_x = \text{id}_{\bb Y_0}$. Therefore, $f_x:\bb Y_0\to\bb Y$ is a Borel \emph{injection}. The Lusin-Souslin theorem then implies that the set $f_x(\bb Y_0)$ is Borel and that the map $T_{x_0,x}\!\!\restriction_{f_x(\bb Y_0)}:f_x(\bb Y_0)\to\bb Y_0$ is also Borel measurable \citep[][Corr.~15.2 ]{kechris2012classical}. 

Now, we will extend $T_{x_0,x}\!\!\restriction_{f_x(\bb Y_0)}$ appropriately to define a measurable left inverse $f_x^+ : \bb Y \to \bb Y_0$ of $f_x$. To that end, choose $y_0\in\bb Y_0$ and define a total map $f_x^+:\bb Y\to\bb Y_0$ by
\[
f_x^+(y)\;:=\;
\begin{cases}
T_{x_0,x}(y), & y\in f_x(\bb Y_0),\\
y_0,      & y\notin f_x(\bb Y_0).
\end{cases}
\]
As $f_x(\bb Y_0)$ is a Borel set, $f_x^+$ is Borel measurable (with codomain $\bb Y_0$ equipped with the trace
$\sigma$-algebra $\mc B(\bb Y)\!\!\restriction_{\bb Y_0} \;:=\; \{\, \bb Y_0 \cap B : B \in \mathcal B(\bb Y) \,\}$). Moreover $f_x^+\circ f_x=\id_{\bb Y_0}$ by \eqref{eq:NI}. To show the cocycle identity, note for any $x' \in \bb X$ and $y\in f_{x'}(\bb Y_0)$ we have $f_x^+(y) = T_{x_0,x}(y)$. Since $\bb P_{Y|X=x_0}( \bb Y_0) = 1$, \eqref{eq:DA} implies
$\bb P_{Y\mid X=x'}(f_{x'}(\bb Y_0))=1$ and so we can write 
\[
f_x\circ f_{x'}^+
= T_{x,x_0}\circ T_{x_0,x'}
= T_{x,x'} \quad \bb P_{Y|X=x'}\text{-a.s.}
\]
where the last equality holds by \eqref{eq:CPI}. Since $x$ and $x'$ are arbitrary this proves the result.
\end{proof}

\begin{proof}[\cref{thm:iden:group:coboundary}]
    We prove that conditions \ref{eq:id_existence} and \ref{eq:id_uniqueness} of \cref{def:identifiability} hold. For \ref{eq:id_existence} (existence) we construct the required cocycle using a coboundary map from $\mc F_{\bb G}$. To that end, fix $x_0 \in \bb X$ and note by \eqref{eq:NI} and \eqref{eq:CPI} that any $\bb{P}_{Y|X}$-adapted, $\bb G$-valued cocycle $T: \bb X^2 \times \bb Y \to \bb Y$ satisfies 
    \begin{align}T_{x,x_0}\circ T_{x_0,x} = \id, \quad\bb P_{Y|X=x}\text{-a.s.} \label{eq:cocycle_inv}
    \end{align}
    Since $T_{x,x_0} \in \bb G$ it has an exact inverse $T_{x,x_0}^{-1}: \bb Y \to \bb Y$. Applying this inverse to both sides of \eqref{eq:cocycle_inv} gives $T_{x,x_0}^{-1} = T_{x_0,x}\;$ $\bb P_{Y|X=x}$-a.s. Thus for every $x,x' \in \bb X$ we can write
    \begin{align}
        T_{x,x'} = f_x \circ f_{x'}^{-1} \quad \bb P_{Y|X=x}\text{-a.s.}, \quad \text{where} \quad f_x = T_{x,x_0} \;. \label{eq:coboundaryrep}
    \end{align}
    Thus, \ref{eq:id_existence} holds for $f^{\star} \in \mc F_{\bb G}$ defined by $f^{\star}(x,y) = T_{x,x_0}(y)$. Now, for \ref{eq:id_uniqueness} (uniqueness), we first show that $\tilde T$ is another $\bb{P}_{Y|X}$-adapted and $\bb{G}$-valued cocycle if and only if  
    \begin{align} \label{eq:nonunique:coboundary}
       \tilde T_{x,x'} = {f}_x \circ b_x^{-1} \circ b_{x'} \circ {f}_{x'}^{-1} \;, \quad   \bb{P}_{Y|X=x'}\text{-a.s.}, \quad  \forall \; x,x'\in\bb{X} \;,
    \end{align}
    for some $\{b_x\}_{x \in \bb X} \subset \Aut(\bb P_{Y|X=x_0})|_{\bb{G}}$. 
    For the first direction, we fix $\tilde T$ and note that since \eqref{eq:DA} holds, for any $x \in \bb X$ we have 
    \[(\tilde T_{x,x_0})_{\#} \bb P_{Y|X=x_0} = \bb P_{Y|X=x} \implies (\tilde T_{x,x_0}^{-1})_{\#} \bb P_{Y|X=x} = \bb P_{Y|X=x_0}\]
    and the same holds for $T$. Thus, setting $\tilde f_x:= \tilde T_{x,x_0}$ and recalling that $f_x:= T_{x,x_0}$, we have
    \begin{align*}
        \bb{P}_{Y|X=x_0} & = (\tilde{f}_{x}^{-1})_{\#} \bb{P}_{Y|X=x} = ({f}_{x}^{-1})_{\#} \bb{P}_{Y|X=x} \;, \\   
    \implies  \bb{P}_{Y|X=x_0}& =(\tilde{f}_x^{-1} \circ f_x)_{\#} \bb{P}_{Y|X=x_0} = ({f}_x^{-1} \circ \tilde{f}_x)_{\#} \bb{P}_{Y|X=x_0} \;.
    \end{align*}
    So, $b_x :=  \tilde{f}_x^{-1} \circ {f}_x  \in \Aut(\bb P_{Y|X=x_0})|_{\bb{G}}$ and likewise for $b_x^{-1} = f_x^{-1} \circ \tilde{f}_{x}$. Now, note that
    \begin{align}
        \tilde f_x \circ \tilde f_{x'}^{-1} = {f}_x \circ b_x^{-1} \circ b_{x'} \circ {f}_{x'}^{-1}\;,\label{eq:automorph_rel}
    \end{align}
    for any $x' \in \bb X$. Moreover, following the same steps used to prove \ref{eq:id_existence}, we know $\tilde T_{x,x'} = \tilde f_x \circ \tilde f_{x'}^{-1}$ $\bb P_{Y|X=x'}$-a.s. Combining this with \eqref{eq:automorph_rel} and noting $x,x' \in \bb X$ were arbitrary proves \eqref{eq:nonunique:coboundary}. Conversely, note that for any $\{b_x\}_{x \in \bb X} \subset \Aut(\bb P_{Y|X=x_0})|_{\bb{G}}$, the construction $\hat T_{x,x'} := {f}_x \circ b_x^{-1} \circ b_{x'} \circ {f}_{x'}^{-1}$ satisfies \eqref{eq:nonunique:coboundary} and defines a valid $\bb P_{Y|X}$-adapted, $\bb G$-valued cocycle. 

    Thus, the set of $\bb G$-valued, $\bb P_{Y|X}$-adapted cocycles take the form \eqref{eq:nonunique:coboundary} for any  $\{b_x \}_{x \in \bb X} \in \Aut(\bb P_{Y|X=x_0})|_{\bb{G}}$. The cocycle is therefore (a.s.) unique if and only if
        \begin{align}
         f_x \circ  f_{x'}^{-1} & = {f}_x \circ b_x^{-1} \circ b_{x'} \circ {f}_{x'}^{-1}\; \quad \bb{P}_{Y|X=x'}\text{-a.s.} \\
         \implies b_x^{-1}\circ b_{x'} & = \id \; \quad \bb{P}_{Y|X=x_0}\text{-a.s.} \label{eq:bxequalid}
    \end{align}
    for $(\bb P_{X} \otimes \bb P_{X})$-every $x,x' \in \bb X$ and for all possible collections $\{b_x \}_{x \in \bb X} \in \Aut(\bb P_{Y|X=x_0})|_{\bb{G}}$.  The latter holds trivially if $\Aut(\bb P_{Y|X=x_0})|_{\bb{G}} \subseteq  [\id]_{\bb P_{Y|X=x_0}}$. To see that it holds only if  $\Aut(\bb P_{Y|X=x_0})|_{\bb{G}} \subseteq  [\id]_{\bb P_{Y|X=x_0}}$, suppose by contradiction that $\Aut(\bb P_{Y|X=x_0})|_{\bb{G}} \not\subset  [\id]_{\bb P_{Y|X=x_0}}$. Take any Borel set $A \in \mc B(\bb X)$ such that $\bb P_{X}(A) = 1/2$ and $g \in \Aut(\bb P_{Y|X=x_0})|_{\bb{G}} \cap  [\id]_{\bb P_{Y|X=x_0}}^c$. Using these elements, we can set
    \[
    \hat b_x \;:=\;
    \begin{cases}
        g : x \in A, \\[6pt]
    \id : x \in A^c.
    \end{cases} \quad \text{and}\quad  \hat T_{x,x'} := {f}_x \circ \hat b_x^{-1} \circ \hat b_{x'} \circ {f}_{x'}^{-1}, \quad \forall x,x' \in \bb X
    \]
    Then, $\hat T: (x,x',y') \mapsto \hat T_{x,x'}(y')$ is a valid $\bb P_{Y|X}$-adapted, $\bb G$-valued cocycle, but \eqref{eq:bxequalid} does not hold w.r.t. $(\hat b_x,\hat b_{x'})$ for $(\bb P_{X} \otimes \bb P_{X})$-every $x,x' \in \bb X$. Thus, uniqueness of the cocycle holds if and only if $\Aut(\bb P_{Y|X=x_0})|_{\bb{G}} \subseteq  [\id]_{\bb P_{Y|X=x_0}}$. This completes the proof.
\end{proof}

\begin{proof} [\cref{thm:tmi_uniqueness}]
By \cref{thm:iden:group:coboundary}, it suffices to prove that \({\Aut(\bb P_{Y|X=x_0})|_{\bb{G}_{\mathrm{TMI}}} \subseteq [ \id ]_{\bb P_{Y|X=x_0}}}\) for arbitrary $x_0 \in \bb X$. To that end, fix $x_0 \in \bb X$ and {for ease of notation put}
\(
  P:=\bb P_{Y\mid X=x_0}.
\)
To prove the result, we first prove the following lemma. 

\begin{lemma}\label{lem:1D}
Let $\nu$ be a Borel probability measure on $\bb R$.
If $S:\bb R\!\to\!\bb R$ is measurable, non-decreasing and satisfies $S_{\#}\nu=\nu$,
then $S(y)=y$, $\nu$-almost surely.
\end{lemma}

\begin{proof}[\cref{lem:1D}]
Write $F$ for the distribution function of $\nu$. Since $S$ is monotone increasing and preserves $\nu$, we know by the Portmanteau Lemma \citep[e.g.,][Lem.~2.2]{van2000asymptotic} that $S_{\#}\nu(A) = \nu(A)$ for every closed $A \in \mc B(\bb R)$. Since the family of sets $\{(-\infty, t] : t \in \bb R\}$ are closed, we have the following equalities for every $y \in \bb R$:
\begin{align*}\label{eq:CDF-equality}
  F\bigl(S(y)\bigr) = \bb P(Y \leq S(y))  = \bb P(S(Y) \leq S(y)) = \bb P(Y \leq y) =F(y) 
\end{align*}
Where the second inequality follows from the Portmanteau Lemma under measure preserving $S$ and the third inequality follows since $S$ is monotone increasing.
Now, let $Q(\alpha) = \inf \{t \in \bb R : F(t) \geq \alpha\}$ be the generalized quantile function. It is a standard fact that $Q \circ F = \id$ $\nu$-a.s. Since $S$ preserves the null-sets of $\nu$ by definition, we have $Q \circ F \circ S = S$ $\nu$-a.s. Since $Q \circ F \circ S = Q \circ F = \id$ $\nu$-a.s., this immediately implies the result.
\end{proof}

Now we use the Lemma to prove the main result by induction. Let $T=(T_1,\dots,T_p)\in\bb G_{\mathrm{TMI}}$ be measurable and satisfy $T_{\#}P=P$.
We show $T_k(y)=y_k$ $P$-a.s.\ for $k=1,\dots,p$. For the base case $k=1$, the first marginal $P^{(1)}$ is a probability measure on $\bb R$.
Because $T_1$ is non-decreasing and $(T_1)_{\#}P^{(1)}=P^{(1)}$,
Lemma~\ref{lem:1D} forces $T_1(y_1)=y_1$ $P^{(1)}$-a.s. 

For the inductive part, assume $T_j(y)=y_j$ $P$-a.s.\ for each $j\le k$, where $k \in [1,p-1]$ is arbitrary.
Let $P^{(k)} := P \circ \pi_{1:k}^{-1}$ be the projection of $P$ onto the first $k$ coordinates (i.e., $\pi_{1:k}(y) = (y_1,\dots,y_k)$). Fix $y_{\le k}$ and let
$P_{y_{\le k}}$ be the regular conditional distribution of $Y_{k+1}$ given
$Y_{\le k}=y_{\le k}$.
As the first $k$ coordinates of $Y$ are already the identity, $T_{k+1}$ satisfies
\[
  (T_{k+1})_{\#}P_{y_{\le k}}=P_{y_{\le k}},
  \quad
  x\mapsto T_{k+1}(y_{\le k},x)\ \text{non-decreasing}.
\]
Applying Lemma~\ref{lem:1D} to $\nu=P_{y_{\le k}}$ yields
$T_{k+1}(y)=y_{k+1}$ $P$-a.s.
By induction, $T=\id$ $P$-a.s. Therefore, any $T\in\bb G_{\mathrm{TMI}}$ with $T_{\#}P=P$ belongs to the
$P$-a.s.\ equivalence class of the identity, so
\(
  \Aut_{\mathrm{TMI}}(P)\subseteq[\id]_{P}.
\)
Because $x_0\in\bb X$ was arbitrary, the statement holds for every
conditional law $\bb P_{Y\mid X=x_0}$, completing the proof.
\end{proof}

\begin{proof} [\cref{thm:scm_equivalence}]
Assume the counterfactuals satisfy \cref{ass:OSA_counterfactuals} and \eqref{eq:CC} with cocycle $T$. Then by \cref{thm:cocycle_factorization} we have $Y(x) =_{\mathrm{a.s.}} f_x \circ f_{x'}^+(Y(x'))$ with $f_x := T_{x,x_0}\!\!\restriction_{\bb Y_0}$ for any $x,x',x_0 \in \bb X$, and $f_x^+$ the (measurable) left inverse of $f_x$. Now, fix $x_0 \in \bb X$ and set $\xi := f_{x_0}^+(Y(x_0)) \in \bb Y_0$, so that $Y(x) =_{\mathrm{a.s.}} f_x(\xi)$ for every $x \in \bb X$. By  \cref{ass:OSA_counterfactuals}.2 (exchangeability), we have  $X \ind \xi$. We next define $f:  (x,y)\mapsto f_x(y)$ and note since $T: (x,x',y') \mapsto T_{x,x'}(y')$ is Borel measurable so is $f$. By \cref{ass:OSA_counterfactuals}.1 (consistency) we have $Y =_{\mathrm{a.s.}} f(X,\xi)$. Lastly,  by \cref{thm:cocycle_factorization} $f_x:= f(x,\argdot)$ is injective for every $x \in \bb X$.

Conversely, if $Y =_{\mathrm{a.s.}} f(X,\xi)$ with $\xi \in \bb Y_0$, $\xi \ind X$ and $f(x,\argdot) : \bb Y_0 \to \bb Y$ injective for every $x \in \bb X$, then by definition of counterfactuals in an SCM  we have $Y(x) =_{\mathrm{a.s.}} f_{x}(\xi)$, with $f_x := f(x,\argdot)$ \citep{pearl2000models}. Since $f_x$ is injective on $\bb Y_0$, it admits a left inverse $f_x^+$. We construct this left inverse so that the map $f^+: (x,y) \mapsto f_x^+(y)$ is Borel measurable. 

To that end, define the embedding \(
F:\bb X\times\bb Y_0\to\bb X\times\bb Y,\; F(x,e)=(x,f(x,e))
\). Since $F$ is Borel and injective, by the Lusin--Souslin theorem \citep[][Corr.~15.2]{kechris2012classical}, the image
$G:=F(\bb X\times\bb Y_0)$ is Borel in $\bb X\times\bb Y$ and the inverse
$F^{-1}:G\to\bb X\times\bb Y_0$ is also Borel. Now, let $\pi_{\bb Y_0}$ be the projection onto
$\bb Y_0$ and fix $e^*\in\bb Y_0$. We define $f^+:\bb X\times\bb Y\to\bb Y_0$, 
\[
f^+(x,y):=
\begin{cases}
\pi_{\bb Y_0}\big(F^{-1}(x,y)\big), & (x,y)\in G,\\
e^\ast, & (x,y)\notin G.
\end{cases}
\]
Note that $f^+$ is Borel and $\forall\,(x,e) \in \bb X \times \bb Y_0$ we have
$(x,f(x,e))\in G$, so $f^+(x,f(x,e))=e$. Since $\xi \in \bb Y_0$, we have  $f_x^+(Y(x)) =_{\mathrm{a.s.}} f_x^+\circ f_x(\xi) =_{\mathrm{a.s.}} \xi$. Composing with $f_{x'}$ for any $x' \in \bb X$,
\begin{align*}
    Y(x') =_{\mathrm{a.s.}} f_{x'} \circ f_{x}^+(Y(x))  =: T_{x',x}(Y(x))\;,
\end{align*}
which verifies \eqref{eq:CC}. Now, since $X\ind \xi$, and independence is preserved under transformation, it holds that $Y(x) =_{\mathrm{a.s.}} f(x,\xi) \ind X$. Moreover, $Y(X) =_{\mathrm{a.s.}} f(X,\xi)$ which verifies consistency. The measurability of \(T(x',x,y):=f\bigl(x',\,f^+(x,y)\bigr)\) follows from the fact that a composition of Borel maps is Borel.
\end{proof}

\subsection{Proofs for \cref{sec:estimation}}

\begin{proof}[{\cref{thm:cocycle_recovery}}]
First note that by the definition of $T$, we have that 
\begin{align*}
    \bb P_{Y|X=x}(A) = \int_{\bb X} \bb P_{Y|X=x'}(T_{x,x'}^{-1}\{A\})\mu(dx') = \int_{\bb{X}\times\bb{Y}} \bm{1}\{T_{x,x'}(y') \in A\} \bb{P}_{Y|X=x'}(dy') \mu(dx') \;,
\end{align*} 
for every probability measure $\mu \in \mc P(\bb X)$, every measurable set $A \in \mc B({\bb Y})$ (the Borel $\sigma$-algebra on $\bb Y$) and $x \in \bb X$. By the identity $\bb E\| H\|^2 = \|\bb E H \|^2 + \text{Tr}[\text{Cov}[H]]$ for any random $H \in \mc H_{k}$, we can write the following for some $c \in \bb R$,
\begin{align*}
    \ell(T) & = \bb E \| \psi(Y) - \bb E[\psi(T_{X,X'}(Y'))|X] \|^2_{\mc H_k}  
    = \bb E \| \bb E[\psi(Y)|X] - \bb E[\psi(T_{X,X'}(Y'))|X] \|^2_{\mc H_k} +c\\
    & = \bb E_{x\sim \bb P_X}D\left(\bb P_{Y|X=x}(\argdot),  \int_{\bb X} \bb P_{Y|X=x'}(T_{x,x'}^{-1}\{\argdot\})\bb P_X(dx')\right)^2 +c = c\;,
\end{align*}
and $D = \text{MMD}$ is a metric on $\mc P(\bb Y)$ since $k$ is taken as characteristic, we have  $T \in M := \text{arginf}_{\tilde T \in \mc T_{\bb G}}\ell(\tilde T)$ and so $M$ is non-empty. Now, take arbitrary $T^* \in M$. Since its coboundary map $f^*$ lies in $\mc F_{\bb G}$, we have $T^*_{x,x} = f^*_x \circ  f^{*-1}_{x}$ (i.e., \eqref{eq:NI}) and $T^*_{x,x'} = f^*_x \circ  f^{*-1}_{x''} \circ f^*_{x''} \circ  f^{*-1}_{x'} = f^*_x\circ f^{*-1}_{x'}$ (i.e., \eqref{eq:CPI}) for every $x,x',x'' \in \bb X$. Therefore, all that remains is to show $T^*$ satisfies \eqref{eq:DA} $(\bb P_{X} \otimes \bb P_{X})$-almost everywhere. The fact that $D$ is a metric on $\mc P(\bb Y)$ and $\ell(T^*) = 0$ implies the following inequalities for arbitrary $A \in \mc B({\bb Y})$ and $\bb P_{X}$-almost all $x \in \bb X$:
\begin{align*}
    \bb P_{Y|X=x}(A) 
                           = \int_{\bb X \times \bb Y}\bm 1\{T_{x,x'}^*(y') \in A\}\bb P_{Y|X=x'}(dy')\bb P_X(dx') 
                        = \mathbb E[\bm 1\{T_{x,X}^*(Y) \in A\}] \;.
\end{align*}
Therefore, $T_{x,X}^*(Y) := f^*_{ x} \circ f_{X}^{*-1}(Y)  =_d \bb P_{Y|X=x}$, for $\bb P_X$-almost all $x \in \bb X$. Defining $\xi^* :=  f_{X}^{*-1}(Y) \sim \bb P_{\xi}^*$, this implies $\bb P_{Y|X=x}= (f^*_{x})_{\#} \bb P_\xi^*$ for $\bb P_X$-almost all $x \in \bb X$, which immediately implies the result. 
\end{proof}

\begin{proof}[{\cref{prop:MMD_prob_bound}}]
Note that for any cocycle $T: \bb X^2 \times \bb Y \to \bb Y$, $\ell^V_n(T)$ and $\ell^U_n(T)$ are $V$-statistics and $U$-statistics of order three respectively. It is a standard fact (e.g., \cite{serfling2009approximation} Sec 5) that one may replace the kernel of such statistics by its symmetrized version under $\bb S_3$, the group of permutations on $\{1,2,3\}$. That is, we may write
    \begin{align*}
         \ell^V_n(T) = \frac{1}{n^3}\sum_{i,j,k }^n h_T(Z^{(i)},Z^{(j)},Z^{(k)}), \qquad  \ell^U_n(T) = \frac{1}{{n \choose 3}}\sum_{i< j < k}^n h_T(Z^{(i)},Z^{(j)},Z^{(k)}) \;,
    \end{align*}
    where $Z^{(i)} := (X^{(i)},Y^{(i)})$ and $h_T(Z^{(i)},Z^{(j)},Z^{(k)}) = \frac{1}{6}\sum_{\sigma \in \bb S_3}\tilde h_T(Z^{(\sigma(i))},Z^{(\sigma(j))},Z^{(\sigma(k))})$ is the symmetrized version of the original kernel of the statistic:
    \begin{align*}
  \tilde h_T(Z^{(i)},Z^{(j)},Z^{(k)}) = -2k(Y^{(i)},T_{X^{(i)},X^{(j)}}(Y^{(j)})) +k(T_{X^{(i)},X^{(j)}}(Y^{(j)}), T_{X^{(i)},X^{(k)}}(Y^{(k)})) \;.
    \end{align*}
    Now, note that by the boundedness of the kernel, $\tilde h_T$ is uniformly bounded and therefore so is $h_T$. Therefore, under the assumption that  $Z^{(1)},\dots,Z^{(n)}\overset{\mathrm{iid}}{\sim} \bb P_{Z}$, by Hoeffding's inequality for bounded U-statistics (e.g., see Sec 5.6.2. Theorem A in \cite{serfling2009approximation}) we have $\ell^U_n(T) - \mathbb Eh_T(Z, Z', Z'') = \mc{O}_P(n^{-\frac{1}{2}})$. Since $f$ is bounded, it is known that $|\ell^U_n(T) - \ell^V_n(T)| = \mc O_p(n^{-\frac12})$ (e.g., see Lemma 5.7.3. in \cite{serfling2009approximation}), which immediately implies $\ell^V_n(T) - \mathbb Eh_T(Z, Z', Z'') = \mc{O}_P(n^{-\frac{1}{2}})$ also.  Note here $Z,Z',Z''\overset{\mathrm{iid}}{\sim} \bb P_{Z}$ are independent copies.
    
    All that remains is to show that $\mathbb Eh_T(Z,Z',Z'') = \ell(T) + \beta$, where $\beta$ is constant with respect to $T$. Since $\bb E h_T(Z,Z',Z'') = \bb E \tilde h_T(Z,Z',Z'')$, it suffices to show this for the latter (unsymmetrized) function). In what follows, we define $\mu(\bb P_{Y|X}(\argdot|X)) := \bb E[\psi(Y)|X]$, $\mu(\bb P_{Y_T|X}(\argdot|X)) := \bb E[\psi(T_{X,X'}(Y'))|X]$, and $Y_T := T_{X,X'}(Y')$. 
    \begin{align*}
       & \mathbb E\tilde h_T(Z,Z',Z'') 
       = -2\mathbb Ek(Y,T_{X,X'}(Y')) +\mathbb Ek(T_{X,X'}(Y'), T_{X,X'}(Y')) \\
        & = -2\mathbb E \langle \mu(\bb P_{Y|X}(\argdot|X)),\mu(\bb P_{Y_{T}|X}(\argdot|X)) \rangle_{\mc H_k} + \mathbb E \langle \mu(\bb P_{Y_{T}|X}(\argdot|X)),\mu(\bb P_{Y_{T}|X}(\argdot|X)) \rangle_{\mc H_k} \\
        & =  \mathbb E (\lVert \mu(\bb P_{Y_{T}|X}(\argdot|X))\lVert^2_{\mc H_k} -2\langle \mu(\bb P_{Y|X}(\argdot|X)),\mu(\bb P_{Y_{T}|X}(\argdot|X)) \rangle_{\mc H_k} \pm \lVert \mu(\bb P_{Y|X}(\argdot|X))\lVert^2_{\mc H_k}) \\
        & = \mathbb E \lVert \mu(\bb P_{Y|X}(\argdot|X)) - \mu(\bb P_{Y_{T}|X}(\argdot|X))\rVert^2_{\mc H_k} + \beta \\
        & = \ell(T) + \beta
\end{align*}
Where $\beta =  -\mathbb E \lVert \mu(\bb P_{Y|X}(\argdot|X))\lVert^2_{\mc H_k}$. This completes the proof.
\end{proof}

\begin{proof}[{\cref{thm:CMMD_consistency}}]
For notational convenience let $Z := (X,Y)$, $Z^{(1)},\dots,Z^{(n)}\overset{\mathrm{iid}}{\sim} \bb P_{Z}$ and $\ell_n := \ell_n^U$. We prove the result by extending known results for convergence in probability (e.g., Theorem 5.7 in \cite{van2000asymptotic}) to the case of $U$-statistics and a minimizing set $M$ rather than a unique minimizer $\theta_0$. Now, since $\ell_n^U(\theta)$ is an order-3 U-statistic, it is a standard fact \citep[][Sec.~5]{serfling2009approximation} that we can express it as 
\[
  \ell(\theta)=\mathbb{E}[h_\theta(Z,Z',Z'')],\quad
  \ell_n(\theta)=\frac{1}{{n \choose 3}}\sum_{i< j < k}h_\theta(Z^{(i)},Z^{(j)},Z^{(k)}),
\]
where $h_{\theta}(Z^{(i)},Z^{(j)},Z^{(k)}) = \frac{1}{6}\sum_{\sigma \in \bb S_3}\tilde h_\theta(Z^{(\sigma(i))},Z^{(\sigma(j))},Z^{(\sigma(k))})$ is the symmetrized version of the original kernel of the statistic: 
\begin{align*}
 \tilde h_{\theta}(Z^{(i)},Z^{(j)},Z^{(k)})= -2k(Y^{(i)},T_{\theta,X^{(i)},X^{(j)}}(Y^{(j)}))+k(T_{\theta,X^{(i)},X^{(j)}}(Y^{(j)}),T_{\theta,X^{(i)},X^{(k)}}(Y^{(k)})).
\end{align*}

Define $M=\arg\min_{\theta\in\Theta}\ell(\theta)$. We start by showing $\ell$ is continuous and $M$ is compact. By Assumption~\ref{ass2:consistency}.4 the kernel $k$ is continuous and bounded, so the composite $h_\theta$ is uniformly bounded by a constant $C<\infty$ and, by Assumption~\ref{ass2:consistency}.2, continuous in $\theta$. Therefore, for any sequence $(\theta_n)_{n \geq 1} \in \Theta$ such that $\theta_n \to\theta$, we have point-wise convergence $h_{\theta_n}(z,z',z'')\to h_\theta(z,z',z'')$ and the uniform bound $|h_{\theta_n}(z,z',z'')|\le C$. Dominated Convergence then gives
\[
  \ell(\theta_n)=\mathbb{E}[h_{\theta_n}(Z,Z',Z'')]\to\mathbb{E}[h_\theta(Z,Z',Z'')]=\ell(\theta).
\]
Hence $\ell$ is continuous on $\Theta$ and, since $\Theta$ is compact, by the Weierstrass extreme-value theorem $\ell$ attains its minimum. Therefore, $M=\ell^{-1}(\{\min_{\vartheta\in\Theta}\ell(\vartheta)\})$ is the inverse image of a closed set under a continuous map, and so $M$ is closed in $\Theta$ and compact by the Heine–Borel Theorem. Now, for strong consistency of $\hat \theta_n$, we will show the following properties:
\begin{align}
 \text{Well-separatedness:}  \quad &
      \displaystyle\inf_{\theta \in \Theta:\inf_{\theta' \in M}\|\theta - \theta'\|_2\ge\varepsilon}\bigl[\ell(\theta)-\min_{\vartheta\in\Theta}\ell(\vartheta)\bigr]=\delta_\varepsilon>0 \label{eq:ws}\\
\text{Uniform convergence:} \quad & \displaystyle\sup_{\theta\in\Theta}|\ell_n(\theta)-\ell(\theta)|\to0\quad\text{a.s.} \label{eq:uc}
\end{align}

For well-separatedness, fix $\varepsilon>0$ and set
\(A_\varepsilon=\{\theta\in\Theta:d_M(\theta)\ge\varepsilon\}\), \text{where} \( d_M(\theta):=\inf_{\theta' \in M}\|\theta - \theta'\|_{2}.
\)
By the triangle inequality, the map $d_M:\mathbb R^{d}\to[0,\infty)$ is $1$-Lipschitz, hence continuous. Therefore $A_\varepsilon=d_M^{-1}([\varepsilon,\infty))$ is closed in $\mathbb R^{d}$ and is compact since $\Theta$ is compact. Now, define
\[
  \ell^\star:=\min_{\vartheta\in\Theta}\ell(\vartheta),\qquad
  \ell^\varepsilon:=\min_{\vartheta\in A_\varepsilon}\ell(\vartheta).
\]
Continuity of $\ell$ on compact $A_\varepsilon$ guarantees $\ell^\varepsilon$ exists and, as $A_\varepsilon$ is disjoint from $M$, $\ell^\varepsilon>\ell^\star$. Define $\delta_\varepsilon:=\ell^\varepsilon-\ell^\star>0$. For any $\theta\in\Theta$ with $d_M(\theta)\ge\varepsilon$ we have $\ell(\theta)\ge \ell^\varepsilon$, hence \eqref{eq:ws} holds.

For uniform convergence of $\ell_n(\theta)$ to $\ell(\theta)$, we note that $\mc F:= \{h_\theta:\theta\in\Theta\}$ is uniformly bounded, and continuous in $\theta$, and $\Theta$ is compact. Such classes are known to be Glivenko--Cantelli and so have a finite bracketing number $N_{[]}(\epsilon,\mc F, L^2(\bb P_{Z}^3) < \infty$) \citep[e.g.,][Example 19.8]{van2000asymptotic}. Since $\mc F$ is continuous it is also Borel-measurable. This along with the finite bracketing number, means that the uniform strong law of large numbers \eqref{eq:uc} holds for the U-statistic $\ell_n^U(\theta)$ \citep[][Corr.~5.2.5]{de2012decoupling}.

Now we are ready to prove the consistency result. Since uniform convergence holds, for every $\varepsilon>0$ there exists an almost-surely finite random index
$N_\varepsilon$ such that, for all $n\ge N_\varepsilon$,
\begin{equation}\label{eq:uniform_ineq}
  \sup_{\theta\in\Theta}\bigl|\ell_n(\theta)-\ell(\theta)\bigr|
      <\delta_\varepsilon/2 .
\end{equation}

We work on the full-probability event $\{N_\varepsilon<\infty\}$ and fix any
$n\ge N_\varepsilon$.  Suppose, for contradiction, that
$d_M(\hat\theta_n)\ge\varepsilon$, so $\hat\theta_n\in A_\varepsilon$.
Combining \eqref{eq:uniform_ineq} with \eqref{eq:ws} gives
\[
  \ell_n(\hat\theta_n)
        \;\ge\;\ell(\hat\theta_n)-\delta_\varepsilon/2
        \;\ge\;\ell^\star+\delta_\varepsilon/2,
\]
whereas for any minimizer $\theta^\star\in M$ we have
\(
  \ell_n(\theta^\star)\le\ell^\star+\delta_\varepsilon/2
\)
by \eqref{eq:uniform_ineq}, contradicting the minimality of $\hat\theta_n$.
Hence $d_M(\hat\theta_n)<\varepsilon$ for all $n\ge N_\varepsilon$. Since $\varepsilon>0$ is arbitrary, we conclude that \(d_M(\hat\theta_n)\to 0  \text{ a.s.}\). This proves the consistency result for $\hat \theta_n$.

Lastly, we transfer the result to the cocycle $T_{\hat \theta_n}$. Since $M$ is compact and $\theta\mapsto\|\hat\theta_n-\theta\|$ is continuous, by the Measurable Maximum Theorem (Theorem 18.19 in \cite{aliprantis:border:2006}) one can define the measurable function $\eta_n:=\arg\min_{\theta\in M}\|\hat\theta_n-\theta\|$ and a.s. consistency of $\hat \theta_n$ implies $\|\hat\theta_n-\eta_n\|\to0$ almost surely. By continuity of $T$ in $\theta$, this implies 
\begin{align}
T_{\hat\theta_n,x,x'}(y)\to T_{\eta_n,x,x'}(y) \quad \text{a.s.} \label{eq:cocycle_consistency}
\end{align}
    Since $\bb N$ is countable, by Assumption~\ref{ass2:consistency}.3 we have $\{T_{\eta_n,x,x'}(y)=T_{\theta_0,x,x'}(y),\ \forall n\in\mathbb N\}$ $(P_X\otimes P_{X,Y})$-a.s., for any $\theta_0\in M$. Combining this with \eqref{eq:cocycle_consistency} yields point-wise convergence $T_{\hat\theta_n,x,x'}(y)\to T_{\theta_0,x,x'}(y)$ on a full-measure set, which completes the proof.
\end{proof}

\begin{proof}[\cref{thm:cmmd-rate}]
For notational convenience, let $Z := (X,Y)$, $Z^{(1)},\dots,Z^{(n)}\overset{\mathrm{iid}}{\sim} \bb P_{Z}$ and note we can express the gradient of the U-statistic $\ell_n(\theta) := \ell_n^U(\theta)$ as
\[\displaystyle\nabla_\theta\ell_n(\theta)
=\frac{1}{n(n-1)(n-2)}\sum_{i\neq j\neq k}\nabla_{\theta} f_\theta(Z^{(i)},Z^{(j)},Z^{(k)}).\] 
Now, define the symmetrized and centered function
\[
H_{\theta}(z^{(1)},z^{(2)},z^{(3)})
:=\frac{1}{6}\sum_{\pi\in \bb S_{3}}
\nabla_{\theta} f_\theta\bigl(z_{\pi(1)},z_{\pi(2)},z_{\pi(3)}\bigr)
\;-\;\nabla_\theta\ell(\theta) \,.
\]
By standard theory of U-statistics \citep[e.g.,][Sec.~5]{serfling2009approximation}, we can express $\nabla_\theta\ell_n(\theta)-\nabla_\theta\ell(\theta)$ as a centered U-statistic with symmetric kernel $H_\theta$:
\begin{align}
 \nabla_\theta\ell_n(\theta)-\nabla_\theta\ell(\theta) &  
 =
 \tfrac{1}{\binom{n}{3}}\sum_{i< j < k}H_{\theta}(Z^{(i)},Z^{(j)},Z^{(k)}) \;. \label{eq:symmetrized_U_CMMD}
\end{align}
Now, note that if  $\nabla_\theta \ell(\theta) = \bb E [\nabla_{\theta} f_\theta(Z^{(1)},Z^{(2)},Z^{(3)})]$, by the Hoeffding decomposition of $H_\theta$ \citep[][Sec 5.1.5., Lemma A]{serfling2009approximation},  there exist  zero-mean symmetric projections
\[
\begin{aligned}
h_{1,\theta}(z)&=\bb E[\,H_{\theta}(z,Z^{(2)},Z^{(3)})\,],\\
h_{2,\theta}(z^{(1)},z^{(2)})
&=\bb E[\,H_{\theta}(z^{(1)},z^{(2)},Z^{(3)})\,]
 \;-\;h_{1,\theta}(z^{(1)})-h_{1,\theta}(z^{(2)}),\\
h_{3,\theta}(z^{(1)},z^{(2)},z^{(3)})
&=H_{\theta}(z^{(1)},z^{(2)},z^{(3)})
-\sum_{i=1}^3h_{1,\theta}(z^{(i)})
-\sum_{1\le i<j\le3}h_{2,\theta}(z^{(i)},z^{(j)}),
\end{aligned}
\]
which enables us to write 
\[
H_{\theta}(z^{(1)},z^{(2)},z^{(3)})
=\sum_{i=1}^3h_{1,\theta}(z^{(i)})
\;+\;
\sum_{1\le i<j\le3}h_{2,\theta}(z^{(i)},z^{(j)})
\;+\;
h_{3,\theta}(z^{(1)},z^{(2)},z^{(3)}).
\]

We will make use of this representation of $H_\theta$ to split $\nabla_\theta \ell_n(\theta) - \nabla_\theta\ell(\theta)$ into a first-order term and remainder term. To do this, we first show that we can swap the expectation and the gradient so that $\nabla_\theta \ell(\theta) = \bb E [\nabla_{\theta} f_\theta(Z^{(1)},Z^{(2)},Z^{(3)})]$ via the dominated convergence theorem (DCT) (i.e., $\nabla_{\theta} f_\theta < \kappa_\theta : \bb E[|\kappa_\theta(Z^{(1)},Z^{(2)},Z^{(3)})|] < \infty$). In particular, note that by \cref{ass3:normality}.2 we have $\partial k < B$, so by the chain rule
\begin{align*}
    & \nabla_{\theta} f_\theta(Z^{(1)},Z^{(2)},Z^{(3)})  = \nonumber \\
    & \quad \quad \quad -2\nabla_\theta k(Y^{(1)},T_{\theta, X^{(1)},X^{(2)}}(Y^{(2)}) ) + \nabla_\theta k(T_{\theta, X^{(1)},X^{(2)}}(Y^{(2)}),T_{\theta, X^{(1)},X^{(3)}}(Y^{(3)})) \\
   &  \quad \leq 2 B\nabla_\theta T_{\theta, X^{(1)},X^{(2)}}(Y^{(2)}) + B\nabla_\theta (T_{\theta, X^{(1)},X^{(2)}}(Y^{(2)}) + T_{\theta, X^{(1)},X^{(3)}}(Y^{(3)}))
\end{align*}
Similarly, by \cref{ass3:normality}.1, we have $\nabla _\theta T_{\theta, X^{(1)},X^{(3)}}(Y^{(3)}) \leq L_T(X^{(1)},X^{(3)},X^{(3)})$.
Thus, we can set 
\[\kappa_\theta(Z^{(1)},Z^{(2)},Z^{(3)}) = B(3L_T(X^{(1)},X^{(2)},Y^{(2)})+L_T(X^{(1)},X^{(3)},Y^{(3)}))\;,\]
and it is clear by the integrability of $L_T$ that $\bb E[|\kappa_\theta(Z^{(1)},Z^{(2)},Z^{(3)})] < \infty$. Therefore, the DCT applies and $\nabla_\theta \ell(\theta) = \bb E [\nabla_{\theta} f_\theta(Z^{(1)},Z^{(2)},Z^{(3)})]$. Replacing $H_\theta$ in \eqref{eq:symmetrized_U_CMMD} with its representation in terms of the projections, we get the standard ANOVA decomposition
\begin{align*}
& \nabla_\theta\ell_n(\theta)-\nabla_\theta\ell(\theta)
= \\
& \qquad \frac{3}{n}\sum_{i=1}^nh_{1,\theta}(Z^{(i)})
\;+\;
\frac{6}{n(n-1)}\sum_{i<j}h_{2,\theta}(Z^{(i)},Z^{(j)})
\;+\;
\frac{1}{\binom{n}{3}}\sum_{i<j<k}h_{3,\theta}(Z^{(i)},Z^{(j)},Z^{(k)}).
\end{align*}
Since \(\bb E[h_{1,\theta}(Z)]=0\), multiply by \(\sqrt n\) and set
\[
g_{1,\theta}(z)=3\,h_{1,\theta}(z),
\quad
R_{n,\theta}
=\sqrt n\Bigl\{\tfrac{6}{n(n-1)}\sum_{i<j}h_{2,\theta}(Z^{(i)},Z^{(j)})
+\tfrac{1}{\binom{n}{3}}\sum_{i<j<k}h_{3,\theta}(Z^{(i)},Z^{(j)},Z^{(k)})\Bigr\}.
\]
This allows us to obtain the desired split into leading and remainder terms \[
\sqrt n\bigl(\nabla\ell_n(\theta)-\nabla\ell(\theta)\bigr)
=\frac{1}{\sqrt n}\sum_{i=1}^n
g_{1,\theta}(Z^{(i)})
\;+\;
R_{n,\theta}.
\tag{HD}
\]
We now bound each term accordingly. For the remainder term, it is known that, by construction, \(h_{2,\theta},h_{3,\theta}\) are \(\bb P\)-degenerate \citep[][pp.~178]{serfling2009approximation}. Additionally, since $\Theta \subset \bb R^d$ is compact, the packing number is $\mc M(\epsilon, \Theta, \|\argdot \|_2)  \leq D/\epsilon^d$, where $D = Diam(\Theta)$. This polynomial dependence on $\epsilon$ lets us apply Sherman’s maximal inequality for degenerate U-statistics
\citep[][Corr.~4]{sherman1994maximal}:
\[
\sup_{\theta\in U_\delta}\|R_{n,\theta}\|
=O_p\bigl(n^{-3/2}\bigr)
=o_p(1).
\tag{U}
\]
For the first order term, by Assumptions~\ref{ass3:normality}(i)–(ii),
\(\{g_{1,\theta}:\theta\in U_\delta\}\) is a parametric Lipschitz class and so is known to be \(\bb P\)‐Donsker
\cite[Example  19.7]{van2000asymptotic}.  Therefore
\[
\sup_{\theta\in U_\delta}
\biggl\|\frac{1}{\sqrt n}\sum_{i=1}^n
\bigl(g_{1,\theta}(Z^{(i)})-\bb E[g_{1,\theta}(Z)]\bigr)\biggr\|
=O_p(1).
\tag{D}
\]

Combining (HD), (U) and (D) yields
\(
\sup_{\theta\in U_\delta}
\|\nabla\ell_n(\theta)-\nabla\ell(\theta)\|
=O_p(n^{-1/2}).
\) Since \(\hat\theta_n\) minimizes \(\ell_n\),
\(\nabla\ell_n(\hat\theta_n)=0\) implies
\(\|\nabla\ell(\hat\theta_n)\|=\|\nabla\ell_n-\nabla\ell\|(\hat\theta_n)=O_p(n^{-1/2})\).
Since \cref{ass2:consistency} holds, so does  \cref{thm:CMMD_consistency} and so we know that \(\hat\theta_n\in U_\delta\) w.p.\,1. Local strong convexity (Assumption~\ref{ass3:normality}(iii)) then yields
\(\|\nabla\ell(\hat\theta_n)\|\ge c\,d_{\hat\theta_n}(M)\) for all $n \geq N(\delta)$, where $N(\delta) \in \bb N$.  Hence, for all such $n$, we have
\[
d_{\hat\theta_n}(M)\le c^{-1}\|\nabla\ell(\hat\theta_n)\|
=O_p(n^{-1/2})\;.
\]
\end{proof}

\subsection{Proofs for \cref{sec:cocycle:vs:scm}}
\begin{proof}[\cref{thm:min_complexity}]
By definition of \(\bb G_{f^{(g)}}\), one of its generators is
\(f_{x_{0}}\circ g = \id_{\bb Y}\circ g = g\).  Hence
\(
  g, g^{-1} \;\in\;\bb G_{f^{(g)}}\).  Since for each \(x\in\bb X\), we have
\(   f_{x}
  \;=\;
  \bigl( f_{x}\circ g\bigr)\,\circ\,g^{-1},
\) and \(f_{x}\circ g\) is a generator of \(\bb G_{f^{(g)}}\), it follows that \(f_{x}\in\bb G_{f^{(g)}}\) for all \(x \in \bb X\).  Hence
\(  \bb G_{f} \;=\;\bigl\langle f_{x}:x\in\bb X\bigr\rangle
  \;\subseteq\;
  \bb G_{f^{(g)}}\). This means that any $\bb G_f$-valued coboundary map is also $\bb {G}_{f^{(g)}}$-valued,     and so \(\mc F_{\bb G_{f}} \subseteq \mc F_{\bb G_{f^{(g)}}}\). This proves (i). To prove (ii), take by hypothesis \(g\notin\bb G_{f}\). In this case,
\(  \bb G_{f^{(g)}} \;\not\subset\;\bb G_{f}\). Define $\tilde f: (x,y) \mapsto g(y)$ and note since $g \in \bb G_{f^{(g)}}$ it is a $\bb G_{f^{(g)}}$-valued coboundary map. Since $g \notin \bb G_f$ we have $\tilde f \notin \mc F_{\bb G_f}$ and so \(\mc F_{\bb G_{f^{(g)}}} \not\subset\;\mc F_{\bb G_{f}} \). Combining with (i) completes the proof.
\end{proof}

\section{Algorithms} \label{app:algs}

\begin{minipage}{0.5\textwidth}
    \begin{algorithm}[H]
\SetAlgoLined
\caption{Minibatch CMMD-V optimization.}
\KwData{
  samples $\{(x^{(i)},y^{(i)})\}_{i=1}^n$,  
  cocycle $T_\theta$,  
  kernel $k$,
 batch size $B$, epochs $E$, step-size $\eta$
}
\KwResult{Optimized cocycle $T_\theta^*$}

\BlankLine
\For{epoch $=1,\dots,E$}{
  $s \gets 0$ \tcp*{samples counter}
  \While{$s < n$}{
    Sample without replacement a set $\mathcal B\subset\{1,\dots,n\}$, $|\mathcal B|=B$\;
    \ForAll{$i,j\in\mathcal B$}{
      $y_\theta^{(i,j)} \gets T_{\theta,x^{(i)},x^{(j)}}(y^{(j)})$ 
    }
    
      $\ell^V_b(\theta)
      \gets  \frac{1}{B^3}\sum_{i,j,k\in\mathcal B} k\bigl(y_\theta^{(i,j)},\,y_\theta^{(i,k)}\bigr) -\frac{2}{B^2}
        \sum_{i,j\in\mathcal B} k\bigl(y^{(i)},\,y_\theta^{(i,j)}\bigr)$\;
    $\theta \;\leftarrow\;\theta - \eta\,\nabla_\theta \ell^V_b(\theta)$\;
    $s \;\leftarrow\; s + B$\;
  }
}
\Return{$T_\theta^*$}
\label{alg:cmmd}
\end{algorithm}
\end{minipage}
\begin{minipage}{0.49\textwidth}
    \begin{algorithm}[H]
\SetAlgoLined
\caption{Cocycle model selection via K-fold cross-validation on CMMD.}
\KwData{
  samples $\{(x^{(i)},y^{(i)})\}_{i=1}^n$,  
  cocycles $\{T_m\}_{m=1}^M$,  
  folds $K$
}
\KwResult{
  Selected cocycle $T^*$
}
\BlankLine
1. Partition $\{1,\dots,n\}$ into $K$ disjoint folds $\{I_{\text{tr}}^{(k)},I_{\text{val}}^{(k)}\}_{k=1}^K$\;  
2. \For{$m=1$ \KwTo $M$}{
     \For{$k=1$ \KwTo $K$}{
       2.1. Train $T_m$ on $\{i\in I_{\text{tr}}^{(k)}\}$ by minimizing CMMD\;  
       2.2. Evaluate $\ell_{m,k} \leftarrow \text{CMMD}\bigl({T_m},\,\{i\in I_{\text{val}}^{(k)}\}\bigr)$\;
     }
     2.3. Compute $\bar\ell_m \leftarrow \tfrac1K\sum_{k=1}^K \ell_{m,k}$\;
}
3. $m^* \leftarrow \arg\min_{m}\bar\ell_m$\;  
4. Retrain ${T_{m^*}}$ on all data by minimizing CMMD\;  
5. \Return{$T^*$}
\label{alg:CV}
\end{algorithm}

\end{minipage}

\bibliography{references}

\end{document}